\documentstyle[aps,pre,fleqn,epsf,psfig]{revtex}
\newcommand{\hide}[1]{}

\def\bbbq{{\mathchoice {\setbox0=\hbox{$\displaystyle\rm Q$}\hbox{\raise
0.15\ht0\hbox to0pt{\kern0.4\wd0\vrule height0.8\ht0\hss}\box0}}
{\setbox0=\hbox{$\textstyle\rm Q$}\hbox{\raise
0.15\ht0\hbox to0pt{\kern0.4\wd0\vrule height0.8\ht0\hss}\box0}}
{\setbox0=\hbox{$\scriptstyle\rm Q$}\hbox{\raise
0.15\ht0\hbox to0pt{\kern0.4\wd0\vrule height0.7\ht0\hss}\box0}}
{\setbox0=\hbox{$\scriptscriptstyle\rm Q$}\hbox{\raise
0.15\ht0\hbox to0pt{\kern0.4\wd0\vrule height0.7\ht0\hss}\box0}}}}
\def\bbbz{{\mathchoice {\hbox{$\sf\textstyle Z\kern-0.4em Z$}}
{\hbox{$\sf\textstyle Z\kern-0.4em Z$}}
{\hbox{$\sf\scriptstyle Z\kern-0.3em Z$}}
{\hbox{$\sf\scriptscriptstyle Z\kern-0.2em Z$}}}}

\begin{document}
\newcounter{num}
\def\theequation{\arabic{section}.\arabic{equation}} 
\global\firstfigfalse


\title{Quantal-Classical Duality and the 
  Semiclassical Trace Formula}
\author{Doron Cohen, Harel Primack and Uzy Smilansky}
\date{August 1997}
\address{Department of Physics of Complex Systems, \\ 
  The Weizmann Institute of Science, Rehovot 76100, Israel.} 
\maketitle

\hide{ \vspace*{-0.5cm} }
\begin{abstract}
\hide{ \vspace*{-0.5cm} \setlength{\baselineskip}{2mm} }

We consider Hamiltonian systems which can be described both classically
and quantum mechanically. Trace formulas establish links between
the energy spectra of the quantum description, and the spectrum
of actions of periodic orbits in the Newtonian description. This is
the duality which we investigate in the present paper. The duality holds
for chaotic as well as for integrable systems.

Billiard systems are a very convenient paradigms and we use them
for most of our discussions. However, we also show how to transcribe
the results to  general Hamiltonian systems.
In billiards, it is natural to think
of the quantal spectrum (eigenvalues of the Helmholtz equation)
and the classical spectrum (lengths of periodic orbits)
as two  manifestations of the properties of the billiard boundary.
The trace formula express this
link since  it can be thought of as a Fourier
transform relation between the classical and the quantum spectral densities.
It follows that the two point statistics of the quantal
spectrum is related to the two point statistics of the
classical spectrum via a double Fourier transform.
The universal correlations of the quantal spectrum are
well known, consequently one can deduce the classical
universal correlations. In particular, an explicit expression
for the scale of the classical correlations is derived and
interpreted. This allows a further extension of the formalism
to the case of complex billiard systems, and in particular
to the most interesting case of diffusive system. The effects
of symmetry and symmetry-breaking are also discussed.

The concept of classical correlations allows a better
understanding of the so-called diagonal approximation
and its breakdown. It also paves the way towards a
semiclassical theory that is capable of global
description of spectral statistics beyond the breaktime.
An illustrative application is the derivation of the
disorder-limited  breaktime in case of a disordered chain,
thus obtaining a semiclassical theory for localization.
We also discuss other applications such as the two-cell systems,
periodic chains and localization theory in more than one
dimension.

A numerical study of classical correlations in the case
of the 3D Sinai billiard is presented. Here it is
possible to test some assumptions and conjectures
that underly our formulation. In particular we gain
a direct understanding of specific statistical properties
of the classical spectrum, as well as their semiclassical
manifestation in the quantal spectrum.

We also analyze the spectral duality for
integrable systems, and show that the Poissonian statistics of both
the classical and the quantum spectra can be traced to the same origin.
\hide{ \footnote{
Number of manuscript pages: 82, figures: 17, tables: 0. 
}}

\end{abstract}


\hide{

\newpage

{\bf Page 2 - Reference Page}

\ \\ \ \\
{\it Propose running head 
(abbreviated form of the title):} \\
spc

\ \\ \ \\ \ \\
{\it Name and mailing address of the author 
to whom proofs should be sent:} \\
\ \\ 
Doron Cohen \\
Department of Physics of Complex Systems \\
The Weizmann Institute of science \\
Rehovot 76100, Israel. \\
E-mail: fndoron1@wicc.weizmann.ac.il \\
Tel: +972-8-934-3202  \\
Fax: +972-8-934-4157 \\

\newpage
   
}

\section{Introduction}
\setcounter{equation}{0}

Trace formulas  manifest an intimate link between
two seemingly unrelated spectra which pertain to the quantum and the
classical descriptions of a given Hamiltonian system. The quantum spectrum
of eigenenergies is described in terms of the spectral density
\begin{equation}
d_{qm}(E;\hbar) \ = \ 
\sum_{n=1}^{\infty} \delta \left[E-E_n(\hbar)\right]
\label {introd/rhoqm}
\end{equation}
and it depends  parametrically on Planck's constant. 
The classical spectrum of actions characterizes the set of 
periodic manifolds at an energy $E$. These are isolated periodic 
orbits (POs) in case of a chaotic system or periodic (rational) tori
in case of integrable system. The actions are expressed as contour 
integrals in the $2d$ dimensional phase space
\begin{equation}
S_{\alpha}(E) \ = \ 
\oint_{C_{\alpha}} 
\sum_{i=1}^d  p_i {\rm d} q_i
\label {introd/actions}
\end{equation}
where $C_{\alpha}$ is a closed trajectory on the periodic manifold
(PO in the case of a chaotic system).  
Note that the periodic manifolds, as well as their 
actions, depend parametrically on the chosen energy $E$. 
One can define a density of classical actions
\begin{equation}
d_{cl}(S;E) \ = \ 
\sum_{\alpha} A_{\alpha} \delta \left[S-S_{\alpha}(E)\right]
\label {introd/rhocl}
\end{equation}
and with an appropriate choice of the coefficients $A_{\alpha}$ the trace
formulas are expressed as
\begin{equation}
d_{qm,osc}(E;\hbar) \ \sim \  
(2\pi \hbar)^{-f} 
\sum_{\alpha} A_{\alpha} {\rm e}^{iS_{\alpha}/\hbar}
\ = \
(2\pi \hbar)^{-f} 
\int {\rm e}^{iS/{\hbar}} d_{cl,osc}(S;E) \ {\rm d}S
\label {introd/trace}
\end{equation}
where $d_{qm,osc} \equiv d_{qm}-\bar{d}_{qm}$ and 
$\bar{d}_{qm}(E;\hbar)$ is is the smoothed spectral density.
The power $f$ depends on the dimensionality of the classical 
periodic manifold: it is $1$ for isolated POs in 
chaotic systems, and $(d+1)/2$ for periodic tori in integrable 
systems. The symbol $\sim$ indicates that
the right hand side  provides the leading term in an asymptotic expansion
in $\hbar$ of the left hand side. This is the semiclassical
trace formula (SCTF) which is the main object of our discussion.
Explicit expressions for the semiclassical  $A_{\alpha}$
coefficients  for  chaotic and for integrable systems were derived
by Gutzwiller \cite {gutzwiller} and Berry and Tabor \cite{BerryTabor},
respectively.  It should be noted, that there are few cases where 
the symbol $\sim$ can be replaced  by an equality sign. 
One example is the Selberg trace formula for the modular domains in the
hyperbolic plane. A second example is the rectangular billiards 
in any dimension. The higher terms in the $\hbar$ expansion are 
known in some cases \cite{Gaspard} and are introduced by adding 
to $A_{\alpha}$ a power series in $\hbar$. For a general billiard 
there is a theorem by Andersson and Melrose \cite{melrose} that states 
that exist $A_{\alpha}$ for which (\ref{introd/trace}) is an exact 
equality.

It should be emphasized that both sides of the SCTF (\ref {introd/trace}) 
are distributions rather than functions. Therefore one cannot use it 
directly but only after it is integrated over an appropriate test 
or window function. Once this is done one can use the
SCTF to study the quantal spectral statistics of a finite spectral interval 
in terms of the properties of the classical dynamics which are embedded in
$d_{cl}$ \cite{berry_a400,Pcl,thomas}. Consider the
two point spectral form factor, which is calculated for the spectral 
window $w(E-\bar{E})$ whose effective width is $\Delta E$. For $t>0$  
\begin{equation}
K(t;\bar E) \ \equiv \ \frac{1}{\bar{d}_{qm} \Delta E}
\left| \sum _n w(E-E_n) {\rm e }^{iE_nt/\hbar}\right|^2
\label{introd/K(t)}
\end{equation}
The inner summation over $n$ can be replaced by an integration over $E$ 
with the density $d_{qm}(E)$. Substituting the SCTF for $d_{qm}(E)$ and 
assuming that $\Delta E$ is sufficiently small on the
classical scale, but large enough on the quantal scale,
one obtains:
\begin{equation}
K(t; \bar E) \ \approx \ 
\frac{(2\pi \hbar)^{-f}}{\bar{d}_{qm}\Delta E} 
\left| \sum _{\alpha} \tilde{w}(t-T_{\alpha})
A_{\alpha} {\rm e}^{iS_{\alpha}(\bar E)/\hbar} \right|^2
\label{introd/SClK(t)}
\end{equation}
Here $\tilde{w}$ is the Fourier transform of the energy window 
function, and it picks out of the PO sum only those 
POs whose period $T_{\alpha}$ approach $t$ to within
$\hbar/\Delta E$. Note that the sum over POs in
(\ref{introd/SClK(t)}) absolutely converges for a suitably chosen 
window. 
The absolute square of the sum in (\ref {introd/SClK(t)}) is 
composed of two kinds of contributions. The contribution of 
the diagonal terms is 
\begin{equation}
K_D (t,\bar E) = \frac{(2\pi \hbar)^{-f}}{\bar{d}_{qm}\Delta E} 
\sum _{\alpha}  \left[ \tilde{w}(t-T_{\alpha}) \right]^2  
\left| A_{\alpha} \right|^2
\label{introd/KD(t)}
\end{equation}
and the contribution of the non diagonal terms is
\begin{equation}
K_{ND}(t;\bar E) = 
\frac{(2\pi \hbar)^{-f}}{\bar{d}_{qm}\Delta E} 
\sum _{\alpha \ne \beta } 
\tilde{w}(t-T_{\alpha})\tilde{w}(t-T_{\beta}) \
A_{\alpha}A_{\beta}^{\star} \
{\rm e }^{i[S_{\alpha}-S_{\beta}]/\hbar}
\label{introd/KND(t)}
\end{equation}
Berry \cite{berry_a400} was the first to observe that for times much shorter than the
Heisenberg time $t_H(\bar E) = 2\pi\hbar\bar{d}_{qm}$ the
diagonal contribution dominates. 
Using the Hannay and Ozorio de Almeida sum-rule \cite{Hannay} he was 
able to reproduce the expression of $K(t)$ for $t<t_H$ as 
derived from random matrix theory (RMT) 
for generic chaotic systems, and the expression derived for 
a Poissonian spectrum for generic integrable systems. 
This approach has been generalized later \cite{Pcl} 
by observing that the diagonal sum is related to the classical 
probability to return \cite{Pcl}.

The contribution of the non diagonal sum (\ref{introd/KND(t)}) 
is by no means small for $t \ge t_H$. As a matter of fact it should 
contain a term which (for chaotic systems) cancels the monotonically 
increasing $K_D\propto t$ so that the correct asymptotic behavior 
$K(t)\rightarrow 1$ is achieved for $t > t_H$.
This asymptotic behavior reflects the fact that the 
density $d_{qm}(E)$ corresponds to a discrete spectrum.
Note that if $d_{qm}(E)$ were smooth, then $K(t)$ would 
asymptotically vanish. The asymptotic non-zero value is 
therefore a hallmark of the discreteness.      
Argaman et al.\,\cite {argaman} have related $K_{ND}(t)$ to the 
two point correlations of the classical spectrum. 
The distribution of the classical actions differences 
$[S_{\alpha}-S_{\beta}]$, with the restriction that the periods
are confined to the vicinity of the time $t$, 
has been determined by taking an inverse Fourier transform 
of (\ref{introd/KND(t)}) in the variable $1/\hbar$.
For chaotic systems in 2D, the classical two-point correlations 
scales into a universal function, provided 
the action differences are expressed in units of
${1\over t}\left.{\partial\Omega\over\partial E}\right|_{\bar E}$, 
where $\Omega (E)$ is the volume of the phase space with energy $\le E$. 
This prediction was checked numerically for a few chaotic systems, and  
its relation to the Hardy-Littlewood conjecture on pair correlations of 
primes was also discussed \cite{primes}. 

The important new element in \cite{argaman} was that a hitherto 
unnoticed correlations between classical actions were derived 
from the SCTF, and from the phenomenological observation
that the spectral statistics of quantum chaotic systems reproduce the
results of RMT. (See also the discussion by Dahlqvist \cite{dalcqvist}). 
It should be emphasized that it is 
crucial to assume that the SCTF indeed reproduces discrete energy levels
(`convergence' in the sense of distributions), else the argument for the 
existence of `classical correlations' loses its validity.     
Thus, there is an inherent connection between the quest 
for classical correlations, and the wonder concerning to the 
mathematical soundness (`convergence') of the SCTF. This point 
has been raised by Gutzwiller \cite{entropy} who introduced 
the term `third entropy' in order to refer to the statistical 
properties of the actions.

The conjecture of classical action-correlations leads 
naturally to the introduction of some fundamental questions:
\begin{itemize}
\item  What is the dynamical origin of the correlations between
classical actions? In other words, how can one derive these correlations 
from purely classical considerations, without recourse to the SCTF 
and quantum mechanics ?
\item The classical density consists of  $\delta$ functions which are
supported by the actions of the periodic manifolds, and weighted by 
complex coefficients whose magnitude as well as phase differ from orbit 
to orbit. How much of the expected correlations reflect the distribution 
of actions, and what is the r\^ole (if any) of the correlations amongst 
the coefficients themselves or amongst the actions and the coefficients?
\item  Periodic orbits proliferate exponentially with their periods. 
Excluding non generic degeneracies, this implies an exponentially large 
density of actions of POs. Previous numerical checks 
\cite {cl_poiss} have convincingly shown that on the scale of the
(exponentially small) nearest actions separation, the action spectrum is
Poissonian. How is this compatible with action correlations?
\item   What is the classical phase-space significance of the 
classical correlation scale (geometrical interpretation in case of 
billiards)?  What is the way to generalize the scaling relation  
for action correlations beyond 2D ? What are the modifications 
that are required in case of complex (e.g. diffusive) systems ? 
\end{itemize}
Even though the non diagonal contribution to the form factor
attracted much attention in the past few years, all the studies so far
avoided the direct confrontation with action correlations and the
questions listed above. 
Khmelnitskii and coworkers \cite{Khmel}  
as well as Altshuler et al \cite{AAA} computed the form factor 
for disordered (chaotic) systems by employing techniques other than the 
semiclassical theory. Bogomolny and Keating \cite{BogKeat}
effectively replaced the contributions of POs with $t >t_H/2$
by composite POs constructed in a special (synthetic) way.   
Miller \cite{daniel} has deduced action correlations of composite 
POs from unitarity, and derived from them the classical 
correlations of primitive POs.


The purpose of the present paper is to pick up again the subject of 
action correlations from the point where it was left by \cite{argaman}.
We shall introduce an explicit formulation of the 
{\em quantal-classical duality} concept, and will extend it to 
the general case of either integrable or chaotic systems with $d$ degrees 
of freedom. We shall refer to generic as well as to non-generic 
circumstances. At a later stage we shall discuss the modifications that 
are required in the application to complex (e.g. diffusive) systems.
The formulation of quantal-classical duality and the associated 
heuristic understanding of classical correlations thus paves the way towards 
a global semiclassical theory of spectral statistics that goes 
beyond the diagonal approximation.   
We use from now on the term `classical correlations'
or `classical PO-correlations' rather than the term 
`action correlations'. Sometimes, the latter term is misleading.  
Some of the results were already reported by D. Cohen \cite{brk}. 
They are discussed and further developed in the present paper.

Some remarks on the method of presentation are in order. Most of the
discussion will be carried out for billiard systems. Billiards are 
particularly convenient since they are scaling systems. Because of 
this property, the SCTF can be simplified by considering the quantum 
wavenumbers instead of the energies, and in the classical description 
the length of the PO replaces the action and the time, 
(since one can always choose $m=v=1$). Once the theory for
billiards is presented, we shall show how to transcribe it for any
Hamiltonian system.

In {\bf section II} the general formulation of the quantal-classical duality 
is presented. The SCTF is used in order to derive the relation between the
two point statistics of the quantal and the classical spectra. 
In particular, the universal classical correlation scale $\lambda$ is 
identified. This section contains a detailed discussion of the 
statistical procedure and its relation to the energy-time plane.   
We discuss the crossover from the `classical regime' to the 
`quantal regime' and the proper definition of the   
breaktime follows naturally. The applicability of the 
semiclassical theory beyond the (irrelevant) Ehrenfest 
`log'-time is clarified. 
The role of non-universal features is discussed as well.

In {\bf section III} we give an interpretation of the 
classical correlation scale, and introduce a statistical 
model for rigidity in terms of `families'. 
Then we use various arguments and the notions
of `classification' and `similarity' in order to reveal 
the actual structure of the classical spectrum. 
Finally we make some observations concerning the origin 
of classical correlations.

Some of our observations concerning classical correlations  
are further supported by the numerical study of {\bf section VI}.  
There we have carried out a detailed
analysis of the quantum and classical spectra of the Sinai billiard in
3D, for which we possess substantial and immaculate data bases 
\cite{HarelPhD}. This analysis
provides for the first time a clue to the basic problem -- the
derivation of the classical PO-correlations based on a purely
classical argument rather than on the semiclassical duality.

{\bf Section IV} extends the formulation of section II to 
the general case of either chaotic or integrable Hamiltonian 
systems. For generic integrable system we discuss the dual
Poissonian nature of both the quantal and the classical 
spectra. (See also \cite{BerryTabor,Sinai,Bleher}).
The general strategy is further illustrated by 
referring to specific examples, namely, the Riemann zeros and 
the 3D torus. The {\em cubic} 3D torus is particularly 
interesting due to its non-genericity and is analyzed in detailed.

Besides the  mathematically oriented questions that 
concern the {\em existence} of classical correlations,   
there is also the practical hope to extend the 
semiclassical theory beyond the diagonal approximation. 
Classical PO-correlations become most significant in the 
analysis of {\em complex systems} since they lead to relatively 
short breaktime scales that are not related to the 
universal Heisenberg time. 
In {\bf section V} we make use of the physical insight gained in the 
discussion of classical PO-correlations in order to formulate 
the modifications that are required in order to go beyond the 
diagonal approximation. A specific example is the quasi-1D 
classically-diffusive chain. The quantal form factor should correspond 
either to Anderson-localized eigenstates or to band structure, depending 
on whether the chain is disordered or periodic.   
These genuine quantal effects are due to interference, hence 
the non-diagonal terms play a crucial role.  
One may question the feasibility of having PO-theory for 
the spectral properties in such circumstances.  
Thus, our interest in complex systems is twofold. 
On the one hand we develop a semiclassical theory which is 
not limited by the diagonal approximation. On the other 
hand we test the limits of the semiclassical 
approach. We use our present level of understanding of 
the SCTF and of classical PO-correlations in order to demonstrate that 
the SCTF should be effective also in the study of complex systems.  
The emergence of a non-universal volume-independent breaktime in case 
of a disordered systems is an obvious consequence of the formulation.

\section{General Theory}
\setcounter{equation}{0}

The semiclassical trace formula (SCTF) relates the quantal spectrum of 
energy levels to the classical spectrum of periodic orbits (POs). 
As explained in the introduction the SCTF can be utilized in order to 
derive a semiclassical expression for the quantal form factor. 
The purpose of this section, following \cite{brk}, is to introduce an 
explicit formulation of the {\em quantal-classical duality} concept. 
The reader should notice that a clear and explicit 
formulation of the quantal-classical duality requires 
modifications of some common notations. In particular, 
we find it convenient neither to unfold the quantal spectrum nor 
to scale the quantal form factor. It is also most convenient 
to adhere to the standard Fourier transform conventions. 
The new formulation, which will be discussed thoroughly in 
the following section, promotes the understanding of the 
{\em classical correlation scale}, illuminates the crossover 
from the `classical' to the `quantal' regime, and sheds new 
light on the significance of the statistical procedure.

\subsection{The Semiclassical Trace Formula}

The quantal spectrum $\{k_n\}$ consists 
of real positive eigen-wavenumbers that are defined 
by the Helmholtz equation $(\nabla^2+k^2)\psi=0$ 
with the appropriate boundary conditions.
The average (smoothed) density is found 
via Weyl \cite{weyl} law whose leading order 
term is   
\begin{eqnarray}   \label{e2_1}
K_Q(k) \ \equiv \
\left\langle\sum_n2\pi\delta(k{-}k_n)\right\rangle_{\delta k} 
\ \approx \ C_d\Omega \ k^{d{-}1}
\ \ \ \ \ \ \ \ ,
\left[ C_d=\left(2^{d{-}2}\pi^{\frac{d}{2}{-}1}
\Gamma(\frac{d}{2})\right)^{-1} \right] \ \ \ .
\end{eqnarray}
In the above $\Omega$ is the volume of the billiard, 
and $d$ is the dimensionality. The statistical 
averaging $\langle...\rangle$ implies smoothing 
using a window of width $\delta k$. Further 
discussion of the statistical procedure will 
appear at later subsections.
It is natural to define a density $\rho_{qm}(k)$ that 
corresponds to the quantal spectrum, namely 
\begin{eqnarray}   \label{e2_2}
\rho_{qm}(k) \ \equiv \ 
2\pi [d_{qm}(k)-\bar{d}_{qm}(k)] \ \equiv \ 
\left. \sum_n 2\pi\delta(k{-}k_n) \right|_{osc}
\ \ \ \ \ .
\end{eqnarray}
The subscript $osc$ implies that the smooth 
component of the density is being subtracted. This is done
in order to simplify the subsequent formulation.
Furthermore, in order to facilitate the application of 
Fourier transform (FT) conventions we define $\rho(-k)=\rho(k)$
for $k<0$. The factor $2\pi$ has been incorporated in 
the above definition for the same reason.

The classical spectrum $\{L_j\}$ of periodic
orbits (POs) is defined as the set of all the primitive 
lengths $L_p$ and their repetitions $0{<}r$ such 
that $L_j=rL_p$. 
For a simple chaotic billiard, due to ergodicity, the so 
called `Hannay and Ozorio de Almeida sum rule' implies 
\begin{eqnarray}   \label{e2_3}
K_D(L) \ \equiv \  
\left\langle \sum_j |A_j|^2\delta(L-L_j) \right\rangle_{\delta L} 
\ \approx  \ L \ \ \ \ , 
\end{eqnarray}
where a smoothing window of width $\delta L$ is being used.
The generalization of this sum rule to complex systems 
is discussed later in section V-B. 
The weighting factors are $|A_j|^2=L_p^2/|\det(M_p^r{-}I)|$ 
where $M$ is the monodromy matrix. These weighting factors 
decay exponentially with $L$, namely $|A_j|^2\sim L^2 \exp(-\sigma L)$, 
where $\sigma$ is the average Lyapunov exponent. 
Hence (using (\ref{e2_3})) one deduces 
that the actual smoothed density $\bar{d}_{cl}(L)$ of POs grows 
exponentially as $\exp(\sigma L)/L$. 
It is natural to define a weighted density $\rho_{cl}(L)$ that 
corresponds to the classical spectrum, namely 
\begin{eqnarray}   \label{e2_4}
\rho_{cl}(L) \ \equiv \ 
[d_{cl}(L)-\bar{d}_{cl}(L)] \ \equiv \ 
\left. \sum_j A_j \delta(L{-}L_j) \right|_{osc}
\ \ \ \ , \ \ \ \ \ \ \ \ \ 
\left[ \ A_j=\frac{L_p}{\sqrt{|\det(M_p^r{-}I)|}}     
\ \mbox{e}^{+i\frac{\pi}{2}\nu_j} \ \right]
\ \ \ \ \ .
\end{eqnarray}
The instability amplitudes $A_j$ are endowed with a phase
factor that contains the effective Maslov index $\nu_j=r\nu_p$ 
which includes repetitions and the boundary contribution due 
to reflections.
As in the analogous definition of the quantal density, 
we adhere to standard FT conventions: The 
classical density satisfies the symmetry relation 
$\rho(-L)=\rho^*(L)$, and the sign in front of the Maslov 
index is positive. A negative sign would be incorporated   
if time domain FT conventions ($-i\omega t$) 
rather than space domain FT conventions ($+ikx$) 
were used. The former are used in the general derivation 
of Gutzwiller \cite{gutzwiller}.       

The semiclassical trace formula (SCTF) relates the quantal 
density $\rho_{qm}(k)$ to the classical density $\rho_{cl}(L)$. 
The relation is simply 
\begin{eqnarray}   \label{e_SCTF}
\rho_{qm}(k) \ \ = \ \ {\cal FT} \ \rho_{cl}(L) 
\end{eqnarray}
Disregarding the smooth part, the SCTF (\ref{e_SCTF}) states that 
$\sum_n 2\pi\delta(k{-}k_n) = \sum_j A_j \exp(-ikL_j)$.
Its Fourier transformed version is 
$\sum_n \exp(ik_nL) = \sum_j A_j \delta(L{-}L_j)$. 
However, if the two sides of the trace formula  
were multiplied by a window function     
then the Fourier transformed version would be
(See remark \cite{r_SCTF}),
\begin{eqnarray}   \label{e2_6}
\sum_n w(k_n{-}k) \exp(ik_nL) \ = \
\mbox{e}^{ikL}
\sum_j \tilde{w}(L_j{-}L)  A_j \exp(-ikL_j)  \ \ \ ,
\end{eqnarray}
where $\tilde{w}$ is the Fourier transformed window. 
This version of the trace formula is manifestly symmetric 
and is also more convenient for numerical studies.
From a mathematical point of view the latter version
of the SCTF is superior due to its convergence property.

\subsection{Duality of Two Points Statistics}

In order to study the statistical properties of either 
the quantal or the classical spectrum it is natural
to define the corresponding two point correlation functions.
It follows from the SCTF that the two-point statistics 
of the quantal spectrum is related to the two-point statistics
of the classical spectrum via a double Fourier transform, namely 
$ \langle \rho_{qm}(k)\rho_{qm}(k') \rangle \ = \
{\cal F\!T}{\cal F\!T} \ \langle \rho_{cl}(L)\rho_{cl}(L') \rangle$. 
Without the averaging operation $\langle ... \rangle$, 
the latter relation is mathematically trivial and useless.
In the sequel, we shall argue that the equality holds non-trivially 
as a statistical relation, meaning that the averages on both sides 
of the equality can be carried out independently. We postpone for a 
moment both the discussion of this point, and the precise definition 
of the averaging procedure. We turn first to introduce some formal 
definitions. 
 
The two point correlation function of the quantal spectrum 
is defined as follows
\begin{eqnarray}   \label{e_Rqm}
R_{qm} (k,\epsilon) \ & \equiv & \  
\left\langle \rho_{qm}\left(k{-}\frac{\epsilon}{2}\right) 
\ \rho_{qm}\left(k{+}\frac{\epsilon}{2}\right) \right\rangle \\ 
\ & = & \left\langle \ \sum_{nm}
\ \delta\left(\epsilon{-}(k_m{-}k_n)\right)
\ \ \delta\left(k{-}\frac{k_n{+}k_m}{2}\right) \ \right\rangle
\ - \ (2\pi\bar{d}_{qm}(k))^2 \\ 
\ & = & 2\pi K_Q(k)\cdot[\delta(\epsilon)-b(\epsilon;k)] 
\ \ \ .
\end{eqnarray}
It is a function of $\epsilon$ and it depends parametrically 
on $k$. Sometimes it is convenient to notationally suppress the 
parametric dependence. We shall frequently use the brief 
notation $R_{qm}(\epsilon)$. 
The smoothed density squared has been subtracted, 
and therefore $R_{qm}(\epsilon)$ approaches zero 
asymptotically for $\epsilon\rightarrow\infty$. 
The $\delta(\epsilon)$ term is due to the 
self correlations ($n=m$) of the levels. 
The function $b(\epsilon)$ can be 
interpreted as the `probability' density for finding a vacancy 
at a range $\epsilon$ from some reference level. {\em It should 
have normalization equal to one since  the quantal
spectrum is characterized by a finite rigidity scale}, still
it may possess negative parts if there is a clustering effect.         
The spectral form factor 
$K_{qm}(k,L)$ is the FT of $R_{qm}(k,\epsilon)$ 
in the variable $\epsilon{\leadsto}L$.

The two point correlation function of the classical 
spectrum is defined analogously as follows
\begin{eqnarray}   \label{e_Rcl}
R_{cl} (x,L) \ & \equiv & \ 
\left\langle \rho_{cl}\left(L{-}\frac{x}{2}\right) 
\ \rho_{cl}\left(L{+}\frac{x}{2}\right) \right\rangle \\
\ & = & \ \left\langle \ \sum_{ij} 
A_i^*A_j \ \delta(x{-}(L_j{-}L_i)) 
\ \ \delta\left(L{-}\frac{L_i{+}L_j}{2}\right) \ \right\rangle 
\ - \ (\bar{d}_{cl}(L))^2 \\
\ & = & \ K_D(L) \cdot[\delta(x) - p(x;L)] \ \ .
\end{eqnarray}
It is a function of $x$ and it depends parametrically 
on $L$. Sometimes it is convenient to notationally suppress the 
parametric dependence on $L$. The smoothed density squared has  
subtracted, and therefore $R_{cl}(x)$ 
approaches zero asymptotically for $x\rightarrow\infty$. 
The $\delta(x)$ term is due 
to the self correlations of POs. The function $p(x)$ can be 
interpreted as the `probability' density for finding a vacancy 
at a range $x$ from some reference PO. {\em It should 
have normalization equal to one if the
the spectrum is characterized by a finite rigidity scale}.
We shall argue later that this is indeed the case.
Still, it may possess negative parts if there is a clustering
effect. 

With the above definitions the relation between the
the quantal two points statistics and the classical two
points statistics can be written in the following form
\begin{eqnarray}   \label{e2_13} 
R_{qm}(k,\epsilon) \ = \
{\cal F\!T}{\cal F\!T} \   R_{cl}(x,L) \ \ \ ,
\end{eqnarray}
where the first ${\cal F\!T}$ corresponds to 
$x\leadsto k$ and the second ${\cal F\!T}$ 
corresponds to $L\leadsto\epsilon$. Alternatively
\begin{eqnarray}   \label{e_KeqK} 
K_{qm}(k,L) \ = \ K_{cl}(k,L) \ \ \ .
\end{eqnarray}
Thus, on semiclassical grounds, if the function $K(k,L)$ is viewed 
as a function of $L$ then it equals the quantal form factor,
while if it is viewed as a function of $k$ it equals the classical 
form factor. It also follows that if the quantal spectrum is 
characterized by non-trivial correlations, then also the classical 
spectrum should be characterized by some non-trivial correlations.

We turn now to define the meaning of the statistical 
averaging. It is most convenient to do it in the $(k,L)$ plane.
Namely,
\begin{eqnarray} \label{e2_15} 
\langle f(k,L) \rangle_{\delta k \delta L}  
\ \equiv \ \frac{1}{\delta k \delta L}
\int_{k{-}\frac{\delta k}{2}}^{k{+}\frac{\delta k}{2}} 
\int_{L{-}\frac{\delta L}{2}}^{L{+}\frac{\delta L}{2}}
f(k',L') \ dk'dL'   \ \ \ \ \ \ .
\end{eqnarray}  
In the $(k,\epsilon)$ plane the $\delta L$ averaging implies 
multiplying $R_{qm}(\epsilon)$ with a cutoff function whose
width is $2\pi/\delta L$. Analogously, in the $(x,L)$ plane
the $\delta k$ averaging implies multiplying $R_{cl}(x)$ with 
a cutoff function whose width is $2\pi/\delta k$.
The semiclassical relation holds trivially provided the 
{\em same} window parameters are used in both sides of the 
equation, namely   
\begin{eqnarray} \label{e2_16} 
K_{qm}(k,L,\delta k,\delta L) \ = \ K_{cl}(k,L,\delta k,\delta L)
\ \ \ \ .
\end{eqnarray} 
If the further condition $2\pi \ll \delta k\delta L$ is 
satisfied, then this relation can be cast into the form 
\begin{eqnarray} \label{e2_17}
{\frac{2\pi}{\delta k}} \left\langle\left| 
\matrix{\ \scriptscriptstyle \cr \displaystyle \sum\exp(ik_nL) \cr 
\scriptscriptstyle k{-}\frac{\delta k}{2}<k_n<k{+}\frac{\delta k}{2}
\ \ \ } 
\right|^2\right\rangle_{\scriptscriptstyle \delta L} 
\!\! = 
{\frac{1}{\delta L}} \left\langle\left| 
\matrix{\ \scriptscriptstyle \cr \displaystyle \sum A_j\exp(ikL_j) \cr
\scriptscriptstyle L{-}\frac{\delta L}{2}<L_j<L{+}\frac{\delta L}{2}
\ \ \ \ } 
\right|^2\right\rangle_{\scriptscriptstyle \delta k}
\ \ \ \ .
\end{eqnarray} 
The condition $2\pi \ll \delta k\delta L$ guarantees that the 
two points statistics is not sensitive to the exact 
value of the window parameters. It follows that we may choose 
the window parameters that appear in both sides 
of (\ref{e2_16}) independently.
In this sense (\ref{e_KeqK}) becomes a statistically meaningful relation. 
Further discussion of the $(k,L,\delta k,\delta L)$ space 
will appear at later subsections.

The summations in the above statistical relation 
are the same which occur in the calculation of
the SCTF. The number of terms in these summations 
is finite as in (\ref{e2_6}), due to the windows which are used. 
Equation (\ref{e2_17}) is in some sense the `squared' version 
of (\ref{e2_6}), the main difference being 
that in the latter case $\delta L=0$ on the left hand side, 
while $\delta k=0$ on the right hand side. 
Furthermore, in (\ref{e2_6}) the $L$-window 
should be related to the $k$-window by a Fourier transform. 
Later we shall discuss one more aspect of the relation 
of (\ref{e2_17}) to the SCTF.

\subsection{Universal correlations}
  
The quantal spectrum is characterized 
by a universal correlation scale $\Delta(k)$.
For a simple ballistic billiard it 
simply coincides with the average level 
spacing. Random Matrix Theory (RMT) further 
determines the appropriate universal 
scaling function, namely  
\begin{eqnarray} \label{e_hatb}
b(\epsilon;k) \ \ = \ \
\frac{1}{\Delta(k)}\ \hat{b}
\left(\frac{\epsilon}{\Delta(k)}\right)
\ \ \ \ .
\end{eqnarray} 
From this scaling relation one derives the 
corresponding scaling relation of the quantal 
form factor, namely
\begin{eqnarray} \label{e_RMT}  
K_{qm}(k,L) \ = \ L_H(k)\cdot\hat{K}_0(L/L_H(k)) \ \ ,
\end{eqnarray}
where $L_H(k)=2\pi/\Delta(k)=K_Q(k)$ is the 
Heisenberg time. For the Gaussian Unitary Ensemble
(GUE) there is a particularly simple result 
$\hat{K}_0(\tau)=\min(\tau,1)$. If there is a time 
reversal symmetry one should use the result of the
Gaussian Orthogonal Ensemble (GOE).

Using analogous strategy, let us assume that the classical spectrum 
is characterized by some universal correlation scale $\lambda(L)$ 
such that 
\begin{eqnarray} \label{e_hatp}
p(x;L) \ = \ 
\frac{1}{\lambda(L)}\hat{p}\left(\frac{x}{\lambda(L)}\right)
 \ \ \ \ ,
\end{eqnarray}  
where $\hat{p}(s)$ is some scaling function. The justification 
for assuming the above scaling behavior will be discussed 
in subsection F. For simplicity of presentation it has been 
assumed that the classical spectrum has no degeneracies.
We note however that if time reversal symmetry is taken 
into account, where for the generic non self tracing orbits 
the degeneracy is $g=2$, then the scaling 
relation should be modified by writing 
$p(x)=-(g{-}1)\delta(x)+(g/\lambda)\hat{p}(x/\lambda)$ so that 
$R_{cl}(x)=g[\delta(x)-(1/\lambda)\hat{p}(x/\lambda)]$. One 
may say that this degeneracy constitutes a trivial type
of classical correlations.  
   
\begin{figure} 
\begin{center}
\leavevmode 
\epsfysize=1.5in
\epsffile{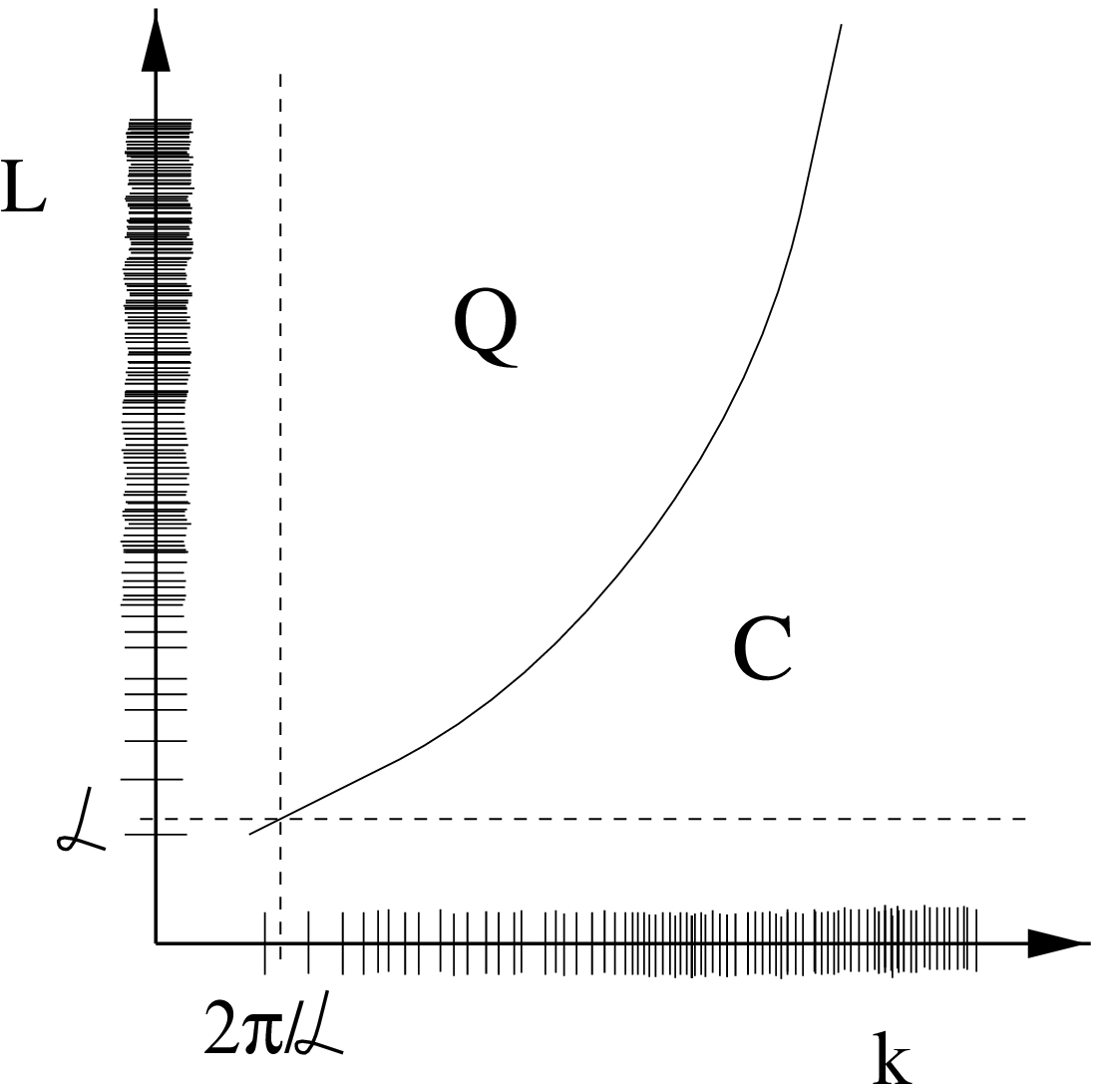}
\epsfysize=1.5in
\epsffile{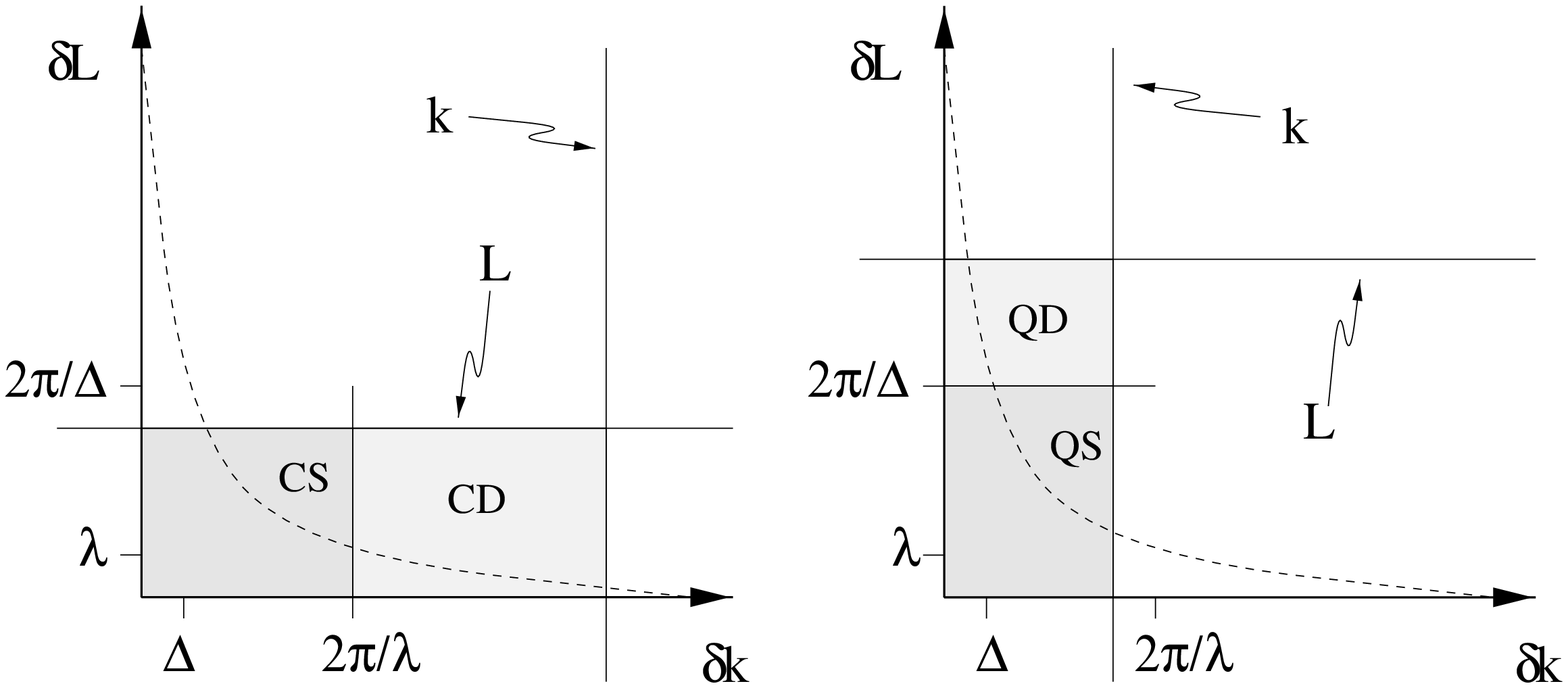}
\end{center}
\caption{\protect\footnotesize Regions in the $(k,L,\delta k,\delta L)$
space. The classical (C) and the quantal (Q) regimes, as well as the  
classical-diagonal (CD), classical-statistical (CS), quantal-diagonal (QD)
and quantal-statistical (QS) sub-regimes are indicated. The mean chord
is ${\cal L}$. Individual POs and energy levels are represented 
in the left drawing by small bars. The dashed line is the 
$\delta k \delta L = 2\pi$ border for statistical stability.}    
\label{f_regs}
\end{figure}

Taking the Fourier transform of $R_{qm}(\epsilon)$ we have 
deduced that $K_{qm}(L)$ should have a crossover at the  
Heisenberg time $L=2\pi/\Delta(k)$. Similarly, taking the 
Fourier transform of $R_{cl}(x)$ it is obvious 
that $K_{cl}(k)$ should have a crossover at $k=2\pi/\lambda(L)$.
Alternatively, one may say that the crossover should occur      
when $\lambda(L)$ equals the De-Broglie wavelength $2\pi/k$.    
However, due to the semiclassical relation (\ref{e_KeqK}) 
the latter condition should coincide with the former condition 
for having a crossover at the Heisenberg time. 
It follows that $K(k,L)$ has two regimes in the $(k,L)$ plane
which are separated by a crossover line whose equation is
\begin{eqnarray} \label{e_regs}
L=\frac{2\pi}{\Delta(k)} 
\ \ \ \ \ \ \Leftrightarrow \ \ \ \ \ \ 
k=\frac{2\pi}{\lambda(L)}
\ \ \ \ .
\end{eqnarray}
Assuming that $\Delta(k)$ and $\lambda(L)$ are monotonic 
decreasing functions, it follows that the $(k,L)$ plane is divided into 
a {\em classical} low-$L$--large-$k$ regime (C), 
and a {\em quantal} large-$L$--low-$k$ regime (Q).
These are illustrated in the left drawing of Fig.\ref{f_regs}.
Using Weyl law $2\pi/\Delta(k)=C_d\Omega k^{d{-}1}$, 
it follows from (\ref{e_regs}) that for a simple ballistic billiard
the classical correlation scale should be 
\begin{eqnarray}   \label{e_lambda}
\lambda_0(L,\Omega) \ = \ {2\pi}
\left(C_d\frac{\Omega}{L} \right)^{1/(d{-}1)}
\end{eqnarray}
Note that this prediction is
independent of any detailed RMT result.
For a 2D system, it coincides with the 
scaling relation of \cite{argaman} 
that we have mentioned in the introduction.   
The validity of (\ref{e_lambda}) follows from the 
validity of the SCTF. 
The subscript $0$ indicates that the
universal (rather than some non-universal)
correlation scale were considered here.

%
\begin{figure}
\begin{center}
\leavevmode 
\epsfysize=1.1in
\epsffile{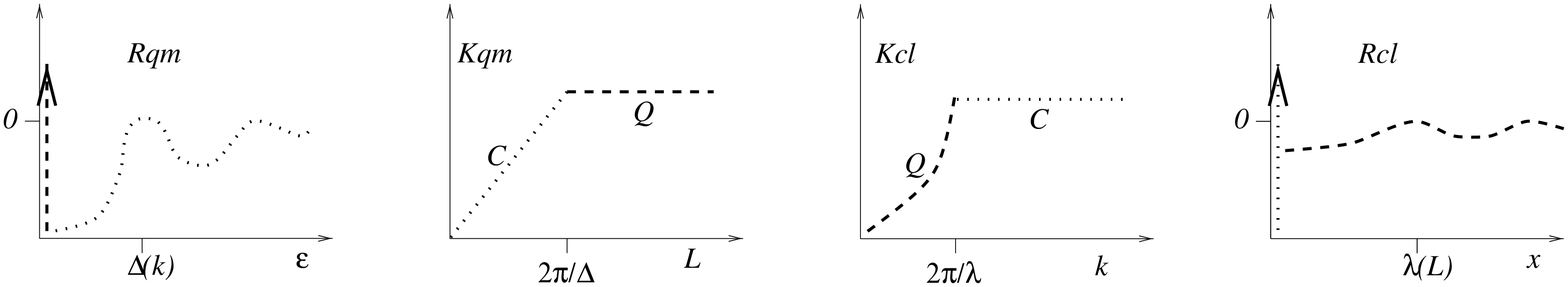}
\end{center}
\caption{\protect\footnotesize 
Schematic plots of $R_{qm}(\epsilon)$ and $K_{qm}(L)$ 
for $k=const$, and of $K_{cl}(k)$ and $R_{cl}(k)$ for 
$L=const$. The dashed lines reflect the quantal diagonal 
approximation, and the dotted lines reflects the 
classical diagonal approximation. $K_{qm}(L)$ is obtained 
from $R_{qm}(\epsilon)$ via FT, $K_{cl}=K_{qm}$ 
from semiclassical considerations, and $R_{cl}$ is 
obtained from $K_{cl}$ via an inverse FT. Both the 
quantal (Q) domain and the classical (C) domain are
indicated in the plots of the form factor. The origin 
of the vertical axis in the plots of the $R_{qm}$ 
and $R_{cl}$ is shifted upwards since we subtract 
the smooth background component. }    
\label{f_corr}
\end{figure}


\subsection{The Diagonal approximation}

We turn now to discuss the so called ``diagonal approximation''
for the form factor. By diagonal approximation it is
meant that the non trivial correlations of either the 
quantal or the classical spectrum are ignored. In the 
absence of degeneracies only self-correlations are taken 
into account. 
For the quantal form factor, the diagonal approximation is
\begin{eqnarray} \label{e_Qdiag}
K_{qm}(k,L)\approx K_Q(k) \ \ \ \ 
[2\pi/\Delta(k){\ll}L] \ \ \ 
\mbox{Diagonal approximation, deep quantal regime} \ \ \ .
\end{eqnarray}
The analogous diagonal approximation that refers 
to the classical form factor is 
\begin{eqnarray} \label{e_Cdiag}
K_{cl}(k,L)\approx K_D(L) \ \ \ \ 
[2\pi/\lambda(L){\ll}k] \ \ \ 
\mbox{Diagonal approximation, deep classical regime } \ \ \ .
\end{eqnarray}
If the classical spectrum has an effective degeneracy 
$g$ due to some symmetry then one should 
write $K_{cl}(k,L)\approx gK_D(L)$. 
Thus, upon using the semiclassical relation (\ref{e_KeqK}), 
the asymptotic behavior of the function $K(k,L)$
is determined both in the deep classical regime (C) 
and in the deep quantal regime (Q). This is illustrated 
schematically in Fig.\ref{f_corr}.      

The above approximations may become an exact
identities if appropriate windows are used.  
In the deep quantal regime one may use a window 
such that $2\pi/\Delta \ll \delta L$. 
It is equivalent to multiplication of $R_{qm}(\epsilon)$ 
by a cutoff window which is much narrower than the 
correlation scale $\Delta$. It implies that the 
non-trivial correlations that are represented by 
$b(\epsilon)$ are washed out, and therefore 
$K_{qm}(k,L)$ becomes {\em identical} with the 
one-point function $K_Q(k)$. The sub-regime in
$(k,L,\delta k,\delta L)$ space where the latter 
observation applies will be called the Quantal-Diagonal (QD) 
regime. Similarly, we may define a Classical-Diagonal (CD) 
sub-regime. The latter occurs at the deep classical regime 
if $2\pi/\lambda \ll \delta k$. In this latter sub-regime 
one may argue that $K_{cl}(k,L)$ becomes identical 
with $K_D(L)$. The different sub-regimes in 
the $(k,L,\delta k,\delta L)$ space are demonstrated 
in Fig.\ref{f_regs}.  

In both the QD and the CD regimes the 
semiclassical relation (\ref{e_KeqK}) obviously holds. 
However, it looses some of its statistical significance since 
it does no longer represent a two-way statistical relation.
Rather it reduces to a `one-way' statistical relation 
where one side of the equation is a one-point entity. 
An extreme case would be such that either $\delta k$ is 
smaller than the average level-spacing $\Delta k$, 
or $\delta L$ is smaller than the average 
length-spacing $\Delta L$. The latter conditions are  
sometimes assumed while analyzing non-universal long range 
correlations in the quantal spectrum. 

It is interesting to note the behavior of
the classical form factor $K_{cl}(k,L)$ in the 
quantal diagonal (QD) regime.   
The summation in the right hand side of (\ref{e2_17}) is the same 
which occurs in the calculation of the SCTF. It will 
yield delta functions of width $2\pi/\delta L$. These 
delta functions are resolved (at least in the statistical
sense). After squaring, it should give a semiclassical estimate 
for the one-point density of the quantal spectrum.
Analogous behavior is found for the 
quantal form factor in the CD regime.

\subsection{Breaktime Concept}  

The semiclassical description of quantum-dynamics consists
of several time regimes. In this subsection we wish to 
make a conceptual distinction between time scales which are 
associated with the breakdown of the stationary phase approximation, 
and those which are associated with interference effects.
The latter are important while discussing the diagonal 
approximation for the form factor.    

{\bf Important Note:}
From now on we shall frequently translate 
lengths into times by using $L=vt$ where $v$ 
is the velocity. Similarly one may translate
$\Delta k$ into conventional energy units 
by using $\Delta E = v \Delta k$. Formulas 
and relations such as $2\pi/\Delta<L$ are 
easily transformed into the corresponding 
time domain version $2\pi/\Delta<t$ provided 
the proper `units' are used for $\Delta$, 
or any other relevant object. As far as the semiclassical 
limit is concerned $k$ plays the role of $1/\hbar$. 
Using different terminology one may say that 
classical mechanics is concerned with lengths of 
rays, while quantum mechanics associates with these 
rays a wavenumber $k$. The mass and the velocity 
of the particle by themselves are insignificant 
physically. The velocity $v$ is used merely for 
converting units of length into units of time and 
therefore can be set equal to $1$.

The correspondence principle implies that quantal evolution 
should follow the classical one on short time scales. 
Deviation due to breakdown of the 
{\em stationary phase approximation} are expected after the 
time $t_{scl}$. It has been argued 
\cite{heller} that $t_{scl}\sim k^{1/3}$.
Further deviations from the leading order 
semiclassical expansion due to e.g. diffraction effects
are discussed in \cite{harel}. 
One should be careful not to confuse these deviations 
which are associated with the accuracy of the stationary phase
approximation with the following discussion of the breaktime
concept. We assume in this paper that the leading order 
semiclassical formalism constitutes a qualitatively good 
approximation also for $t_{scl}<t$ in spite of these deviations.

Interference effects lead to further deviations from 
the classical behavior. Well isolated classical paths 
that are involved in common semiclassical calculations  
may give rise to either constructive or destructive 
interference effects. In particular, interference contribution 
may be significant in the semiclassical computation of 
the form factor (right side of (\ref{e2_17})). 
The diagonal elements of the 
sum $\sum_{ij}A_iA_j^*\exp(ik(L_j{-}L_i))$ represent the classical 
contribution due to self-correlations of POs, while the off diagonal 
part of the sum constitute the interference effect. The 
actual contribution of interference is determined by the 
statistical properties of the classical spectrum. If the 
classical spectrum is of Poisson type, then the interference 
contribution is self-averaged to zero. We shall use the term 
breaktime ($t^*$) in order to denote the relevant time scale 
for the manifestation of such interference effect.      

If only the universal classical correlations are considered 
than $t^*$ should be identified with the Heisenberg time 
scale $t_H$. 
Recall that the Heisenberg time is $2\pi/\Delta(k)$,
hence $t_H\sim k^{d{-}1}$ which is semiclassically much
larger than $t_{scl}$.  For $t \ll t_H$ the diagonal 
approximation (\ref{e_Cdiag}) should hold.
Deviations from the diagonal approximation may be  
apparent in the classical (C) regime $t<t_H$, and depend 
on the actual functional form of $p(x)$. For GUE, for some 
unknown reason, the diagonal approximation actually holds over 
the whole classical (C) regime. In the quantal (Q) regime $t_H<t$,
the diagonal approximation fails completely.   
On time scales $t_H \ll t$ the {\em recurrent} quasiperiodic 
nature of the dynamics is revealed.

If one confused the statistical scale $\lambda_0$ with the 
classical spacing $\Delta L\sim L\exp(-\sigma L)$, then one 
would obtain a false condition $2\pi \ll k\Delta L$ for 
the validity of the diagonal approximation. Such confusion 
would lead to the wrong conclusion that there should be 
a breaktime at the `log' time $t_E \sim \ln(k)$, also known 
as the Ehrenfest time.  
The time scale $t_E$, over which classical POs 
proliferates on the uncertainty scale, is semiclassically 
much shorter than both $t_{scl}$ and $t_H$.
The false condition $2\pi \ll k\Delta L$ is neither necessary 
nor sufficient for the validity of the diagonal approximation. 
This point is further discussed in the introduction for section III, 
and subsection III-F in particular. 
Summarizing, the message is that the `log' time $t_E$ has no physical 
significance as far as the spectral form-factor is concerned. 
It is neither related to the breakdown of the stationary 
phase approximation, nor to the interference that leads to the breakdown 
of the diagonal approximation. 

In case of complex (non-generic) billiard system, 
the classical spectrum may be characterized by a  
non-universal shorter correlation scale $\lambda_{-}$.
Such occurrences will be discussed in later sections. 
As a result there may appear a distinct breaktime $t^*$ which 
is shorter than $t_H$. The crossover to quasiperiodic behavior 
may occur either at $t^*$ or still at $t_H$. In the former 
case $t_H$ may loss its physical significance, as in 
the theory of quantum localization \cite{brk}.

\subsection{Beyond the diagonal approximation}
\label{subsec:ckl}

In order to describe the departure from the diagonal 
approximation we shall define a correlation 
factor $C(k,L)$ via the relation 
\begin{eqnarray}   \label{e2_24}
K_{cl}(k,L) \ = \ C(k,L) \cdot K_D(L) \ \ .  
\end{eqnarray}
Hence
\begin{eqnarray}   \label{e_Ckl}
C(k,L) \ = \  
1-\int_{-\infty}^{\infty}p(x;L) \ \mbox{e}^{-ikx}dx \ \ .  
\end{eqnarray}
If the classical spectrum has no degeneracies then 
the correlation factor should equal {\em one} in the deep classical 
regime $L \ll 2\pi/\Delta(k)$. There, the classical 
diagonal approximation is expected to hold. The correlation 
factor should go to zero in the deep quantal regime  
$2\pi/\Delta(k) \ll L$. This simply follows from (\ref{e2_24})
using the fact that $K_{qm}(k,L)\rightarrow K_Q(k)$ is finite,  
while $K_D(L)\sim L\rightarrow\infty$. It follows also 
(via (\ref{e_Ckl})) that $p(x)$ should have the normalization 
{\em one}, implying finite rigidity scale. Thus it is natural 
to introduce a scaling function $\hat{p}(s)$ with scaling 
behavior that reflects normalization as in (\ref{e_hatp}), 
and with scaling constant $\lambda=\lambda_0(L,\Omega)$.
It follows that there is a related scaling function such 
that $C(k,L)=\hat{C}_0(k\lambda_0(L,\Omega))$. More generally, 
if the the classical spectrum has degeneracy $g$ than the proper 
scaling relation is $C(k,L)=g\hat{C}_0(k\lambda_0(L,\Omega))$.
This will be further discussed in section V-A.

The considerations above actually determines 
the functional form of the scaling function $\hat{C}_0(\kappa)$ 
both in the deep classical regime $1{\ll}\kappa$ and also in 
the deep quantal regime $\kappa{\ll}1$. 
The interpolation requires further information. 
Using the RMT result (\ref{e_RMT}) it follows  
that the classical scaling function $\hat{C}_0(\kappa)$ 
is related to the quantal RMT scaling function $\hat{K}_0(\tau)$.  
For GUE statistics the scaling relation is 
$C(k,L)=\hat{C}_0(k\lambda_0(L,\Omega))$, 
and the scaling function is 
\begin{equation}   \label{e2_27} \label{e_GUE} 
\hat{C}_0(\kappa) =  
\left\{ \begin{array}{lll}
(\frac{\kappa}{2\pi})^{d{-}1}  & 
\mbox{for} \ \ \kappa{\ll}2\pi  & 
\mbox{[corresponds to discrete quantal spectrum]}  \\
(\frac{\kappa}{2\pi})^{d{-}1}  & 
\mbox{for} \ \ \kappa{<}2\pi  & 
\mbox{[RMT based interpolation - GUE ensemble]}  \\
1 &
\mbox{for} \ \ 2\pi{<}\kappa  & 
\mbox{[RMT based interpolation - GUE ensemble]}  \\
1 & 
\mbox{for} \ \ 2\pi{\ll}\kappa  & 
\mbox{[based on the diagonal approximation]}   
\end{array}
\right.
\end{equation} 
where the subscript $0$ has been added in order to 
emphasize that universal correlations are discussed. 
For GOE statistics the scaling relation is 
$C(k,L)=g\hat{C}_0(k\lambda_g(L,\Omega))$, 
with $\lambda_g(L)=\lambda_0(L,\Omega/g)$
and the scaling function is 
\begin{equation}   \label{e5_2} \label{e_GOE} 
\hat{C}_0(\kappa) =  
\left\{ \begin{array}{lll}
(\frac{\kappa}{2\pi})^{d{-}1}  & 
\mbox{for} \ \ \kappa{\ll}2\pi  & 
\mbox{[corresponds to discrete quantal spectrum]}  \\
2(\frac{\kappa}{2\pi})^{d{-}1}
-\frac{1}{2}\ln\left(\frac
{1{+}({\kappa}/{2\pi})^{d{-}1}}
{1{-}({\kappa}/{2\pi})^{d{-}1}}\right) & 
\mbox{for} \ \ \kappa{<}2\pi  & 
\mbox{[RMT based interpolation - GOE ensemble]}  \\
1-\frac{1}{2}\ln(({\kappa}/{2\pi})^{d{-}1}{+}1)  & 
\mbox{for} \ \ 2\pi{<}\kappa  & 
\mbox{[RMT based interpolation - GOE ensemble]}  \\
1 & 
\mbox{for} \ \ 2\pi{\ll}\kappa  & 
\mbox{[based on the diagonal approximation]}   
\end{array}
\right.
\end{equation} 
The fact that we can deduce the explicit scaling functions  
((\ref{e_GUE}) and (\ref{e_GOE})) for the classical correlations
does not imply that their dynamical origin is understood. We propose 
some possible mechanisms in the next section. 
It should be also emphasized that the GUE scaling function 
is as little understood as the GOE scaling function, 
in spite of its apparent simplicity.
See further considerations in section V.

Let us assume for a moment that the scaling function 
possess actually a smooth crossover at $\kappa{=}2\pi$.
If this assumption were true then the tail of 
$\hat{p}(s)$ would be determined by the singularity 
at $\kappa \sim 0$ via the asymptotic relation 
$\hat{C}(\kappa)\sim\int_{1/\kappa}^{\infty}\hat{p}(s)ds$
that follows from (\ref{e_Ckl}). 
Therefore the asymptotic behavior of classical 
correlations would correspond to $\hat{p}(s)\sim{1}/{s^d}$.  
This argument concerning the asymptotic behavior 
of $\hat{p}(s)$ is not-valid if a non-smooth crossover 
at $\kappa=2\pi$ occurs, as suggested by RMT.   
Assuming that $K_{qm}(k,L)$ actually obeys RMT 
prediction, one may perform an inverse Fourier transform 
of $\hat{C}(\kappa)$ in order to find a complete expression 
for $\hat{p}(s)$ via (\ref{e_Ckl}). For $d{=}2$ assuming 
GUE statistics one obtains simply \cite{argaman}
\begin{equation} 
\hat{p}^{d{=}2}(s) \ = \ 
\left(\frac{\sin(\pi s)}{(\pi s)}\right)^2 
\ \ \ \ .  
\end{equation}
This subject is discussed further in subsection III-F.

\subsection{non-universal statistics}

\begin{figure} 
\begin{center}
\leavevmode 
\epsfysize=1.0in
\epsffile{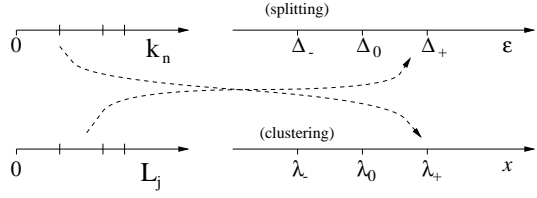}
\end{center}
\caption{\protect\footnotesize Illustration of the various 
correlation scales of the quantal spectrum (upper right axis)
and of the classical spectrum (lower right axis). The relation 
of the generic non-universal correlations to either 
short POs or low-laying levels is schematically indicated.} 
\label{f_nonu}
\end{figure}

For a generic ballistic billiard, due to level repulsion, 
the shortest quantal correlation scale is simply the average spacing
$\Delta_0$. It has been found that the short-range correlations  
on the scale $\Delta_0$ are well described by RMT. These 
quantal correlations are both generic and universal.  
If the billiard is characterized by a non-trivial
structure, then one may find a shorter quantal correlation 
scale $\Delta_{-}$ due to splitting effect.
(For example, one may consider the energy splitting 
in case of a double-well system).
Such quantal correlation scale is neither generic nor universal. 
From semiclassical considerations one deduces that there should 
be also generic but non-universal quantal correlations. The latter     
should be found on large energy scales and are semiclassically 
related to the shortest POs. In order to deduce these long-range 
quantal correlations we should consider the basic statistical relation 
(\ref{e_KeqK}) in the CD regime where it reduces to $K_{qm}(k,L)=K_D(L)$. 
The averaging window $\delta L$ should be small enough in 
order to resolve individual POs, which is equivalent to 
the obvious requirement that $R_{qm}(\epsilon)$ should not
be multiplied by a cutoff window which is narrower than 
the corresponding correlation scale. The various correlation scales 
are illustrated in Fig.\ref{f_nonu}.     

Similar considerations apply upon analyzing the 
classical spectrum. The actual length-spacing
$\Delta L\sim L\exp(-\sigma L)$ has no statistical 
significance. The universal correlation scale 
$\lambda_0$ is much larger. Still, there may be 
a shorter classical correlation scale $\lambda_{-}$ due to 
e.g. clustering of POs. Such classical correlation scale 
is neither generic nor universal. A specific example
will be discussed in section VI. There should be also 
generic but non-universal classical correlations on scales 
larger than $\lambda_0$, which are related to the 
low-lying eigen-energies. These are deduced via the 
semiclassical relation $K_{cl}(k,L)=K_Q(k)$ that 
holds in the QD regime. One should use a sufficiently
small $\delta k$ in order to resolve these long-range
classical correlations. This is equivalent to saying 
that the SCTF is capable of resolving individual 
energies levels of the quantal spectrum.

\subsection{Duality Within the Scattering Approach}

The classical dynamics of  billiards can be conveniently  described
in terms of the Poincar\'e maps $M_{\sigma}$ induced on appropriately chosen
sections $(\sigma)$ in the billiard phase space.
A complete correspondence between the flow and the discrete dynamics is
achieved when the section consists of the entire
boundary surface and the tangential momenta at impact. Other sections,
which are obtained by hyper-planes which cut through the billiard volume, 
and the components of the momentum in the
hyper-planes may capture only a part of the billiard dynamics.

The quantum mechanical analogue of the billiard Poincar\'e  map is
the scattering matrix $S(k;\sigma)$ \cite{DoronUS,Bogomol}, which depends 
in a non trivial way on the particular section $\sigma$ and on  
the wavenumber $k$. In the semiclassical approximation the $S(k)$ matrix 
is of finite dimension, which can be related to the volume 
$\ell_{\sigma}^{d{-}1}$ of the $d{-}1$ dimensional section by
\begin {equation}
 N_{\sigma}(k)  = \left [ \left( { k\ell_{\sigma} \over \pi}\right)
^{d-1} \right]
\label {dimension}
\end{equation}
Where $[x] $ stands for the integer part of $x$. It was realized long time
ago that if the map $M_{\sigma}$ induces chaotic dynamics on the section, 
then the statistical properties of the $S(k;\sigma)$ matrix are well 
reproduced by the predictions of RMT for the 
{\em Circular Ensembles} \cite{BlumelUS,DoronUS}. In particular,
\begin {equation}
 {1\over N} \left < |{\rm tr} S^n |^2 \right > \equiv {1\over \Delta k}
\int_{\bar k -\Delta k/2}
^{\bar k +\Delta k/2}|{\rm tr} S^n(k;\sigma) |^2 {\rm d}k = \hat{K}_{\beta}
\left({n \over N}\right)
\label {rmt}
\end{equation}
Here, the integration is carried out over a $k$ interval over which
$N_{\sigma}(k)$ is constant $N$, so that 
$\bar k = \pi \left ( {(N+1/2)^{1 / (d{-}1)} / \ell_{\sigma}} \right)$ 
and $\Delta k = \pi / (2(d{-}1)) N^{-(d{-}2)/(d{-}1)} / \ell_{\sigma}$. 
The function $\hat{K}_{\beta} \left({n / N}\right)$ is known explicitly 
for the Circular ensembles ($ \beta = 1,2,4$). See e.g. \cite{Haake}. 
In (\ref{rmt}) the $k$ average replaces the ensemble average of 
RMT.  This is justified since the correlator
$\left <{\rm tr}[S(k) S^{\dagger} (k{+}\delta k)]\right>$
vanishes once $\delta k$ exceeds the correlation length $\delta _{\sigma}$
which is inversely proportional to the mean length of the classical orbit
between successive encounters with the section $\sigma$. 
For ergodic billiard $\Delta k \gg\delta _{\sigma}$.

The semiclassical expression for ${\rm tr} S^n(k;\sigma)$ 
is \cite{Tabor,BlumelUS}
\begin {equation}
{\rm tr} S^n(k;\sigma) \ \approx \ 
\sum_{p\in P_{\sigma}(n)}g_p n_p (A_p)^r
{\rm e}^{i k r L_p}
\label {scltrS}
\end{equation}
where $P_{\sigma}(n)$ is the set of primitive POs of
$M_{\sigma}$ of period $n_p$ which divides $n$ so that $n=n_p r$. 
POs which are conjugate by an exact symmetry are represented only 
once in $P_{\sigma}(n)$ and their multiplicity
$g_p$ appears explicitly in (\ref {scltrS}).
The stability amplitudes (including the Maslov and
boundary indices) and the lengths are denoted by $A_p$ and $L_p$
respectively. Substituting (\ref {scltrS}) in (\ref {rmt}) and separating 
the diagonal and non diagonal contributions we get
\begin{equation}
-{N_{\sigma} \over n} \left[ \hat{K}_{\beta}\left({n \over N_{\sigma}}\right)-
\langle g \rangle {n \over N_{\sigma}}\right] \ \ = \ \ \int {\rm d}x \ 
\cos \left (\left [{N_{\sigma}+1/2 \over n}\right ]^{{1\over d-1}}\pi x
\right ) \ \hat{p}_{\sigma}(x;n)
\label {duality}
\end{equation}
where $\langle g \rangle$ is the mean multiplicity of symmetry 
conjugate orbits and $\hat{p}_{\sigma}(x;n)$ is the classical POs 
correlation function
\begin{equation}
\hat{p}_{\sigma}(x;n) \ \equiv \ 
-\!\!\sum _{p\ne p' \in P_{\sigma}(n)}g_p g_{p'}
(A_p)^r (A^*_{p'})^{r'} \ {n_p n_{p'} \over n} \ 
\delta \left (x- {L_p-L_{p'}\over \lambda_{\sigma}(n)}\right) \ 
{\sin [\Delta k (L_p-L_{p'}) / 2] \over 
{[\Delta k (L_p-L_{p'}) / 2]}}
\label {corrfunc}
\end{equation}
where $\lambda_{\sigma}(n) = \ell_{\sigma} / n^{1/(d{-}1)}$ is the 
classical correlation scale. The last factor limits the summation to 
pairs of POs whose lengths differ by 
$(d{-}1) \ell_{\sigma} N^{(d{-}2)/(d{-}1)}$ at
most. Equations (\ref{duality}-\ref{corrfunc}) express the statistical 
duality between the spectra of eigenphases of the {\it quantum} $S$ matrix 
and the {\it classical} POs of the
corresponding map. Note that the correlation scale $\lambda_{\sigma}(n)$
can be interpreted as the mean linear distance between successive impacts 
of the orbit on the section $\sigma$. Therefore it is {\em independent} 
of $\sigma$. It should be emphasize that the same result for $\lambda$ 
is obtained as in the analogous flow approach in spite of the fact that 
the POs are restricted to have a specific `topological' period $n$.

The construction above can be now used to further reveal the statistical 
properties of the classical spectrum. Let us denote by ${\sigma}$ the  
section which is built on the complete boundary,
and by $\rho$ any other section. Consider the POs 
of $M_{\sigma}$ with period $n_{\sigma}$. Only a fraction of them
traverse ${\rho}$, and they belong to the set of POs of $M_{\rho}$ 
with  periods $n_{\rho} \le n_{\sigma}$. On the average
${n_{\sigma} / \ell_{\sigma}^{d{-}1} } \approx  
{n_{\rho} / \ell_{\rho}^{d{-}1} }$ implying again that the 
classical correlation scale $\lambda_{\sigma}(n)$ is independent 
of $\sigma$. On the quantum mechanical level, the scattering 
matrix $S(k;\rho)$ will be of  smaller dimension 
$N_{\rho}= N_{\sigma}({\ell_{\rho}/\ell_{\sigma}})^{d{-}1}$. 
Thus 
\begin{equation} \label{ratio}
\frac{N_{\rho}}{N_{\sigma}} \ \approx \ 
\frac{n_{\rho}}{n_{\sigma}} \ \approx \
\left(\frac{\ell_{\rho}}{\ell_{\sigma}}\right)^{d{-}1}
\end{equation}
Let us compare  the duality relations for the two sections. If
$n_{\sigma}$ and $n_{\rho}$ satisfy (\ref {ratio}) then the left 
(quantum) hand side of (\ref{duality}) remains unchanged.
The classical  functions $\hat{p}_{\sigma}(x;n_{\sigma})$ and
$\hat{p}_{\rho}(x;n_{\rho})$ express the same statistical reality, 
but the latter uses only a fraction of the
POs which contribute to the former. That is why
$\hat{p}_{\rho}(x;n_{\rho})$ is rescaled by a factor 
$({n_{\sigma}/n_{\rho}})$ which follows from (\ref {corrfunc}). 
Thus, we can construct {\em finer classes} of POs which traverse 
smaller (but nested) sections, and which manifest the same POs 
correlations as contained in the original set of POs. 
However, this refinement is quite limited by the purely classical 
condition $\lambda_{\sigma}(n)\ll\ell$. 
The argument for this condition is as follows. 
Semiclassical considerations are valid as long as $2\pi\ll k\ell$,
else the scattering matrix becomes of dimension {\em one}.  
On the other hand, in order to deduce non-vanishing $p(x;n)$ from 
the duality relation (\ref{duality}) the left hand side should 
be non-trivial, meaning $N<n$ or equivalently $k\lambda\ll 2\pi$.
Combining the two restrictions one obtains the $k$-independent
condition $\lambda_{\sigma}(n)\ll\ell$. 
Recalling the exponential proliferation of POs as a 
function of $n$, this condition implies that an exponentially 
large number of POs will be included in those classes. 
       
\section{The Origins of PO-correlations: Observations and Speculations}
\setcounter{equation}{0}

It should be emphasized from the outset that we do not have 
a {\em direct classical derivation} for any of the results 
obtained above concerning the statistical two-point correlations 
of the classical spectrum. 
In this section we try to acquire, at least on a heuristic level, 
some insight concerning the origins of classical correlations. 

The first subsection introduces a simple statistical model 
for classical correlations using the notion of {\em families}. 
This family-structure is introduced on a purely formal grounds, 
without considering the question whether it is realized in 
actual dynamical systems. Still it provides an explanation for 
one of the puzzling features of the classical spectrum, namely, 
that its correlations are not apparent on the (exponentially small)
level spacing, but rather on a much larger length intervals.    

In order to reveal the actual (statistical) structure of 
the classical spectrum we use various arguments. Some of the 
arguments constitute a variation of the quantal-classical duality 
argument for having classical correlations.  
In subsection B we introduce the notion of {\em classification} \cite{brk}. 
In subsection C we argue that one can find classifications that reflect 
the actual structure of the classical phase space. 
Thus it becomes plausible that some variation of a family-structure is 
indeed realized.  
Actually, we believe that a more complicated picture of 
{\em hierarchical structure} is appropriate (subsection E). 
The last subsection clarifies the r\^{o}le of resurgence in 
the theory of classical correlations. 

When considering POs of length $L_j\sim L$, one may 
{\em define} the {\em geometrical scale} $\lambda_0$ as  
the average distance between points where a typical 
PO of length $L$ intersects a given surface of section. 
This scale appears naturally in the discussion of 
subsection C. There it appears as a limiting 
{\em resolution scale} for the classification of POs.
In subsection D, we point  that $\lambda_0$ also 
has the meaning of {\em variation scale} 
whenever parametric deformation of the billiard is 
considered. Then we argue that it should show up as 
a {\em correlation scale} in the two point statistics 
of the classical spectrum. Thus we 
obtain some heuristic understanding for the origin 
of classical correlations. 

We also try to speculate a formal approach for 
a future derivation of classical correlations. 
We suggest that an appropriate coarse-graining 
procedure may be used in order to approximate 
any generic billiard by a ``graph''. 
This way metric correlations are transformed into 
a comparatively simple degeneracy structure.

\subsection{families}
\label{subsec:families}

In order to gain some insight into the classical correlations, 
let us ignore the fluctuations of the instability amplitudes, 
and let us further assume that correlations between POs 
that correspond to different Maslov index are not important.
These assumptions reflect our belief that the statistical 
correlations of the {\em weighted} classical spectrum are 
actually present also in the {\em bare} spectrum of lengths.
To support this point of view one should notice that our
result (\ref{e_lambda}) for the classical correlation 
length $\lambda_0$ possess      
a very simple geometric interpretation: It is the average 
distance between neighboring points where a PO of length $L$  
bounced from the $(d{-}1)$ dimensional billiard surface.

Using loose notations, the two point statistics of 
the classical {\em bare} spectrum is
\begin{eqnarray} \label{e} 
\delta(x)-p(x;L) \ = \
\left\langle\sum_{j}
\delta\left(x-(L_j{-}L_{\mbox{\tiny PO}})
\right)\right\rangle_{\mbox{\tiny PO}}
 - \mbox{const}
\end{eqnarray}
where the angle brackets denote averaging over  
reference POs whose length is $L_{\mbox{\tiny PO}}\sim L$, 
and $const$ refers to the smoothed density.  
Note that the delta-function arises due to    
the diagonal term $j=\mbox{\tiny PO}$ 
(we assume no degeneracies).   
If the rest of the lengths $L_j$ are independent with respect 
to the reference PO, then $p(x)\sim 0$, which corresponds to 
$\lambda\rightarrow\infty$.
However, if there is a small repulsion, then $p(x)$ can be 
interpreted as the probability for finding a vacancy in 
the vicinity of the reference PO. 

A more detailed interpretation which seems to be relevant is 
as follows. Let us assume that the POs can be grouped 
into families such that each family constitutes a rigid 
spectrum (`ladder') by itself, with average level spacing $\lambda$ 
and strong repulsion. For clarity let us further assume
that there are no cross-correlations between different families. 
If each family is characterized by the same two-point correlation 
function $p(x)$, then also their union will have the same $p(x)$.
The latter interpretation makes it clear that the spectrum will 
look like Poisson on scales which are smaller than $\lambda$.
It is important to notice that the total density of 
POs grows exponentially with $L$ due to 
their proliferation. Still, the average density of those 
POs that actually contribute to $p(x)$ is 
$p(0)\sim L^{1/(d{-}1)}$. It Implies that the overwhelming 
majority of periodic orbits $j$ is uncorrelated with the 
reference PO. 
The rigidity of the spectrum manifests itself only on 
scales which are larger than $\lambda$. 

This interpretation of classical correlations settles the apparent 
contradiction between the observation of Poisson statistics on small 
scales \cite{cl_poiss} and the existence of non-trivial two-point 
statistics on the other hand.  
At this stage it is useful to draw an {\em analogy} with 
the statistical properties of a quantal spectrum that 
corresponds to a system with localization. The spectrum 
looks like Poisson on small energy scales, since neighboring 
levels correspond to well separated eigenstates. However,
on large energy scales the rigidity of the spectrum is 
manifest. This rigidity reflects the fact that the spectrum 
constitutes a union of many local sub-spectra. Eigenstates that 
dwell in the same localization-volume constitute a family.

\subsection{classes}

In an attempt to understand the dynamical origins 
of the PO-correlations, it is useful to classify the 
POs according to some dynamical criteria,  
and to study the statistical properties of POs 
which share some common feature. 
One can consider for example a class of POs having 
the same number of bounces on the boundaries. A more 
refined classification can be obtained when symbolic 
codes are available. In this case one can classify 
the POs according to certain requirements on the 
codes. The most simple classification is obtained by 
gathering all the POs which are conjugate to each 
other by a symmetry operation such as time reversal
or reflection. Such classifications may reflect
certain statistical features of the classical spectrum. 

The classical spectrum is rigid since $p(x)$ has a finite 
range $\lambda$, and it satisfies the normalization 
$\int p(x) dx = 1$. 
If the classification is arbitrary, then one will find that POs 
of a given class constitute a non-rigid spectrum. 
From now on we shall use the notion of `class' and `classification'  
in a more restricted fashion. {\em By definition, 
POs of a class should constitute a rigid spectrum
by themselves}. A classification of POs into (mutually-exclusive) 
classes is an important tool for revealing the statistical 
properties of the classical spectrum. A classification is 
particularly useful if there are no cross correlations 
between different class-spectra. In some cases we can argue 
an a-priori existence of a classification scheme. 
At some other cases, as in section \ref{sec:numerics}, 
the classification can be guessed and verified a-posteriori.
The existence of classes, with some restriction, 
has been argued in subsection II-H using the scattering approach 
point of view. In subsection C we shall argue that the notion of 
classification is appreciable to any billiard system.  
 
The most obvious example for useful classification 
of POs occurs while analyzing a disordered chain of 
cells. Such a complex billiard constitutes a quasi 
one dimensional diffusive system. POs whose 
period $t$ is less than the ergodic time may be 
classified by the volume which they explore. 
POs that belong to different volumes cannot be 
correlated, since if the shape of one of the cells
is modified, it will affect only those POs that 
explore it.   

If the statistical model of subsection A were realized
in actual dynamical systems, then  
the finest classification of POs would be into families. 
Further classification would not be possible.  
A direct understanding of classical 
correlations implies in some cases that a classification 
of POs into families has been indeed identified. An analytical  
theory requires further derivation of the two points 
statistics that characterizes these families, including 
the study of cross-correlations between families. 
It should be emphasized that different classifications
may be found appropriate while analyzing 
non-universal rather than universal correlations.
   
Assuming that cross-correlations between 
classes are negligible, it follows that 
the form factor can be decomposed in the 
following way
\begin{eqnarray} \label{e_Kalpha}
K(k,L) \ = \ \sum_{\alpha} K_{\alpha}(k,L) \ = \ 
\sum_{\alpha}C_{\alpha}(k,L) K_{\alpha}(L) \ \ \ ,
\end{eqnarray}
where $K_{\alpha}(L)$ is the diagonal sum that 
corresponds to POs of the $\alpha$-class, while 
$C_{\alpha}(k,L)$ are the corresponding correlation
factors. This formula will constitute the 
basis for our later study of complex systems.
In case of a complex system, different 
classes may have different statistical 
properties.

\subsection{similarity}
 
If a the statistical model of subsection A 
were indeed realized in actual dynamical systems, 
then we would expect that POs of the same family 
would be geometrically related in some intimate way. 
In the present section we wish to use a variant of 
the quantal-duality argument in order to develop 
a notion of similarity. We do it by extending an idea 
due to Fishman and Keating \cite{KF}. 

Consider a simple ballistic planar billiard.  
A magnetic flux line $\phi_0$ penetrates 
the plane at a point $r_0$ which is located 
inside the billiard. The form factor 
$K(k,L)$ with $\phi_0\ne0$ will 
correspond to GUE statistics.     
Now, assume that additional flux lines 
$(\phi_1, \phi_2, ...)$ penetrate the plane
at the points $(r_1, r_2, ...)$. The form 
factor $K_{\{\phi\}}(k,L)$ still corresponds 
to GUE statistics, as well as its average
$\bar{K}(k,L)$, where the average is over
$(\phi_1, \phi_2, ...)$. However, it is a straightforward 
exercise to prove that the semiclassical expression 
for the averaged form factor is given by (\ref{e_Kalpha}), 
where $\alpha$ denotes a class of POs that share the {\em same}
set of winding numbers $(\nu_1, \nu_2, ...)$ with respect 
to the flux lines. 
Assuming that all the classes are characterized 
by the same statistics, It follows 
that all the $C_{\alpha}(k,L)$ are identical with 
the GUE correlation factor (\ref{e2_27}). Thus, each of the 
classes constitute a rigid spectrum by itself.

One may add more and more flux lines, thus 
achieving finer and finer classification of 
POs. Let us denote the typical distance 
between flux lines by $\ell$. Semiclassical 
considerations are valid as long as $2\pi \ll k\ell$,
where $2\pi/k$ is the De-Broglie wavelength.
Manifestation of rigidity is expected only in 
the quantal (Q) regime where $k\lambda_0(L) \ll 2\pi$.
It follows that the semiclassical argumentation 
for rigidity of the $\alpha$ classes is restricted 
by the $k$-independent condition $\lambda_0(L)\ll\ell$.
Thus we deduce that $\lambda_0(L)$, (which is essentially 
the mean distance between chords of a typical PO whose
length is L), actually determines the similarity-resolution.

It is plausible that in some sense, by using $\ell$ of order 
$\lambda_0$, one will obtain the desired classification 
of POs into families.  Literally, one may object this point 
of view using the argument that the locations $(r_1,r_2,...)$
are arbitrary, therefore the classification into families 
is ill-defined. This objection can be dismissed by noting 
the analogy with localization theory (see end of subsection A).
There, eigenstates may be classified by partitioning 
the physical space into `blocks'.  The partitioning into 
blocks is arbitrary. The limit of fine partitioning into small 
blocks, whose linear dimension is of the order of localization 
length, is problematic. Still, this limit is conceptually 
meaningful and leads to the correct identification of the 
family-structure.

\subsection{Bifurcations}

In this subsection we point out that the 
{\em geometrical scale} $\lambda_0$ has 
also the meaning of a {\em variation scale} whenever 
parametric deformation of a billiard is considered.   
We propose that this variation scale 
should show up as a {\em correlation scale}
in the two-point statistics of the classical spectrum. 
We use quantal-classical duality in order to 
further support this conjecture.  

Consider the effect of some parametric deformation of 
a billiard system on the energy levels. We focus our 
attention on levels that are contained is some window 
around $k$. 
Changing a parameter $b$ is considered to be a small 
perturbation if $k_n(b){-}k_n(b{=}0) \ll \Delta_0$. 
Thus we can define the perturbative regime as $b<b_c$.  
For $b\sim b_c$ the difference 
$\delta k \equiv k_n(b){-}k_n(b{=}0)$ is of the 
order $\Delta_0$, and one expects to have gone via an 
avoided crossing. Thus we deduce that the 
{\em correlation scale} $\Delta_0$ can be identified 
with the {\em variation scale} $\delta k$.  
The latter can be computed by extrapolating the 
leading order perturbative calculation up to the point 
where we expect that it will loose its validity.  

Now we wish to speculate an analogous property 
for the classical spectrum. Consider POs that 
are contained in some window around $L$. 
The conjecture is that the {\em correlation scale} $\lambda_0$ 
should be identified with the {\em variation scale} $\delta L$.
By definition, in order to determine the 
variation scale $\delta L$ we should consider some 
parametric deformation of the billiard.
Rather than having `avoided crossings' we 
shall have `avoided bifurcations'. The 
variation scale is calculated by applying  
linear analysis up to point where it is expected 
to loose its validity. Rather than considering 
a general deformation it is most convenient to
analyze a specific deformation that takes us 
right to the bifurcation. Note the analogy 
between `bifurcations' and `crossings'.      

We consider a local deformation of the boundary. 
The normal distance of this deformation is $b$, 
and it involves a small surface area. We further 
assume that the linear dimension of the deformed surface 
is of the order $\lambda_0(L)$. At this stage 
$\lambda_0(L)$ is {\em defined} as 
the average distance between points where a typical 
PO of length $L$ hits the surface. It follows that
each PO of length $L$ hits the deformed surface 
only once (on the average). If the deformation is 
small, meaning no bifurcation, then a PO keeps its
identity, but its length is changed by 
$\delta_j=2b\cos(\theta_j)$
where $\theta_j$ is the incidence angle 
(see appendix A for proof). 
The PO will disappear if $b$ becomes 
of the order $\lambda_0(L)$. The deformed 
surface hits then one of the segments of the PO.
Thus we conclude that the {\em variation scale}
is $\lambda_0(L)$. It follows from our conjecture 
that this variation scale will show up in the two 
point statistics. 

It is interesting to analyze the significance 
of bifurcations from the quantal-classical duality 
point of view. Here we present a consistency 
argument that further substantiate the 
identification of the classical correlation scale 
$\lambda$ with the geometrical scale $\lambda_0$.       
Considering the specific deformation of 
the previous  paragraph, it is  
expected that the $\delta_j$ are uncorrelated.   
The distribution of $y=\delta_i{-}\delta_j$ will be 
denoted by $G(y)$. The average $y$ is zero, and
the standard-deviation is of order $b$. 
In appendix A we prove that if a classical 
spectrum is modified as in the present 
case, namely $L_j \rightarrow L_j+\delta_j$, then the 
new correlation factor is  
$C(k) \rightarrow [1 - \tilde{G}(k)\cdot(1{-}C(k))]$.
Here $\tilde{G}(k)$ is the Fourier transform of $G(y)$.
However, assuming that no symmetry breaking is
involved, the new correlation factor  should correspond 
to the same universal statistics as the old one. 
This will be true if either $C(k)=1$ and/or $\tilde{G}(k)=1$.
The function $G(k)$ has a crossover at $k \sim 1/b$, 
while the function $C(k)$ has a crossover at $k \sim 1/\lambda$, 
where $\lambda$ is the classical correlation scale
(yet to be determined).  
Hence we obtain the $k$-independent condition 
$b \ll \lambda$. If the latter condition is 
not satisfied then our considerations will lead 
to inconsistency. On the other hand the only essential 
assumption for our considerations to hold is having 
no bifurcations. Therefore, it is plausible 
that the condition $b \ll \lambda$ actually coincides 
with the condition $b \ll \lambda_0$ leading to the 
identification of the universal correlation scale $\lambda$ 
with the geometrical scale $\lambda_0$. 


\subsection{Quantum Graphs and Shadowing} 

Quantized graphs (networks) \cite{networks}
display spectral statistics which are reproduced
to a high level of accuracy by the predictions of RMT \cite{networks}. At
the same time, the spectral density can be expressed in terms of an exact
trace formula which involves summation over POs (loops).  
With an appropriate definition for the instability amplitudes, this 
trace formula looks formally the same as the SCTF of section II.  
Still, the graphs are sufficiently simple to enable a clear 
understanding of the origins of PO-correlations as well as their 
implications on the quantal spectrum \cite{networks}.

A graph is a strictly  one dimensional system. It
is composed of a finite number of bonds (wires) which are connected at 
junctions (vertices). Classical dynamics is mixing because at 
each junction a particle takes a Markovian choice of the next 
bond on which it moves. The multiple connectivity of the graph induces
the mixing nature of the system. The graph dynamics is Bernoulli,
and as such it is characterized by
an exponential proliferation of POs, to which
positive Lyapunov coefficients can be assigned according
to the probability to remain on the graph.
The smooth part of the quantal spectral density  is
given by Weyl's  law for a $d=1$ system. Thus the Heisenberg 
time is $L_H=2\sum l_b$, where $l_b$ are the lengths of the bonds 
(assumed to be non-commensurate to avoid non generic degeneracies).
The  oscillatory part of the spectral density is a sum of contributions
from POs. Each contribution is a complex number whose amplitude is
the square root of the classical probability to remain on the
graph, and its  phase is the action $k L_p$ where $L_p$ is the orbit  
length. A topological phase factor
${\rm e}^{i\pi \mu_p}$ plays a similar role to the Maslov index. 
$\mu_p$ counts the number of time the orbit back-scatters from a vertex. 
(that is, the number of sequences of the type "$aba$" which appear in 
the code of the PO).

The classical length spectrum for the graph has a simple structure -  
it is obtained by taking the linear combinations $L_p = \sum_{b} n_b l_b$, 
which are consistent with the connectivity of the graph, and $n_b$ are 
natural numbers.  This structure of the length spectrum is revealed only 
if $L_p > L_H = 2 \sum L_b $ because  for such lengths the orbit must 
be  reducible to combinations of lengths of shorter orbits.
We see again that the Heisenberg length $L_H$ plays a natural role in the
discussion of the morphology of the PO-spectrum. This is in
complete accordance with the previous discussion of the duality concept.

The fact that the Heisenberg length is constant ($k$-independent) implies 
that one can generate the spectral two point function using an arbitrarily 
large spectral interval ($\delta k$). This way, one can reach the domain 
where the function $K_{qm}(L)$ is composed of arbitrarily sharp spikes which 
resolve completely the length spectrum for lengths which are both smaller 
and larger than $L_H$. The smoothed $K_{qm}(L)$ 
was compared with the predictions of RMT and was found to be in excellent 
agreement \cite{networks}. The only correlations of POs 
which survive this treatments are correlations between orbits which have 
the {\it same lengths}. One may consider this to be a complicated 
(generalized) case of degeneracy. Note that the instability amplitudes 
that correspond to degenerated POs should not have the same phase. 
For $ L < L_H$ there are hardly any length 
degeneracies, and therefore  $\bar K(L) \approx L$ for $L<L_H$. For
$L > L_H$ the degeneracies of lengths increase rapidly with $L$, and the
correlations between the POs which bring about the correct $L$ dependence 
of the form factor are the correlations between the back-scatter indices
$\mu_p$. This result demonstrates one important features of the POs
correlations, namely, that POs carry both 
{\it metric} information (action integrals or just lengths) and 
{\it topological} information (Maslov indices and degeneracies).  
The correlations which are responsible to the quantum-classical
duality sometimes reside in the metric information exclusively (Sinai
Billiards) and sometimes in the topological information as was discussed 
above for quantum graphs.

One of the most important recent developments in our understanding of the
connection between classical chaos and quantum statistics was recently put 
forward by Bogomolny and Keating (BK) \cite{BogKeat}. They showed how to 
derive the non diagonal part of the quantal form factor  
by using information about short POs with $L_p < L_H$ exclusively, 
and moreover, they assumed that these short orbits are statistically
independent. We would like to discuss now the BK construction and 
show how it is related to the correlations which are 
discussed here. It is convenient to discuss this issue in the context 
of quantized graphs, since this system is both simple and exact.

BK define an approximate spectral counting function $N(k, L^*)$ where the
SCTF is truncated to include POs of length $L< L^*$, and $L^*$ is yet to 
be determined. It is assumed that $N(k, L^*)$
provides a reasonable first approximation to the counting function, and it 
is improved by identifying the points $k_n^*$ where $N(k, L^*)$ assumes the 
values $n+{1\over 2}$ for all integers $n$. The resulting spectral density is
\begin{eqnarray}
d(k) \ = \ \sum _{n} \delta\left (N(k, L^*) -n-{1\over 2}\right ) {\partial
N(k, L^*) \over \partial k} 
\ = \ d(k,L^*) \sum_{m=-\infty}^{\infty} (-1)^m {\rm e}^{2\pi i m N(k, L^*)}
\label {B-K/1}
\end{eqnarray}
The main purpose of this construction is to ensure the $\delta$ function
structure of the spectral density. This is done, however, at the cost of 
producing a spectral density which might differ from the real one in detail. 
This can be easily understood when  (\ref {B-K/1}) is expanded and the 
length spectrum is deduced. The length spectrum beyond $L^*$  consists of 
all the sums $\sum m_p L_p$ where here the $m_p$ are integers, and the 
basic lengths $L_p < L^*$. There are no restrictions on the
combinations of the $L_p$, which for quantized graphs implies that the rules 
of the symbolic dynamics are not obeyed! In other words, even the topological 
entropy of the synthetic POs is not that of the original system.  
BK used their procedure to obtain averaged spectral correlations which 
might be less sensitive to the problems inherent to the method.

Using the intuition we derive from the work on quantized graphs,
we would like to propose a heuristic understanding of classical 
correlations in billiard systems with $d>1$. For billiards $L_H$ 
is $k$-dependent as in Fig.\ref{f_regs}.
Suppose that we need to perform a {\em classical} summation
of the type $\sum_{j}\exp(ikL_j)$. We
wish to approximate the true lengths $L_j$ by some
new set of lengths that corresponds to a  coarse grained
phase space. Given $k$ one can define a spatial
resolution scale $2\pi/k$. It is plausible that
this coarse graining scale is appropriate in order to
obtain a faithful approximation for the classical sum.
However, we do not have a purely classical argumentation for
this speculation. If this speculation is true, then we may
approximate the length of long POs by a sum of the
form $\sum_{b} n_b l_b$, where $\{l_b\}$  is a set
of comparatively
{\em short POs that form a skeleton structure}.
The idea that long POs are supported by short POs is
called `shadowing' \cite{shadowing}.
We can estimate the sum $\sum_{b} l_b$  for the skeleton
which is required for a {\em faithful} representation
of {\em all} the long POs of the coarse-grained phase
space. We do it as follows: due to the coarse graining
each PO of the skeleton has a thickness $(2\pi/k)^{d{-}1}$.
The total volume of the skeleton structure is
$\sum_{b}(2\pi/k)^{d{-}1}l_b$, and it should be equal
to $\Omega$. The billiard is now approximated by a ``graph''.
The Heisenberg time for this graph is
$L_H=2\sum_b l_b\sim\Omega k^{d{-}1}$, which is the
generic quantal result.

Using the shadowing concept we can therefore obtain a heuristic
understanding for the origin of classical correlations.
By improving gradually the coarse-graining resolution
of phase space, the hierarchical structure of the classical
spectrum is revealed. POs that are identical in length
on a given resolution scale, will become a rigid family
if the classical spectrum is probed with better resolution.

\subsection{Resurgence}

It turns out that the statistics of composite
periodic orbits (CPOs) is simpler that the 
statistics of primitive periodic orbits (PPO) \cite{daniel}. 
The CPO-spectrum 
$L_{(r_1,r_2,...)} = \sum r_j L_j $ is obtained 
from the PPO spectrum, where the $r_j$ are integer 
repetitions. There is a strong analogy
PPO $\Leftrightarrow$ primes, and CPO $\Leftrightarrow$ 
natural-numbers (see section IV-F).    
Actually, it turns out that the two point statistical 
correlations of CPOs is essentially the same as that 
of the natural numbers, meaning that the scaling 
function is universal and $d$-independent. Namely,
\begin{eqnarray}   
\hat{p}^{CPO}(s)=\frac{\sin(2\pi s)}{\pi s} \ \ \ .
\end{eqnarray} 
The analogy to our formulation has been 
extended $\cite{daniel}$. The role of the density  
$\rho_{qm}(k)\sim\sum\delta(k{-}k_n)$ is taken 
by the Zeta function $\zeta_{qm}\sim\prod(k{-}k_n)$. 
The two points correlations of the former are 
$R_{qm}(k,\epsilon)$, while the two points
correlations of the latter are $C_{qm}(k,\epsilon)$.
The latter auto-correlation function is 
essentially {\em real} and therefore the corresponding 
form factor $K_{qm}^c(k,L)$ is symmetric, namely 
satisfies the functional relation
\begin{eqnarray} \label{e_funcrel} 
K^c(k,L) \ = \ K^c(k,L_H(k)-L) \ \ \ .
\end{eqnarray} 
The quantal form factor $K_{qm}^c(k,L)$ is 
related to the two points statistics 
$R_{cl}^{CPO}(x,L)$ of the CPO spectrum.
Therefore (\ref{e_funcrel}) can be 
re-interpreted as a {\em classical} relation 
between the statistics of long POs and the 
statistics of short POs. The existence 
of such relation is hardly surprising, it should 
follow from the observation that long POs are 
supported by shorts ones.  
A classical derivation of (\ref{e_funcrel}) 
would be a major step in the understanding of 
classical correlations. In \cite{daniel} 
the functional relation has been re-expressed
in terms of PPO-statistics, leading to the 
well known relation \cite{AAA,BogKeat} between the diagonal 
behavior of $K(t)$ for short times, and its 
`off-diagonal' behavior in the vicinity of 
Heisenberg time.

\section{Applications, Generalizations and Examples of Duality}
\setcounter{equation}{0}

The formulation of quantal-classical duality will be extended 
to the case of either general chaotic system (subsection A)
or general integrable system (subsection C). We also 
discuss specific adaptations of the general formulation:
The introduction of magnetic field and the analysis of 
periodic chains (subsection B); The theory of the 3D torus 
billiard (subsection D); The theory of the 3D Sinai billiard 
with mixed boundary conditions (subsection E). 
Finally (subsection F) we discuss, 
using the same approach, the statistical properties of the primes.  

Both, the classical spectrum of the cubic 3D torus billiard, 
and the `classical' spectrum of the primes, are characterized by 
a non-trivial correlation scale. These two examples, together with 
the generic case of chaotic billiard, demonstrate our claim 
that there is no a-priori relation between the classical correlation 
length $\lambda$ and the average (or typical) spacing between 
actions $\Delta L$. Indeed, if we summarize these three cases we get:
\begin{eqnarray}
\lambda & \ll & \Delta L 
\; \; \; \; \mbox{for the primes} \nonumber \\
\lambda & \approx & \Delta L 
\; \; \; \; \mbox{for the cubic 3D torus} \nonumber \\
\lambda & \gg & \Delta L 
\; \; \; \; \mbox{for a generic billiards} \, . \nonumber
\end{eqnarray}
Note that both $\lambda$ and $\Delta L$ are functions of $L$. 
These functions have no a-priori inter-relation.

\subsection{General Chaotic Systems}

For general time-independent Hamiltonian system the SCTF is:
\begin{eqnarray} \label{e4_1} 
\left.\sum\delta(E{-}E_n) \right|_{osc} \ = \
\frac{1}{2\pi\hbar}\sum_p\sum_{r} 
\frac{T_p}{\sqrt{|det(M_p^r-I)|}}
\exp\left(i\frac{rS_p(E)}{\hbar}-i\frac{\pi}{2}r\nu_p\right)
\end{eqnarray}
%
%
where $\{ E_n \}$ are the eigenenergies. The actions 
$S_j=rS_p$ correspond to the primitive POs 
and their repetitions (both positive and negative).
The primitive periods are $T_p$, and the Monodromy 
matrix is $M_p$. The effective Maslov index $\nu_j=r\nu_p$
is incorporated with the appropriate (negative) sign  
that corresponds to time-domain FT conventions.  
%

In order to apply the formulation of section II we 
may use one of two strategies. The first strategy is to
consider the relation between the $\hbar$-spectrum
and the $S$-spectrum. The former is obtained by fixing 
the energy $E$ and varying $\hbar$. It is convenient 
to define the quantal variable $\alpha=1/\hbar$, and to 
cast (\ref{e4_1}) into the following form
\begin{eqnarray} \label{e4_2} 
\left.\sum\frac{2\pi}{v_n}
\delta(\alpha{-}\alpha_n)\right|_{osc} \ = \
\sum_j A_j \exp\left(i\alpha S_j\right)
\end{eqnarray}
where $v_n=\alpha(dE_n/d\alpha)$ are the so-called 
velocities. The dependence of the actions $S_j$ on the 
energy $E$ is not emphasized in the above formula since the 
value of $E$ is fixed to some constant value. With this notations the 
formulation of the relation between the quantal $\alpha$-spectrum
and the classical $S$-spectrum is exactly as for billiards, 
the only difference being that now also the quantal 
spectrum is weighted.

If one insist to fix $\alpha$ and to consider the 
statistical properties of the $E$-spectrum, then 
a different strategy is required.  
Picking an energy $\bar{E}$, one should assume that it
belongs to an energy interval $[E_1,E_2]$ where the POs 
are structurally stable (See remark \cite{r_window}), meaning 
that there are no bifurcations of POs while $E$ is varied. 
Thus $S_j(E)$ is a well defined function in this range.    
The time periods are $T_j=(dS_j(E)/dE)$. 
The general statistical theory of  section II applies 
with small modifications.  It is also essential  
to make the approximation $S_p(E)=S_p(\bar{E}){+}(E{-}\bar{E})T_p$,
and to define a scaled energy variable $\omega=(E{-}\bar{E})/\hbar$.
Note that $(E_2{-}E_1)$ is classically a very small energy interval, 
while $\omega_{max}\equiv(E_2{-}E_1)/\hbar$ is semiclassically 
very large. The SCTF is now 
\begin{eqnarray} \label{e4_3} 
\left.\sum 2\pi\delta(\omega{-}\omega_n)\right|_{osc} \ = \ 
\sum_{j}A_j\exp\left(i\alpha S_j+i\omega T_j\right) \ \ \ , 
\end{eqnarray}
The Two point correlation function 
$R_{cl}(\alpha,\epsilon)$  of 
the classical spectrum will be considered, where $\epsilon$ 
corresponds to difference in the energy variable $\omega$, 
and $\omega=0$ without loss of generality. 
In order to establish the basic statistical 
relation $K_{qm}(\alpha,t)=K_{cl}(\alpha,t)$,  it turns
out that the proper definition of the classical 
two point correlation function should be
\begin{eqnarray} \label{e12} 
R_{cl}(x,t) \ = \ \left\langle\sum_{ij}A_iA_j
\ \delta(x{-}(S_j{-}S_i))\ \delta(t-T_i) \right\rangle 
\ - \ const\ \ \ ,
\end{eqnarray}   
where $const$ is the subtracted smooth component.  
This two point correlation function does not
correspond to any one-point density unless there
is a scaling relation such that $S_j \propto T_j$.
Actually it is sufficient to assume a 
strong statistical correlation within families, 
rather than an exact global scaling relation.  

There is a better formulation of the general 
quantal-classical duality, which enables to 
treat {\em on equal footing} both the classical 
and the quantal spectra, as in the special case 
of billiard systems. This formulation
is based on a time-domain version of the SCTF. 
The validity of standard FT procedure is 
somewhat more subtle, and therefore we shall  
only sketch the main idea.      
The time domain version of the SCTF is
\begin{eqnarray}  
\sum\exp\left(-i\frac{E_n}{\hbar}t\right) \ = \
\frac{1}{(i2\pi\hbar)^{1/2}}\sum_p\sum_{r} 
\left(\frac{dE_p}{dt}\right)^{1/2}
\frac{t}{\sqrt{|det(M_p^r-I)|}}
\exp\left(i\frac{r{\cal A}_p}{\hbar}-i\frac{\pi}{2}r\nu_p\right)
\end{eqnarray}
where ${\cal A}_j=r{\cal A}_p$ denotes the {\em full} action
(integral of the Lagrangian) of POs whose period is $t$. 
With the appropriate definitions, this relation can be cast 
into the form  
\begin{eqnarray} \label{e_ts} 
\left(\frac{2\pi}{\alpha}\right)^{1/2} 
\sum_n\exp(-it\omega_n(\alpha)) \ = \
\sum_{j}A_j\exp(i\alpha{\cal A}_j(t)) \ \ \ , 
\end{eqnarray}
where $\alpha=1/\hbar$. An optional way to write the same is  
\begin{eqnarray} 
{\cal FT}_{[\omega\leadsto t]} \: \rho_{qm}(\alpha,\omega) 
\ \ \ = \ \ \
{\cal FT}_{[{\cal A}\leadsto \alpha]} \: \rho_{cl}({\cal A},t)
\ \ \ \ .
\end{eqnarray}
Note that the classical density $\rho_{cl}({\cal A},t)$ 
depends parametrically on $t$, while the quantal density
$\rho_{qm}(\alpha,\omega)$, depends parametrically on 
$\alpha$. Note also that the smooth $\alpha$ dependent 
factor  $({2\pi}/{\alpha})^{1/2}$  of equation (\ref{e_ts}) 
should be absorbed into the definition of the quantal 
density. Both sides of the equation are functions of 
$(\alpha,t)$. Squaring them, and performing statistical 
averaging, one obtains the basic statistical relation 
$K_{qm}(\alpha,t)=K_{cl}(\alpha,t)$ in complete analogy
to the billiard formulation of the quantal-classical duality.

\subsection{Billiards with magnetic field, Periodic chains}

For a billiard in a uniform magnetic field
one may use the general trace formula (\ref{e4_3}) 
with $\alpha$ replaced by $k$, and
$T_j$ replaced by $L_j$, and 
$S_j = L_j+b{\cal A}_j$.  
Here ${\cal A}_j$ is 
the net area that is enclosed by the PO, 
while $b=(qB/c)/(mv)$ is the scaled magnetic 
field. For simplicity 
units of length are used also for the 
action.
If the magnetic field is concentrated 
in one flux line then
\begin{eqnarray} \label{e4_5} 
S_j = L_j+b\nu_j
\end{eqnarray}      
where $\nu_j$ is the winding number and
$b=(q\phi/c)/(mv)$ is the scaled magnetic flux
(Note that it has now the dimension of length).
In particular one may consider Aharonov-Bohm 
ring geometry. The same formulation applies. 

Periodic chains can be treated using 
the same formulation that is applied for the  
semiclassical study of Aharonov-Bohm ring. 
The quasi-momentum 
$q$ is a constant of the motion, and therefore 
can be regarded as a parameter. Eigenstates
that correspond to $q$ should satisfy 
the Bloch condition $\psi(x+a)=\exp(iqa/\hbar)\psi(x)$. 
This is equivalent to solving one-cell problem 
with periodic boundary conditions and a scaled
magnetic flux $b=(qa)/(mv)$.  
The winding number $\nu$ of the folded PO characterizes
the periodicity of the corresponding unfolded PO. 
 
The form factor $K(b,k,L)$ depends parametrically on $b$.
If one averages over $b$, then one obtains $\bar{K}(k,L)$.
It is a simple exercise to prove that the corresponding 
classical two-point function satisfies
\begin{eqnarray} \label{e4_6} 
\bar{K}_{cl}(k,L) \ = \ \sum_{\nu}K_{\nu}(k,L) \ = \ 
\sum_{\nu} C_{\nu}(k,L) K_{\nu}(L) \ \ , 
\end{eqnarray}
meaning that cross-correlations between classes of POs
that are distinguished by their winding number 
should be ignored in computation of $b$ averaged 
two-point statistics. The last equation would become a 
special case of (\ref{e_Kalpha}) if cross-correlations
were negligible.

\subsection{General Integrable Systems }

For general time-independent integrable Hamiltonian 
system with $d$ degrees of freedom the SCTF is:
\begin{eqnarray} \label{e4_7} 
\left. \sum\delta(E{-}E_n) \right|_{osc} \ = \
\frac{1}{2\pi\hbar^{(d{+}1)/2}}\sum_p\sum_{r}
\frac{T_p}{\sqrt{K_p}} |r\vec{m}_p|^{(d{-}3)/2} 
\exp\left( i\frac{rS_p}{\hbar}
-i\frac{\pi}{2}\vec{\alpha}{\cdot}r\vec{m}_p
+i\frac{\pi}{4}\beta_p \right)
\end{eqnarray}
Here the sum is over rational tori $p$, whose 
frequency vector is $\vec{\omega}_p=(2\pi/T_p)\vec{m}_p$,
where $m_p$ is an integer vector of relatively 
prime integers and $T_p$ is the primitive period. 
The scalar curvature of the torus is $K_p$, and 
the Maslov-Morse indices are $\vec{\alpha}$ and $\beta$.  
Using the same strategy as in subsection A, and using 
analogous notations, we can cast this formula into the form
\begin{eqnarray} \label{e4_8} 
\frac{1}{\alpha^{(d{-}1)/2}} \left.\sum 
2\pi\delta(\omega{-}\omega_n)\right|_{osc} \ = \ 
\sum_{j}A_j\exp\left(i\alpha S_j+i\omega T_j\right) \ \ \ , 
\end{eqnarray}
where $\alpha=1/\hbar$. The quantal form factor 
$K_{qm}(\alpha,t)$ that correspond to the left 
hand side, should be equal to the classical 
form factor $K_{cl}(\alpha,t)$ that corresponds
to the right hand side, defined as in subsection A.   

The asymptotic behavior of the quantal form factor
in the deep quantal domain (Q) is determined by the 
quantal diagonal approximation, namely 
$K_{qm}(t)\approx K_Q(\alpha)$. It reflects the 
self-correlations of the levels, and therefore, 
if we assume no degeneracies, it is determined 
merely by Weyl law. One obtains
\begin{eqnarray} \label{eq:kqint} 
K_{qm}(\alpha,t) \ \approx \ K_Q(\alpha) \ = \ 
\frac{1}{(2\pi)^{d{-}1}} \frac{d\Omega(E)}{dE}  
\ \ \ \ \ \ \ \mbox{deep quantal regime}. 
\end{eqnarray}
Similarly, in the deep classical regime (C) one 
may apply the classical diagonal approximation. 
Here one should utilize the version of the Hannay 
and Ozorio de-Almeida sum rule that applies to 
integrable systems \cite{Hannay}. See also   
appendix B of \cite{thomas}. Assuming no 
degeneracies one obtains
\begin{eqnarray} \label{eq:kdint} 
K_{cl}(\alpha,t) \ \approx \ K_D(t) \ = \ 
\frac{1}{(2\pi)^{d{-}1}}
\frac{d\Omega(E)}{dE}  
\ \ \  
\ \ \ \ \ \ \ \mbox{deep classical regime}. 
\end{eqnarray}
Quite surprisingly $K_Q(\alpha)$ does not depend 
on $\alpha$, $K_D(t)$ does not depend on $t$, and 
both are equal. The quantal diagonal approximation 
is valid in the deep Q-regime and the classical 
diagonal approximation is valid in the deep 
C-regime of the $(\alpha,t)$ plane. It is therefore 
very tempting to assume that the form factor 
$K(\alpha,t)$ is constant over the entire $(\alpha,t)$
plane, and it does not experience any actual 
crossover. We shall show below that this is indeed 
the typical case: For a generic non-degenerated spectrum that 
corresponds to a generic integrable system both the quantal  
and the classical spectra are of Poisson type.     

The time domain formulation of the quantal-classical 
duality can be extended also to the case of integrable 
systems. Rather than considering actions $S_j$ of 
rational tori on the energy surface $E$, 
one should consider the actions ${\cal A}_j$ of rational 
tori with a period $t$. Then one obtains a 
SCTF that can be cast into the form of (\ref{e_ts}) with 
the $\hbar$ dependent factor $({2\pi}/{\alpha})^{1/2}$
replaced by $({2\pi}/{\alpha})^{d/2}$, where $d$ 
is the number of degrees of freedom. 
  
The Poissonian nature of both the quantal spectrum 
and the (dual) Poissonian nature of the classical spectrum,  
can be deduced using the following argumentation.
The number of energies  up to an energy $E$ equals essentially 
to the number of lattice points $\vec{I}=\hbar\vec{n}$, such that 
$E(\vec{n})<E$.  
Analogously, the number of actions is given by the
number of lattice points $\vec{\omega}=(2\pi/t)\vec{m}$, 
such that ${\cal A}(\vec{m})<{\cal A}$.   
The quantal and the classical counting functions 
grow by $1$ whenever the respective constant $E$ surface
or the constant ${\cal A}$ surface cross a discrete 
lattice point.
The crossings occur at values of $E$ (or ${\cal A}$)
which depend in a very complex way on the properties 
of the moving surface, and introduce the spectral fluctuations 
which decorate the smooth counting functions.
Berry and Tabor \cite{BerryTabor} where the first to relate 
the spectral statistics of integrable systems to the problem of
counting the number of lattice points within a surface. 
They have shown that the fluctuations for generic surfaces 
are Poissonian. From this point of view, the
constant $E$ or constant ${\cal A}$ surfaces are similar, and therefore,
the statistics for both counting-functions must be Poissonian.
The argument should be modified somewhat if one considers 
the actions $S_j$ rather than the actions ${\cal A}_j$.   
Each point in the $\vec{\omega}$ space labels a torus, and 
the energy surface is represented by some closed surface. 
Tori that correspond to the same frequency ratio $\vec{m}$ are 
located along a ray. The intersection of the ray with the 
energy surface determines the period $t$, and the corresponding 
torus is such that $\vec{\omega}=(2\pi/t)\vec{m}$. 
Let us associate with the ray a length that correspond to 
the action $S_j$ of that rational torus. Locally the tips 
of the rays will form a rectangular grid structure in any 
direction. However, globally they form an irregular array 
of points. The number of actions $S_j<S$ is equal to the 
number of tips that are contained in a sphere whose radius 
is $S$. Rather then having an irregular surface that moves 
in a regular grid we have a `regular' sphere that moves
in an irregular grid. Therefore the same considerations 
that lead to the deduction of Poisson statistics apply also 
in this case.

\subsection{Example: The 3D Torus Billiard}

In this subsection we discuss the 3-dimensional torus billiard. It
will serve as an interesting illustration of the duality relations
presented in the previous subsection. We discuss first the general 
3D torus with unequal edges $(a_1, a_2, a_3)$. The quantal spectral 
density is given explicitly as:
\begin{equation}
  d_{qm}(k) = \sum_{\vec{\rho} \in \bbbz^3} \!\!\!
  \delta \left( k - k_{\vec{\rho}} \right) \; , \; \; \;
  k_{\vec{\rho}} \equiv 
  2 \pi \sqrt{ 
    ( \rho_1 / a_1 )^2 + ( \rho_2 / a_2 )^2 + ( \rho_3 / a_3 )^2 } \, .
\end{equation}
The classical density is:
\begin{equation} 
  d_{cl}(L) \equiv 
  \sum_{\vec{\rho} \in \bbbz^3, \vec{\rho} \neq \vec{0}} \!\!\!\!\!\!
  \delta \left( L - L_{\vec{\rho}} \right) \; , \; \; \;
  L_{\vec{\rho}} \equiv 
  \sqrt{ (\rho_1 a_1)^2 + (\rho_2 a_2)^2 + (\rho_3 a_3)^2 } \, .
  \label{eq:3torus-dcl}
\end{equation}
The Berry-Tabor semiclassical trace formula for the quantal density of
states is obtained by performing Poisson summation of $d_{qm}$ which
can be carried out exactly:
\begin{equation} \label{e_brtb}
d_{qm}(k) = \frac{V k^2}{2 \pi^2} + 
\frac{V k}{2 \pi^2} \!\!\!
\sum_{\vec{\rho} \in \bbbz^3, \vec{\rho} \neq \vec{0}} 
\!\!\! \frac{\sin (k L_{\vec{\rho}})}{L_{\vec{\rho}}} =
\bar{d}(k) + \frac{V k}{2 \pi^2} \int_{-\infty}^{+\infty} d_{cl}(L) 
\frac{\sin(kL)}{L} {\rm d}L \; , \; \; \;
V \equiv a_1 a_2 a_3 \, .
\end{equation}
In order to cast (\ref{e_brtb}) into the explicitly dual form
$\rho_{qm}={\cal FT}\rho_{cl}$ one should define
\begin{equation}
\rho_{qm}(k) \equiv \frac{2 \pi}{k} 
                    \left[ d_{qm}(k) + d_{qm}(-k) \right]_{osc} \; , 
\; \; \; \; \; \;
\rho_{cl}(L) \equiv \frac{i V}{2 \pi L} 
                    \left[ d_{cl}(L) + d_{cl}(-L) \right] \ .
\end{equation}
From these expressions we observe that both the quantum and classical
spectra are supported by 3D rectangular lattices, and hence they
should have the same statistics. The argument at the end of the last
subsection further suggests that this statistics is Poissonian.
Moreover, both the classical and the quantal asymptotes (diagonal
approximations) of the form factor have the same value:
\begin{equation}
K_D(L) = K_Q(k) = \frac{V}{\pi} \, ,
\end{equation}
which is indeed independent of both $k$ and $L$, confirming
(\ref{eq:kqint}, \ref{eq:kdint}). All these arguments strongly suggest
that the form factor is a constant for all $k, L$:
\begin{equation}
  K(k, L) = \frac{V}{\pi} \, .
\end{equation}
We note that the conventional form factor which is calculated from
$d_{qm}$ rather than from $\rho_{qm}$ is $V k^2 / \pi$ and is not
purely classical for small $L$ values (classical diagonal regime).

We now turn to the special and interesting case of the cubic 3-torus:
$a_1 = a_2 = a_3 = a$. In this case, both the quantum and the
classical spectra are directly proportional to the vectors of the
cubic lattice $\vec{\rho} \in \bbbz^3$. The vectors $\vec{\rho} = (l,
m, n)$ exhibit a geometrical degeneracy of their lengths
$\rho=\sqrt{l^2+m^2+n^2}$ due to permutations of the indices and/or
sign changes. Much more interesting is the number-theoretical 
degeneracy of $\bbbz^3$, for example $5^2+6^2+7^2=1^2+3^2+10^2=110$. 
Denoting by $g(\rho)$ the overall degeneracy of $\rho$:
\begin{equation}
g(\rho) \equiv 
\# \left\{ \vec{\kappa} \in \bbbz^3 : \ 
|\vec{\kappa}| = \rho \right\} \, ,
\end{equation}
one have the following results \cite{KeaPC,RudPC}:
\begin{equation}
\frac{\langle g(\rho)^2 \rangle}{\langle g(\rho) \rangle} =
\beta \rho \; , \; \; \; \beta = {\rm const}.
\end{equation}
Applying this result to the classical and quantal diagonal
approximations give:
\begin{equation}
K_D(L) = \frac{a^2 \beta}{\pi} \, L \; , \; \; \;
K_Q(k) = \frac{a^4 \beta}{2 \pi^2} \, k \, .
\end{equation}
These asymptotes are in sharp contrast with the Poisson (generic
integrable) case and are actually reminiscent of RMT predictions
(generic chaotic) in 2 dimensions! This unexpected behavior of
``chaotic'' form factor for an integrable system is due to the
number-theoretical peculiarities of the cubic 3-torus. This is an
interesting example, when the classical-quantal duality is explicit,
and yet a statistics which is more interesting then Poisson is
obtained.

The expected break time $L^*$ occurs when the two asymptotes
intersect:
\begin{equation}
L^*(k) = \frac{a^2}{2 \pi} \, k \, .
\end{equation}
This equation defines the crossover line in the $(k, L)$ plane between
the classical and the quantal regimes as discussed in (\ref{e_regs}).
Thus we deduce that the classical correlation scale is given by:
\begin{equation}
\lambda(L) = \frac{a^2}{L}
\end{equation}
while the quantal correlation scale is:
\begin{equation}
\Delta(k) = \left( \frac{2 \pi}{a} \right)^2 \frac{1}{k}.
\end{equation}
We can relate the scale $\lambda$ to the minimal distance between
lengths of POs. Indeed, due to the structure of the
POs $L_{\vec{\rho}} = a \rho = a \sqrt{\rm integer}$, the
minimal difference between lengths near length $L$ is
\begin{equation}
\Delta L \approx \frac{a^2}{2 L}
\end{equation}
and consequently
\begin{equation}
\lambda(L) = 2 \Delta L \ .
\end{equation}
This indicates that the breakdown of the diagonal approximation is
related to the ``number-theoretical rigidity'' of spacing $1$ between
integers. Moreover, it indicates that there is only {\em one} underlying
family of POs in the sense that was discussed in
subsection~\ref{subsec:families}. Analogously, we get for the quantal
scale:
\begin{equation}
\Delta(k) = 2 \Delta k \ ,
\end{equation}
where $\Delta k$ is the minimal spacing between consequent 
energy levels.

\subsection{Mixed boundary conditions and Sinai billiards}
\label{subsec:mbc}

\begin{figure}[htb]
  \begin{center}
    \leavevmode
    \psfig{figure=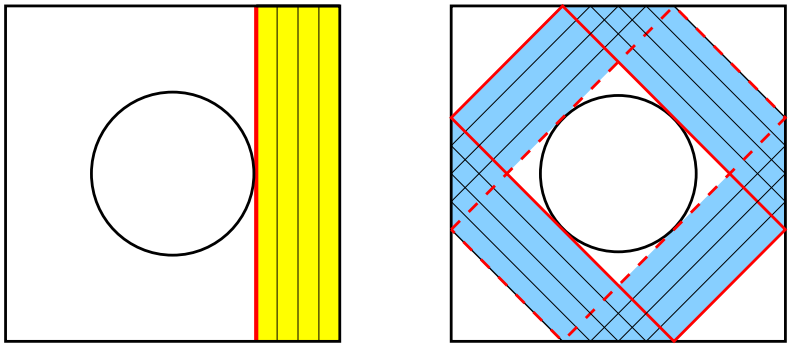,height=5cm,angle=270}
    \caption{Example of two bouncing ball families in 2D Sinai billiard. 
      The families constitute shaded area, and a few representative
      orbits are demonstrated by the thin lines. The marginal
      tangential orbits are denoted by bold lines.}
    \label{fig:bb-sb2d}
  \end{center}
\end{figure}

We present in this subsection a class of billiard systems which will
be central to our numerical studies. The most general boundary
conditions under which the quantum billiard problem is self-adjoint
can be written as \cite{SPSUS95}:
\begin{equation}
\kappa \cos\beta \, \psi(\vec{r}\,) + 
\sin\beta \, \partial_{\vec{n}} \psi(\vec{r}\,) = 0 \; , \; \; \;
\vec{r} \in (\mbox{boundary of the billiard}) \; , \; \; \;
{\vec{n}} = \mbox{normal pointing outside} \; ,
\label{eq:mbc}
\end{equation}
where the angle $\beta$ interpolates smoothly between Dirichlet
($\beta = 0$) and Neumann ($\beta = \pi / 2$) cases and $\kappa$ is a
constant with the dimension of a wavenumber. It was shown in
\cite{SPSUS95} that the Gutzwiller trace formula for billiards can be
modified to incorporate mixed boundary conditions:
\begin{eqnarray}
N(k ; \kappa, \beta) 
& \equiv &
\sum_{n} \Theta \left[ k - k_{n}(\kappa, \beta) \right] 
\nonumber \\ 
& = &
\bar{N}(k ; \kappa, \beta)+
\sum_{PO} A_j \exp \left\{ i r \left[
    k L_p - \frac{\pi}{2} \nu_p + \Phi_p(\kappa, \beta) 
  \right] \right\} \; , \\
\Phi_p(\kappa, \beta) 
& \equiv & 
\sum_{i = 1}^{n_p} \phi ( \theta_{i}^{p}; \kappa, \beta ) \; , \; \; \;
\phi = (-2) \arctan \left( \frac{\kappa \cot \beta}{k \cos \theta} \right)
\, .
\label{eq:gutzmbc}
\end{eqnarray}
In the above, $A_j = 1/(2 \pi i r \sqrt{|M^{|r|}_p - I|})$ are the
instability amplitudes, $p$ is the index of primitive POs,
$r \in \bbbz$ is the repetition number, $\nu_p$ is the Maslov index
due to focusing only, $n_p$ is the number of reflections from the
billiard`s boundary of the $p$'th primitive PO, and
$\theta_{i}^{p}$ are the reflection angles with respect to the inwards
normals. Explicit expressions for $\bar{N}(k ; \kappa, \beta)$ appear
in \cite{SPSUS95}. The parameter $\kappa$ can be generalized to become
a piecewise constant function on the boundary, with the obvious
modifications of $\Phi_p$. This is a very powerful tool to selectively
filter semiclassical contributions of POs. If we change
the boundary conditions only on a segment of the billiard (e.g.\ the
disc in the 2D Sinai billiard) then orbits are selected according to
the number of reflections from this segment only. Let us proceed by
taking the derivative of (\ref{eq:gutzmbc}) at $\beta = 0$ (near
Dirichlet) and get the expression
\begin{eqnarray}
\left. \frac{\partial}{\partial \beta} N(k ; \kappa, \beta)
\right|_{\beta = 0}
& = &
\sum_n \left. \frac{\partial}{\partial \beta}  k_n(\kappa, \beta)
\right|_{\beta=0} \delta(k - k_n) \nonumber \\
& = &
\left. \frac{\partial}{\partial \beta} \bar{N}(k; \kappa, \beta)
\right|_{\beta=0} +
\sum_{PO} A_j B_j \exp \left[ i r 
\left( k L_p - \frac{\pi}{2} \nu_p - n_p \pi \right) \right] \ , 
\label{eq:derivmbc} \\
k_n 
& \equiv & 
k_n({\rm Dirichlet}) \; , 
\; \; \; \; \; \;
B_j \equiv \frac{2 i r k}{\kappa} 
\sum_{i = 1}^{n_p} \cos\theta_{i}^{p} \nonumber \ .
\end{eqnarray}
Equation (\ref{eq:derivmbc}) is similar in structure to the standard
trace formula, but for a few modifications. The quantal part is
essentially the density of states for Dirichlet boundary conditions
dressed by the ``velocities'' $v_n \equiv \partial k_n(\kappa, \beta)
/ \partial \beta |_{\beta=0}$. The phases in the semiclassical
summation over POs are the same as for the Dirichlet case,
and the semiclassical amplitudes are modified relative to the standard
case: $A_j \rightarrow A_j B_j$. The most important application of
this formula is to efficiently filter the non-generic contributions of
the ``bouncing-balls'' in the 2D and 3D Sinai billiards. The
bouncing-balls are families of neutral POs which are of
classical measure 0, whose effects completely shadow the contributions
of generic chaotic POs in the semiclassical analysis
\cite{SSCL93,PS95,SS95}. Instead of subtracting their contributions
explicitly which is a difficult task in 3D, we use
(\ref{eq:derivmbc}): Taking the derivative with respect to $\beta$ is
essentially a subtraction of two spectra with different boundary
conditions on the sphere. Since the bouncing balls never reflect from
the sphere, their leading contribution is trivially eliminated.
Moreover, $B_j \propto \sum_i \cos\theta_{i}^{j} = 0$ if and only if
the PO is wholly tangential to the sphere, thus we
eliminate also the closures of the bouncing ball manifolds (see figure
\ref{fig:bb-sb2d}) which can otherwise overwhelm the generic
contribution. Equation (\ref{eq:derivmbc}) therefore relates the
spectra of Sinai billiards with contributions of generic periodic
orbits only.

Finally, to cast (\ref{eq:derivmbc}) into the standard from
$\rho_{qm}(k) = {\cal FT} \rho_{cl}(L)$ we divide both sides by $k$:
\begin{equation}
\sum_n \left. \tilde{v}_n \delta(k-k_n) \right|_{osc} =
\sum_{PO} \tilde{A}_j \exp ( i k L_j) \; , 
\; \; \; \; \; \;
\tilde{v}_n \equiv \frac{v_n}{k_n} \;, 
\; \;
\tilde{A}_j \equiv
\frac{A_j B_j}{k} 
\exp \left( - i \frac{\pi}{2} \nu_j - i n_j \pi \right) \ .
\label{eq:mbcft}
\end{equation}
The left hand side is a weighted density which is purely quantal, and
it is a Fourier transform of a purely classical density $\sum_j
\tilde{A}_j \delta(L-L_j)$.

\subsection{Theory of the primes}  

It is instructive to test our formalism in the 
special limit  ``$d\rightarrow 1$'' which corresponds
to the spectrum of the primes \cite{primes}. 
The ``classical spectrum'' consists of the 
real numbers $L_j=r\log(p)$ where $p$ are the primes
and $r$ is the repetition index. The ``quantal spectrum'' 
is defined by the non trivial zeros of the Riemann
Zeta function, namely $\zeta(1/2+ik_n)=0$. By the Riemann
hypothesis the spectrum $\{k_n\}$ consists of real numbers. 
The ``quantal'' spectrum is related to the ``classical''
one by a trace formula which is formally identical with  
the SCTF (\ref{e_SCTF})  with $A_j=L_p\exp(-rL_p/2)$. 
The average density of the classical spectrum is  
$\langle\sum\delta(L{-}L_j)\rangle = \exp(L)/L$ 
and therefore the Haney-Ozorio sum rule (\ref{e2_3}) is satisfied.
The average density of the quantal spectrum is 
$\langle\sum 2\pi\delta(k{-}k_n) \rangle = \log({k}/{2\pi})$.
This functional form corresponds in some sense to
the dimensionality $d{\searrow}1$.  From here (via (\ref{e_regs}))  
it is expected that the classical spectrum is characterized 
by a correlation scale $\lambda_0=\exp(-L)$. 
This correlation scale expressed 
in {\em natural} units (via the relation $L=\log(n)$) 
is simply $\Delta n = 1$. Thus $\lambda_0$ has a trivial number 
theoretical interpretation. It simply reflects the fact that 
the classical spectrum is supported by an underlying 
grid that is formed by the natural numbers.   
The manifestation of the $\lambda_0$ scale in two-point 
correlations in this case is not a ``conjecture'' but rather 
a simple mathematical fact. 

Rather than going on ``backwards'' as in section II-C, let us
use at this point the so called Hardy-Littlewood conjecture.
Following Keating \cite{primes} the statement is that if $n$ is a prime, 
than the probability that $n{+}\Delta n$ is also a prime is
reduced, the factor being $(1{-}1/(2\Delta n))$. The statement holds 
asymptotically, but we shall assume that it actually holds
for any $1{<}\Delta n$. Thus the two point correlation function 
of the classical spectrum is characterized by repulsion with  
$p(x)=(1/(2\Delta n))\cdot\langle\sum\delta(L{-}L_j)\rangle$ 
with $x=\Delta\log(n)$. It follows that that 
$p(x)=1/(2Lx)$ with lower cutoff at $\lambda=\exp(-L)$.

Having ``derived'' an expression for the classical correlations, 
we should verify now that indeed the basic relation $K_{qm}=K_{cl}$
is satisfied. To be more precise, we wish to examine whether
using $p(x)$ one can calculate semiclassically the departure 
from the diagonal approximation, and whether the correct asymptotic 
behavior $K_{qm}(L)\rightarrow K_Q(k)$ is achieved. 
Else, one should conclude that the trace formula would not converge 
to a density that corresponds to the spectrum of Riemann zeros.
Obviously, for $k\ll{2\pi}/{\lambda}$ the diagonal 
approximation $C(k,L)\approx 1$ is valid.
For ${2\pi}/{\lambda}\ll k$ one easily 
find via (\ref{e_Ckl}) that  $C(k,L)\approx\log({k}/{2\pi})/L$.
Hence $K(k,L)\rightarrow K_Q(k)$
as is expected. Note that the normalization of $p(x)$
is satisfied in an asymptotic sense, and is  
reflected via (\ref{e_Ckl}) by the observation that 
$C(k,L)\rightarrow 0$ in the deep quantal 
regime (large $L$).   

The spectrum of the primes is an example for system
where the correlation scale $\lambda$ is smaller 
than the average level spacing. This should be contrasted 
with $d$-dimensional systems where the mean level spacing 
is very small due to the fact that the spectrum is 
apparently a union of many independent sub-spectra (families).         
This observation is important since it demonstrate
that $2\pi\ll k \Delta L$ is neither sufficient
nor necessary condition for the validity of the 
diagonal approximation 
$|\sum_jA_j\exp(ikL_j)|^2 \approx \sum_j A_j^2$.  
In case of the primes $\lambda\sim\exp(-L)$ which 
leads to $L\sim\log(k)$.
If one, by mistake, considered $\Delta L \sim L\exp(-L)$ 
as the relevant scale, one would deduce a {\em longer}
breaktime (still it will be roughly logarithmic in $k$).   
The same wrong time scale would be deduced in case  
of billiards. However, for billiards it would be 
much {\em shorter} than the actual Heisenberg breaktime, 
as discussed in section II-E.  

The non-trivial statistical properties of the classical
spectrum, in case of the primes, are implied by the fact 
that the spectrum is supported by an underlying grid that 
is formed by the natural numbers. In analogy, one may 
speculate that the non-trivial statistical properties of 
the classical spectrum, in case of billiards, is due to 
the `fact' that long POs are supported by short 
ones \cite{shadowing}.

\section{Complex billiards}
\setcounter{equation}{0}

We use the term `complex billiard' when the chaotic dynamics 
of the billiard is characterized by a non-trivial approach  
to ergodicity. Such `non-universality' should have manifestations 
in the one-point and in the two-point statistical properties of 
both the classical and the quantal spectrum. {\em It is associated 
with having non-trivial classification of POs}. 
As a first example we mention the 3D Sinai billiard. 
We shall see (section VI) that it is essential to classify 
the POs by the number of times that they are reflected by the 
inscribed sphere. In the following subsections we discuss 
the complexity that arise due to either spatial proximity 
or global symmetry.

It is important to understand that the existence of  
`classical correlations' is not restricted to 
simple ballistic billiards. On the contrary, the 
effect of PO-correlations should be much more 
significant when complex systems are involved,  
since relatively short breaktime-scales may show up.
Our interest in complex systems is twofold. On the one 
hand we want to develop a semiclassical theory which is 
not limited by the diagonal approximation. On the other 
hand we want to test the limits of the semiclassical 
approach. There is a long standing question concerning the 
feasibility of having PO-theory for the spectral properties 
of system with either band structure or Anderson localization.
We want to use our present level of understanding of the SCTF 
and of classical PO-correlations in order to give an 
affirmative answer to this question. 
For clarity of presentation we shall sometimes use the 
symbol $t$ rather than $L$ in order to denote lengths. 
This is done in order to adhere to common notations and 
terminology.

\subsection{Symmetries and Symmetry Breaking } 

If a system is characterized by a symmetry, for 
example a reflection symmetry, then it is natural 
to consider first the POs of the folded system, 
and then, from them, to generate the POs of the 
unfolded system. Most of the POs of the 
unfolded system will not be conjugate to 
themselves under symmetry.   
Therefore, they will have some degeneracy $g$.
One may regard this degeneracy as the simplest   
example for a classification of POs.
The corresponding two points statistics will
be characterized by a correlation factor
\begin{eqnarray} \label{e5_1} 
C(k,L)=g\hat{C}_0(k\lambda_g(L,\Omega)) \ \ \ \ .
\end{eqnarray}
It is assumed here that further classification 
of POs is not required, else an explicit 
class index $\alpha$ should be introduced. 
The scaling function $\hat{C}_0$ refer to the 
bare length spectrum which is obtained 
by ignoring the degeneracy. The scaling 
constant should correspond to the folded 
billiard, therefore 
$\lambda_g(L,\Omega)=\lambda_0(L,\Omega/g)$.
It turns out that the latter scaling
strategy is useful also in the case of 
time reversal symmetry, though the argumentation 
for having $\Omega/2$ is less transparent.
As a general rule phase-space volume
rather than physical volume should be relevant. 
The scaling functions $\hat{C}_0(\kappa)$ for 
GUE statistics as well as for GOE statistics 
has been already displayed in section II-F.

\begin{figure} 
\begin{center}
\leavevmode 
\epsfysize=3.0in
\epsffile{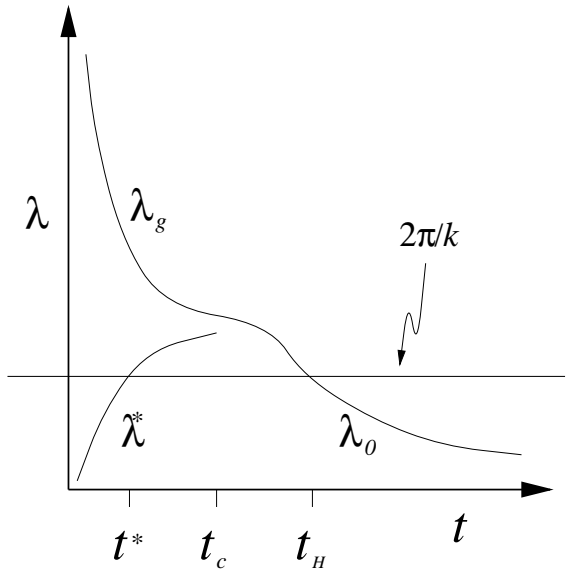}
\end{center}
\caption{\protect\footnotesize 
The universal correlation scales $\lambda_0$ and $\lambda_g$ 
and the non-universal correlation scale $\lambda^*$ are plotted 
against $t$. The horizontal line facilitate the 
determination of the various time-regimes for a given $k$.} 
\label{f_lambda}
\end{figure}

Next, we should discuss the effect of 
symmetry breaking due to some deformation 
of the billiard. We are still interested in 
the two-point statistics of POs whose 
periods are $L_j \sim L$. We assume a small 
deformation $b$ of the surface, that does 
not involve any bifurcations of them. 
Accordingly, the POs change their length    
$L_j \rightarrow L_j+\delta_j$ because of 
the deformation. Our main concern is the statistics 
of the differences $y=\delta_j{-}\delta_i$. Due to 
the central limit theorem we expect 
Gaussian statistics with dispersion 
$\sqrt{{\cal D}L}$ where ${\cal D}$ is 
some constant that formally resembles 
a diffusion coefficient. It is convenient to 
define the deformation parameter via
the relation ${\cal D}=b^2/{\cal L}_0$, 
where ${\cal L}_0$ is the mean chord 
(linear dimension of the billiard). 
The parameter $b$, which has the dimension of length, 
constitutes the natural measure for the 
surface deformation. With these notations      
the distribution of $y$ is 
\begin{eqnarray} \label{e5_3}
G(y)=\frac{1}{ \sqrt{2\pi(L/{\cal L}_0)} b }
\exp\left(-\frac{y^2}{2(L/{\cal L}_0)b^2}\right)
\ \ \ \ .
\end{eqnarray}  
Besides the universal correlations of 
the $L_j$, one may define a non-universal 
correlation scale 
\begin{eqnarray} \label{e5_4}
\lambda^*(L)=\sqrt{(L/{\cal L}_0)} b.
\end{eqnarray}
Unlike the universal correlation scale $\lambda_0(L)$, 
the non-universal correlation scale  $\lambda^*(L)$ is a 
monotonic ascending function of $L$, as illustrated 
in Fig.\ref{f_lambda}.
Following the discussion in section III-D it is 
conjectured that the validity of (\ref{e5_4}) is restricted 
by the condition $\lambda^*(L)\ll\lambda_0(L)$.
This determines a classical length scale $L_c$ and 
a corresponding time scale $t_c$ beyond which the effect 
of bifurcations should be taken into account. 
It is natural to associate this time scale with 
the crossover $\lambda_g \rightarrow \lambda_0$  
of the universal correlations. From now on we shall use 
the notation $t$ rather than the notation $L$ in order 
to denote lengths. Accordingly, we shall use the notion 
of `time' as a synonym for the notion of `length'.

Recall that the Heisenberg time $t_H(k)$ is determined 
by the condition $k\lambda_0(t) \sim 2\pi$, (See section II-C).  
If $t_H(k)\ll t_c$, then there will be no significant 
manifestation of the symmetry breaking. 
On the other hand, if $t_c \ll t_H(k)$ as in Fig.\ref{f_lambda}, 
then there will arise a non-universal breaktime 
$t^*(k)$ that is determined by the condition
$k\lambda^*(t)\sim 2\pi$. It should be emphasized that 
unlike the universal breaktime scale ($t_H$), the 
additional breaktime $t^*$ does not signify a crossover 
to a quasiperiodic behavior in the present context. 
(This should be contrasted with a  situation which will 
be discussed in subsection E). 
The correlation factor (\ref{e5_1}) is modified due to 
the symmetry breaking as follows (see appendix A)
\begin{eqnarray} \label{e5_5} 
C(k,L)=1+(g{-}1){\cdot}\tilde{G}(k) 
\ \ \ \ \ \  
\mbox{[Classical regime, where $C_0(k,L)=1$]} \ \ \ .
\end{eqnarray}   
Above $\tilde{G}(k)=\exp(-t/t^*)$ is the Fourier 
transform of $G(y)$, and $t^*(k)=2{\cal L}_0/(bk)^2$.

We wish now to discuss whether there are 
circumstances in which a more complete
version of (\ref{e5_5}) exists, which describes the 
actual crossover which is associate with 
the symmetry breaking. 
Taking into account the similarity of 
statistically related POs, one should 
specify whether the deformation preserves the 
integrity of families. If the answer is 
negative, then the $\delta_j$ may be considered 
uncorrelated random variables, and then we have 
as in section III-D the result 
$C(k,L)=[1+\tilde{G}(k){\cdot}(gC_0(k\lambda_g)-1)]$.
However, as discussed in section III-D the 
validity of the latter expression should be limited 
to the regime where it coincides with (\ref{e5_5}) . 
However, if integrity of families is preserved, 
then one may assume that $\delta_j$ for POs of 
the same family are the same, but otherwise they 
are uncorrelated random variables. Then one obtains  
\begin{eqnarray} \label{e5_6}
C(k,L) \ = \ [1+(g{-}1)\tilde{G}(k)]{\cdot}C_0(k\lambda_g)
\end{eqnarray}
The validity of the latter formula is not 
restricted by bifurcations, since
it has been assumed that the family 
structure is essentially preserved. 
Still, the global description of the  
symmetry breaking is not satisfactory since the 
$\lambda_g \rightarrow \lambda_0$ crossover 
is missed. The actual reason for this 
failure is further illuminated in the next 
paragraph.

Breaking of time reversal symmetry deserves a particular 
attention. It is simplest to discuss Aharonov-Bohm 
ring geometry, since changing the magnetic field does 
not involve bifurcations of POs. Rather, 
$L_j \rightarrow L_j+\delta_j$, where $\delta_j=b\nu_j$. 
Here the POs are classified by their winding 
number $\nu$ and $b$ is the scaled magnetic field
(see (\ref{e4_5})). 
The distribution of $\nu$ is known to be Gaussian 
with dispersion $\sqrt{{\cal D}L}/a_0$, where 
${\cal D}$ is the coefficient for diffusion along 
the ring (one should imagine an unfolded representation 
of the ring as a periodic multi-cell structure), 
and $a_0$ is the length of the ring (length of 
unit cell in the unfolded structure). The distribution
of $y$ is accordingly given by (\ref{e5_3}), where 
${\cal L}_0\equiv a_0^2/{\cal D}/4$.
Each winding-number class constitute a rigid spectrum 
by itself with some correlation 
factor $\hat{C}_0(k\lambda_0)$. This can be deduced 
by considering the averaged form factor (\ref{e4_6}). 
The scaling function $\hat{C}_0$ should correspond 
to GUE statistics.    
Assuming that there are no cross-correlations, one 
obtains the result (\ref{e5_6}). This result is consistent 
with GUE whenever $\tilde{G}(k)\sim 0$. However, 
if $\tilde{G}(k)=1$, then the correct GOE result is 
not obtained, implying that cross-correlations  between
POs of different winding-numbers are important.  
The identifications of the mechanism that leads 
to these cross-correlations is probably related 
to the geometry characteristics of diagrams which 
are considered in the field-theoretical approach. 
For example, one may consider a PO that follows closely
a loop-shaped orbit (say clockwise) and then 
a second loop-shaped orbit (say anti-clockwise) thus 
creating an $8$ shape. However, due to the time reversal 
symmetry there may be a relate PO that follows 
the second loop clockwise rather than anti-clockwise. 
Thus, the two related POs will have different winding 
numbers, implying the conjectured cross-correlations.        
In this respect the GOE $\rightarrow$ GUE crossover should be 
similar to symmetry breaking of spatial symmetry.  
However, the GOE $\rightarrow$ GUE crossover 
cannot be described merely in terms of change of 
the scaling constant $\lambda_g \rightarrow \lambda_0$, 
since it involves also a change of the scaling 
function. Using different words, unlike spatial symmetries,
the GOE $\rightarrow$ GUE crossover is involved in changing 
universality class.

\subsection{Multi-cell billiards}

Complexity is associated with having non-trivial classification 
of POs. Disregarding symmetries, the simplest
classification is induced by proximity. In the subsequent 
subsections we 
shall discuss the significance of spatial proximity
for the analysis of multi-cell billiards. It should be 
emphasized that the main ideas can be generalized to   
cases where phase-space rather than spatial proximity 
is relevant.      

A multi-cell billiard is a structure that is 
composed of ballistic cavities (cells). 
See for example \cite{thomas,pchain}.
The cells are connected by small openings (holes).
A one-cell billiard means a simple ballistic 
billiard. A two-cell billiard is constructed of two 
connected cavities. An $N$-cell billiard with
$N\rightarrow\infty$ is a prototype for a
1D diffusive system. It is most convenient 
to assume that the last cell is connected to 
the first one, thus creating a closed chain. 
More generally one may discuss $d$-dimensional
structures. Note that the dimensionality of the 
cells-structure may be lower from the 
actual dimensionality of space. For example, one 
may discuss a 1D chain that is composed of 3D cells. 
For the time being we assume that there 
are no special spatial symmetries, neither 
reflection nor translational.  

For a simple ballistic billiard the only 
classical time scale is the ballistic one
(average time between reflections).   
There is no distinct ergodic time scale.   
POs which are longer than a few reflections 
explore the {\em whole} volume. 
In case of a multi-cell billiard
POs may be classified by the 
{\em volume which is explored}. 
This classification implies that 
there is a distinct time scale $t_{erg}$ 
for ergodicity. For POs whose period is  
$t<t_{erg}$ the explored volume is 
typically smaller than the total volume.
For $t_{erg}\ll t$ all the POs are 
ergodic as for a simple ballistic billiard.
The non-universal features that are 
associated with the above classification 
are manifest in the one-point
statistical properties of the spectrum 
and we shall argue that they should 
also be manifest in its two points 
characteristics.

First we discuss one-point properties of 
of the classical spectrum that corresponds
to a multi-cell billiard. These are 
best described in terms of the so called 
classical probability to return $P_{cl}(t)$. 
By definition $P_{cl}(t)/\Omega$ is the 
{\em average} probability density for finding 
the particle in the vicinity 
of its initial point, where $\Omega$ is the 
physical volume. For simplicity we refer here to 
configuration space rather than to phase-space.  
Note that $P_{cl}(t)$ is related, 
essentially by a Laplace transform, to the 
Ruelle Zeta function. The trivial 
eigenvalue corresponds to the ergodic behavior
$P_{cl}(t)=1$. For our multi-cell billiards 
there is a distinct time scale $t_{erg}$, such 
that for $t<t_{erg}$ the classical probability 
to return is larger than $1$. More specifically, 
from the definition it follows that 
$P_{cl}(t)\approx\Omega/\Omega_d(t)$ where $\Omega_d(t)$ 
is the volume over which an initial distribution
spreads after a time $t$. In particular $P_{cl}(t)=N$
initially ($t \ll t_0$), where $N$ is the total 
number of cells, and $t_0$ is the time it takes to 
escape into adjacent cells. 
The relation of the one-point properties of 
the classical spectrum to $P_{cl}(t)$ follow    
from the observation that the latter can be 
expressed as a sum  over POs. Disregarding 
repetitions one obtains \cite{Pcl}
\begin{eqnarray} \label{e5_7}
K_D(t)=tP_{cl}(t) \ \ \ \ .
\end{eqnarray}   
Note that the so-called Hannay and Ozorio de Almeida 
sum rule (\ref{e2_3})
is a special case that corresponds to the ergodic 
behavior. For $t \ll t_0$ this sum rule is 
approximately satisfied separately for each of the cells, 
and summation over the POs of all the cells indeed 
gives the correct result $P_{cl}(t)=N$.  

We turn now to discuss two-point  
correlations of a complex system. The key 
observation is that POs may be classified 
by the {\em explored volume}. This claim 
is especially trivial if $t \ll t_0$. 
Each PO explores one of the volumes that 
corresponds to one of the cells. The POs
may be classified by the cell which is 
explored. There cannot be any cross correlations 
between POs that belong to different cells, 
since a deformation of one cell does not 
affect POs that belong to a different cell.
The same argument can be extended to 
longer times ($t<t_{erg}$). For a given  
cluster of connected cells, corresponds 
a class of POs that explore all of these 
cells. Any other class of POs will have 
a statistically independent spectrum. 
It is sufficient that the corresponding 
clusters will differ by one cell in order 
to extend the above argument. Namely, 
deformation of this one cell will affect 
only one of the corresponding class-spectra, 
implying no cross-correlations.

It follows from the above discussion that 
the form factor can be written as a sum of 
statistically independent contributions, 
namely
\begin{eqnarray} 
K(k,t) \ = \ \sum_{\alpha} \ K_{\alpha}(k,t) \ \equiv \  
\sum_{\alpha} \ \hat{C}_{\alpha}(k,t) \ K_{\alpha}(t)
\end{eqnarray} 
Where $K_{\alpha}(t)$ is the (partial) diagonal sum over POs 
that belong to the $\alpha$-class, $K_{\alpha}(k,t)$ is the 
contribution of these POs to the form factor, and by definition 
$\hat{C}_{\alpha}(k,t)$ is the corresponding correlation factor.
At this stage it is appropriate to suggest  
a conjecture and a further working-hypothesis 
that constitute the basis for applying our formalism to the 
case of a multi-cell billiard. 
\ \\ \ \\
\framebox{
\parbox{12cm}{
\ \\
{\em Ballistic-Like Correlations (BLC):} 
\ \\ \ \\
{\bf The BLC conjecture} - A class
of POs that explore a volume $\Omega_e$ 
is characterized by non-trivial correlation 
scale which is $\lambda=\lambda_0(t,\Omega_e)$.
\ \\ \ \\
{\bf The BLC working hypothesis} -  A class
of POs that explore ergodicly a given volume 
is characterized by non-trivial correlations 
with the same scaling function $C_0(\kappa)$ 
as for a ballistic billiard. 
\ \\
} } \\ 
\ \\ \ \\ 
The BLC working hypothesis leads to the following 
explicit expression for the form factor
\begin{eqnarray} \label{e5_8}
K(k,t) \ = \ \sum_{\alpha} \
\hat{C_0}(k\lambda(t,\Omega_{\alpha})) \ K_{\alpha}(t)
\end{eqnarray} 
Here $\Omega_{\alpha}$ is the volume which is explored by POs
of the $\alpha$-class. The {\em BLC conjecture} seems plausible 
due to the general way of thinking which has been introduced 
in section III. Accepting the claim that POs of a generic ballistic 
billiard are characterized by the universal correlation 
scale $\lambda_0(t,\Omega)$, it is most natural to 
claim that the total $\Omega$ should be replaced by the 
explored $\Omega$ in the more general case of a complex system.    
On the other hand, the {\em BLC working-hypothesis} has no 
a-priori justification. 
Actually we would like to understand the circumstances 
and the way in which it fails.

The strategy of the subsequent subsections is as follows.
Subsections C and D demonstrate how the BLC working-hypothesis 
can be used in order to make a semiclassical computation of 
the spectral form factor {\em beyond the diagonal approximation}. 
This is done mainly for pedagogical reasons. 
The applications in subsections E and F serve as a real 
testing ground for our approach. We consider there a 
quasi-1D classically diffusive chain. In the limit 
$N\rightarrow\infty$ the quantal form factor should correspond 
either to localized eigenstates or to band structures, 
depending on whether the chain is disordered or periodic. 
Our main purpose is to give an answer to the long-standing 
question, whether these results can be derived, at least in 
principle, from the SCTF. We demonstrate that the BLC 
conjecture by itself is valuable, since it gives the correct 
time scales. Furthermore, we try to make a quantitative theory, 
using the BLC working-hypothesis. The final result is not 
completely satisfactory.  However, it demonstrates the 
feasibility of performing effectively such a computation 
in the future.

\subsection{One cell with a hole}       
   
This subsection constitutes an introduction for
later analysis of multi-cell billiards. 
The cell is a $d_0$-dimensional simple chaotic 
cavity ($2\le d_0$). Its volume is $\Omega_0=\ell_0^{d_0}$. 
There is a small opening whose area is $a_0^{d_0{-}1}$.
A generic orbit will escape eventually out of  
the cavity. Each chord of the orbit is roughly 
of length $\ell_0$, and the probability to 
escape, once hitting the surface, is 
$(\ell_0/a_0)^{d_0{-}1}$. Therefore the average 
escape time is 
$t_0=L_0=\Omega_0/a_0^{d_0{-}1}$.
The particle is further characterized by its
wavenumber $k$. Its velocity $v=1$ is 
insignificant physically and it is used merely in 
order to translate units of length into 
units of time. Disregarding the small opening 
the Heisenberg time is 
$t_H^0=L_H^0\approx\Omega_0 k^{d_0{-}1}$.
The dimensionless conductance $g_0$ is defined in the 
usual way as the ratio of the Thouless energy 
$2\pi\hbar/t_0$ to the level spacing $2\pi\hbar/t_H^0$. 
Hence
\begin{eqnarray} \label{e5_9} 
g_0=t_H^0/t_0=(ka_0)^{d_0{-}1} \ \ \ \ .
\end{eqnarray}   
Thus $g_0$ is simply related to the ratio of 
De-Broglie wavelength $2\pi/k$ to the hole size $a_0$.
The De-Broglie wavelength is assumed 
to be much shorter than $\ell_0$ as to allow semiclassical 
considerations. Actually, in order to have a valid 
semiclassical result, it should be smaller or at 
most equal to $a_0$ .

The classical probability to stay is 
found by solving the differential 
equation $dp(t)/dt=-(1/t_0)p(t)$ yielding
$p(t)=\exp(-t/t_0)$. The corresponding 
diagonal sum is
\begin{eqnarray} \label{e5_10}
K_1(t)=\sum_j A_j \delta(t{-}L_j) = t{\cdot}p(t) \ \ \ .
\end{eqnarray}   
The above sum includes those POs that 
survive when the `hole' is opened.

\subsection{Two cell billiard} 

Here the classical probability to return 
is found by solving the coupled equations
$dp_l(t)=-(1/t_0)(p_l(t){-}p_r(t))$ and  
$dp_r(t)=-(1/t_0)(p_r(t){-}p_l(t))$, where 
$p_l$ and $p_r$ are the probabilities to 
find the particle in the left or in the 
right cells correspondingly \cite{Tomsovic}. Taking into account 
the normalization by the total volume, the classical 
probability to return equals $2p_l(t)$ where 
the initial conditions are $p_l=1$ and $p_r=0$. 
Thus one obtains the result
\begin{eqnarray} \label{e5_11}
P_{cl}(t) = 1+\exp(-2t/t_0) = 1+p(t)^2 \ \ \ .
\end{eqnarray} 
There are three classes of POs, those that 
explore the left cell ($l$), those that explore 
the right cell ($r$), and those that explore 
both cells ($b$). The partial diagonal sums  
should satisfy $K_l(t)+K_r(t)+K_b(t)=tP_{cl}(t)$, 
and also $K_l(t)=K_r(t)=tp(t)$. Hence we 
deduce that $K_b(t)=t(1{-}p(t))^2$.
The ergodic time for a two-cell system 
is simply $t_{erg}=t_0$. Non universal 
features are expected whenever $t$ is 
less or of the same order. 

At this stage we wish to test the 
{\em BLC working hypothesis} of subsection B. 
Using (\ref{e5_8}) one obtains the following 
result for the form factor
\begin{eqnarray} \label{e5_12}
K(k,t) \ = \
2\times\hat{C_0}(k\lambda(t,\Omega_0))\cdot tp(t)
\ + \ \hat{C_0}(k\lambda(t,2\Omega_0))\cdot t(1{-}p(t))^2
\end{eqnarray}   
We wish to discuss the validity of this result. 

Consider first the case of large 
conductivity $1\ll g_0$. The non-trivial 
behavior is manifested in the classical regime 
where $\hat{C}_0=1$ and $K(k,t)=tP_{cl}(t)$. 
Disregarding the small time regime $t_0$
the form factor looks like that of a 
ballistic billiard whose volume is $2\Omega_0$.   
However, one important feature is missed.
The non-universal peak around $t_0$ should 
be compensated later, else  
$R_{qm}(k,\epsilon\rightarrow 0)$ will be  
larger (compared with the universal result)
implying that there is no complete repulsion 
for small energy differences.
Thus we conclude that $\hat{C}_0$ should be 
slightly different from the ballistic one.
One may consider such an effect to be a  
consequence resurgence (see section III-F). 

We take now the other limit $g_0\ll 1$. Here 
the semiclassical formulation should not be trusted, 
since the De-Broglie wavelength is much larger
than the hole size. Still, it is interesting to 
observe whether the qualitative behavior of 
the form factor makes sense. On the Heisenberg 
time scale the form factor 
looks like a union of two independent spectra 
that corresponds to the two cells. 
Note that such form factor implies that the 
repulsion on small energy scales is not complete. 
Due to the small hole that connects the two cells, 
there is a non trivial behavior of the form factor 
on the classical time scale $t_0$.  This relatively 
large time scale corresponds to very small energy 
differences. Specifically there is a deep whose 
functional form is $K(k,t)\propto [1-p(t)(1{-}p(t))]$. 
This deep implies a strong repulsion on small energy scale. 
Indeed, this is the expected behavior. However, 
quantitatively the result does not make sense. 
Firstly, the classical time scale $t_0$ should be 
modified as in the case of tunneling. Secondly,  
the above formula predict over-repulsion (negative 
density for finding energy levels).

\subsection{Disordered Chain} 

We consider a closed disordered chain
that consist of $N$ cells.  
The escape time out of a cell will be 
denoted as before by $t_0$. The number 
of cells that are explored after time 
$t$ is simply $\sqrt{t/t_0}$. The ergodic 
time (known also as the Thouless time in 
the present context) is $t_{erg}=N^2t_0$. 
We shall assume from now on that $t \ll t_{erg}$, 
meaning that the finite size of the chain 
should not play any role in the dynamics.
The classical probability to return is 
found by solving a diffusion equation.
The detailed calculation is presented in 
appendix B. One obtains 
$P_{cl}(t)=N/\sqrt{2\pi(t/t_0)}$. 
The POs may be classified by the volume 
which is explored. We denote by $K_n(t)$
the partial diagonal sum that corresponds 
to POs that explore $n$ specified cells. 
Looking on POs of length $t$, the probability 
for finding a PO that explores $n$ cells
should be proportional to $K_n(t)$.
It is obvious that this 
probability should depend on the scaled 
variable $n/\sqrt{t/t_0}$. In appendix B 
we derive the scaling relation 
\begin{eqnarray} \label{e_Kn}
K_n(t) \ = \ \frac{t_0}{\sqrt{2\pi}} 
f\left(\frac{n}{\sqrt{t/t_0}}\right)
\ \ \ \ \ .
\end{eqnarray}   
Above, $f(\tau)$ is the scaled probability function. 
It should have the normalization $1$. The 
function $f(\tau)$ is determined analytically
in appendix B, and it is plotted in Fig.\ref{f_explore}.
However, the main features are quite obvious. It 
should be peaked at $\tau=1$, with a Gaussian-like  
asymptotic profile $f(x)\sim\exp(-x^2)$. 
For small $x$ one should obtain $f(x)\sim\exp(-1/x^2)$,   
which implies an exponential decay of $K_n(t)$ as a function 
of $t$ for small $n$. In particular for $n{=}1$ it should 
be in accordance with the one-cell result (\ref{e5_10}).

\begin{figure} 
\begin{center}
\leavevmode 
\epsfysize=3.0in
\epsffile{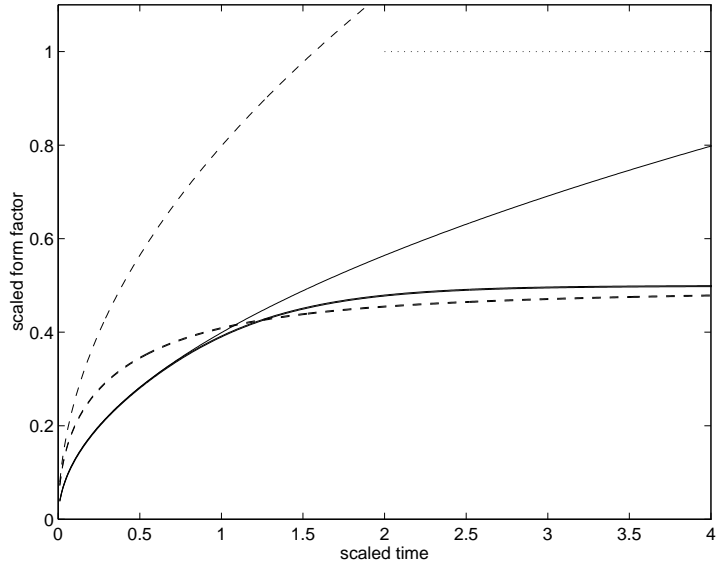}
\end{center}
\caption{\protect\footnotesize 
The scaled form factor for a disordered 
infinite chain. Solid line - GUE result,
Dashed line - GOE result. 
The thiner curves are obtained by employing 
the diagonal approximation, while the thicker
curves are obtained by employing the BLC 
working hypothesis. The dotted line 
illustrates the correct asymptotic behavior. } 
\label{f_chainff}
\end{figure}

At this stage we wish to test the {\em BLC working hypothesis}  
of subsection B. One obtains the following result for the 
form factor
\begin{eqnarray} 
K(k,t) \ = \ N\sum_{n=1}^{\infty} 
\hat{C}_0(k\lambda(t,n\Omega_0)) \cdot K_n(t)
\ = \ 
N \sum_{n=1}^{\infty} 
\frac{t_0}{\sqrt{2\pi}} f\left(\frac{n}{\sqrt{t/t_0}}\right)
\frac{ng_0t_0}{t}\hat{K}_0\left(\frac{t}{ng_0t_0}\right)
\ \ \ \ .
\end{eqnarray}   
Here ${K}_0(\tau)$ stands for either the GUE or the GOE
scaling function of RMT. Changing from summation to 
integration one obtain the scaling relation
$ K(g_0,t) \ = \ Ng_0t_0 \hat{K}(t/t^*) $ ,
where $t^*=g_0^2t_0$ and
\begin{eqnarray}
\hat{K}(\tau) \ = \ \frac{1}{\sqrt{2\pi}} \int_0^{\infty} dx 
\ f\left(\frac{x}{\sqrt{\tau}}\right) \ \frac{x}{\tau} 
\hat{K}_0\left(\frac{\tau}{x}\right) 
\ = \ 
\left\{ \matrix{
 \sqrt{\tau/2\pi} \ & \ \mbox{for} & \tau \ll 1 , & \mbox{GUE} \cr 
2\sqrt{\tau/2\pi} \ & \ \mbox{for} & \tau \ll 1 , & \mbox{GOE} \cr
1/2               \ & \ \mbox{for} & 1 \ll \tau .  
} \right.   \ \ \ \ ,
\end{eqnarray}   
This scaling function is plotted in Fig.\ref{f_chainff}.
The non-universal breaktime
$t^*$ is neither related to the one-cell 
Heisenberg time $t_H^0=g_0t_0$ nor to the volume 
dependent Heisenberg time $t_H=Ng_0t_0$. Rather, 
it is determined by the disorder. The non-universal 
breaktime $t^*$ rather than the non-relevant
universal time scale $t_H$ signify the crossover 
to a recurrent quasiperiodic behavior. Assuming 
that up to $t^*$ there is roughly a classical 
diffusive behavior, with diffusion coefficient 
${\cal D}=\ell_0^2/t_0$, one obtains the correct result  
$\xi=g_0\ell_0$ for the so-called localization length.

\begin{figure} 
\begin{center}
\leavevmode 
\epsfysize=3.0in
\epsffile{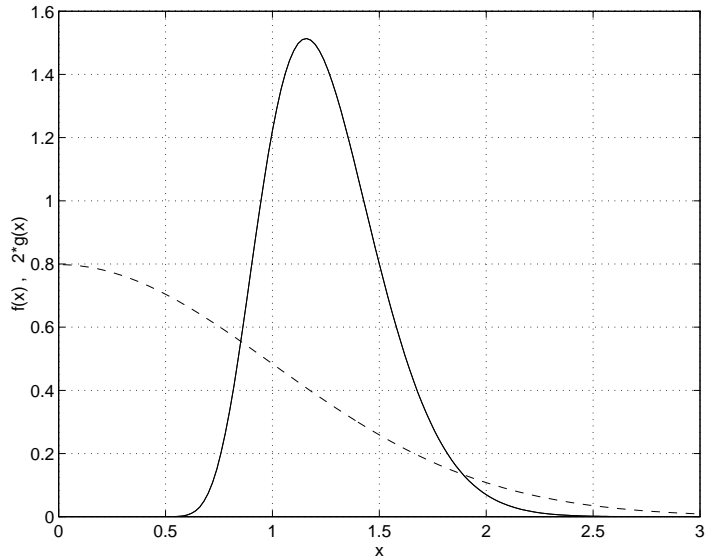}
\end{center}
\caption{\protect\footnotesize 
Solid line - the scaled probability distribution of 
the explored volume for a disordered chain.
Dashed line - the Gaussian distribution of 
the winding number for a folded periodic chain. } 
\label{f_explore}
\end{figure}

The present application demonstrates the 
success of the BLC conjecture, as well as 
the limitation of the BLC working hypothesis.      
The generalization of these ideas to localization 
in higher dimensions is discussed in \cite{brk}.
The BLC conjecture is sufficient in order to 
determine the non-universal breaktime $t^*$ and
the corresponding localization length. Qualitative 
agreement with scaling theory of localization in 
one to three dimensions can be demonstrated. 
In particular, the present theory is capable of
giving a qualitative semiclassical understanding
for the existence of a critical regime in three 
dimensional systems. On the other hand, the 
quantitative analysis is unsatisfactory. The 
BLC working hypothesis overestimates the 
effect of (negative) correlations. 
We believe that the statistics is modified 
due to leaking of POs, 
thus leaving holes in the classical 
spectrum. As a result there is some 
effective clustering that affects
the functional form of the correlations.
See also \cite{brk}.

\subsection{Periodic chain}

It is interesting to discuss the implications 
of translational invariance. Rather than having 
spectral statistics that correspond to localization,  
one should obtain spectral statistics that corresponds 
to a band structure. This problem has been 
discussed by \cite{pchain0,pchain}. Here we wish to clarify the 
present limitations of the semiclassical approach, 
and to illuminate some confusion which may 
arise concerning the classification of POs.

The one point properties of the classical 
spectrum are not changed when the 
disorder $\rightarrow$ order route is 
followed. The relevant classification
of POs is still by the explored volume. 
POs that explore the whole volume ($n=N$)
constitute a negligible minority since we assume 
an effectively infinite chain for which 
$t \ll t_{erg}$. Still, in the case of a periodic 
chain one may further classify the POs   
in the class $n=N$ by their self-symmetry. 
Specifically, There are POs that are invariant 
to translations of $\nu=1,2...$ cells. 
It should be emphasize that as far as the 
{\em unfolded chain} is concerned these self-symmetric
POs constitute a minority within a minority.  
On the other hand, once the classical spectrum of 
the {\em folded chain} is considered, 
the primitive versions of these self-symmetric POs 
become dominant in the classical spectrum. 
The class index $\nu$ in the latter case should be 
interpreted as the winding number. 
The distribution of the folded POs with respect
to $\nu$ is a Gaussian \cite{scharf,pchain}.  
Let us denote by $K_{\nu}(t)$ the diagonal sum that 
corresponds to the class of POs whose winding number 
is $\nu$. Summing over $\nu$ one should obtain 
$tP_{cl}(t)$ where $P_{cl}(t)=1$ as for any simple 
ballistic billiard. Hence we deduce that  
\begin{eqnarray} \label{e_gx}
K_{\nu}(t) \ = \ \sqrt{t_0t} \cdot  
g\left(\frac{\nu}{\sqrt{t/t_0}}\right)
\ \ \ \ \ ,
\end{eqnarray}   
where $g(x)$ is a normalized Gaussian with unit 
dispersion (see Fig.~\ref{f_explore}).
The distribution is peaked at $\nu{=}0$, still
the $\nu{=}0$ POs become a minority with respect 
to $\nu\ne0$ if $t_0 \ll t$. 

From the above discussion it should be clear that
the relevant classification of POs is totally 
different for the folded chain when compared 
to the unfolded chain. If one wishes to study 
the statistical properties of the band spectrum, 
one should consider an unfolded chain, and the 
corresponding classification is the same 
as for the disordered chain. {\em The POs that determine 
the statistical properties of the spectrum for 
a given quasi-momentum, will not be involved in 
the semiclassical analysis of the band spectrum !}   
   
Thus, the only difference between disordered 
and periodic chain is associated with the 
two-point statistics of the classical spectrum. 
For a disordered chain the correlation scale 
is class-dependent, namely 
$\lambda=\lambda_0(t,n\Omega_0)$, while for 
a periodic chain $\lambda=\lambda_0(t,\Omega_0)$
for all the classes. Using the BLC approximation 
scheme, taking into account symmetry, 
one obtains the result
\begin{eqnarray} \label{e5_18}
K(g_0,t)=\frac{N^2}{\sqrt{2\pi}} 
g_0t_0 \sqrt{\frac{t_0}{t}}
\hat{K}_0\left(\frac{t}{g_0t_0}\right)
\ \ \ \ .
\end{eqnarray}
The above analysis essentially coincides 
with that of \cite{pchain} for $t<t_H^0$, 
where $t_H^0=g_0t_0$ is the one-cell 
Heisenberg time. Namely, for 
short times $K(t)\sim\sqrt{t}$. 
Here we have tried to extend the 
semiclassical analysis beyond $t_H^0$  
using the BLC working hypothesis.  
For $t_H^0<t$ we have obtained $K(t)\sim1/\sqrt{t}$, 
which does not agree with the correct $\sim 1/t$ behavior. 
Thus, in order to have a proper semiclassical theory 
for band spectrum we should know the way in which the 
scaling function is modified. Again, we conclude that the 
BLC conjecture is effective in the determination 
of the various time scales involved, whereas the BLC 
working hypothesis is unsatisfactory. 


\section{Numerical Study}
\label{sec:numerics}
\setcounter{equation}{0}

In this section we use our numerical database to study 
{\em classical correlations} in three different billiards: 
the 2D hyperbola billiard, the 2D Sinai billiard and 
the 3D Sinai billiard. 
One possible approach is to compute $p(x)$ directly for various
$L$-windows. A smoothing procedure is then essential in order to get
rid of fluctuations. Our numerical strategy is different. We prefer to
fix the $L$-window and to vary $k$. Our object of study is the
correlation factor $C(k, L) = K_{cl}(k, L) / K_{D}(L)$. In order to
avoid numerical artifacts we have not applied a smoothing procedure.
Rather, the cumulative quantity is considered, namely
\begin{equation}
I(k, L) \ \equiv \ 
\int_{0}^{k} {\rm d}k'
\frac{C(k', L)}{g_{\mbox{eff}}} =
\int_{0}^{k} {\rm d}k'
\frac{|\sum_{pq}A_{pq}\ \mbox{e}^{ik`L_p}|^2}
{\sum_p |\sum_q A_{pq}|^2} \ \ \ .
\end{equation}
Here $\sum_{p}$ is over all the distinct {\em lengths} within a given
$L$-window, and $\sum_{q}$ is over all the POs that correspond to a
given length $L_p$. We have divided $C(k,L)$ by $g_{\mbox{eff}}$ which
is the effective degeneracy within the specified window. This has been
done in order to get rid of the trivial correlation factor 
which appears due to symmetries. This way the integrand always approaches 
$1$ for $k\rightarrow \infty$. Fixing the $L$-window and probing the 
spectrum within the specified window with larger and larger $k$ is both
mathematically appealing and numerically convenient. (It is less
convenient to fix $k$ and to vary $L$). This is because our classical 
databases is limited to a narrow span of $L$ values, while $k$ is a 
free parameter that can be set to any value.

If the form factor complies with RMT predictions, then we expect (see
subsection~\ref{subsec:ckl}):
\begin{equation}
C(k, L) \ \approx \ 
\min \left( 1, \frac{K_{Q}(k)}{K_{D}(L)} \right) \ = \
\left\{ 
  \begin{array}{lll}
    {\rm const} \, k^{d-1}  & , & k < 2 \pi / \lambda(L) \\
    1 & , & k > 2 \pi / \lambda(L)
  \end{array}
\right.
\end{equation}
and consequently
\begin{equation}
I(k, L) \ = \ \left\{ 
  \begin{array}{lll}
    {\rm const} \, k^{d}  & , & k < 2 \pi / \lambda(L) \\
    {\rm const} + k & , & k > 2 \pi / \lambda(L)
  \end{array}
\right. \, .
\end{equation}
The above strictly holds for GUE, and holds approximately for GOE.
\begin{figure}[htb]
  \begin{center}
    \leavevmode
    \psfig{figure=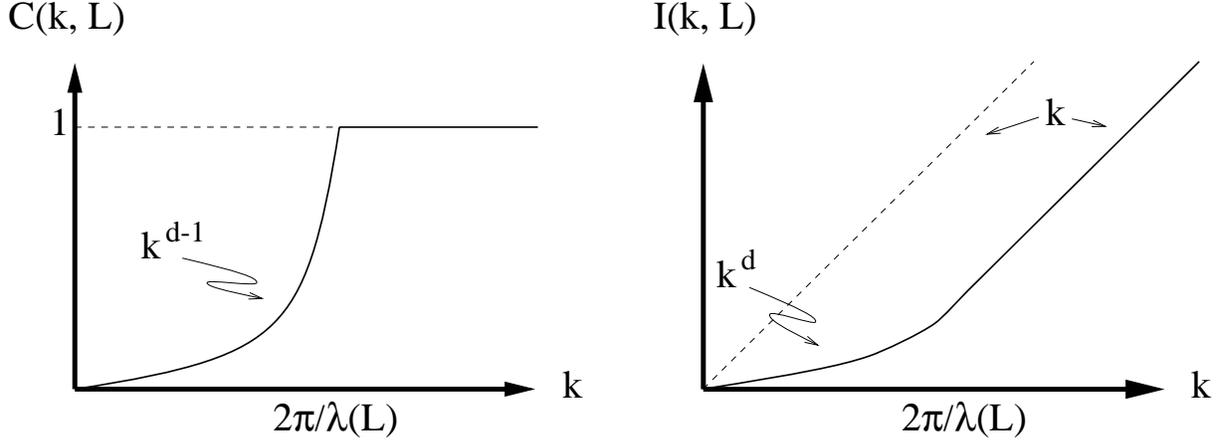,width=16cm}
    \caption{The functions $C(k, L)$ and $I(k, L)$ for RMT GUE. 
      We indicate the expected slope in each regime.}
    \label{fig:ckl-theory-universal}
  \end{center}
\end{figure}

\subsection{The hyperbola billiard}
\label{subsec:hyperbola}

\begin{figure}[htb]
  \begin{center}
    \leavevmode
    \begin{tabular}{cc}
      \psfig{figure=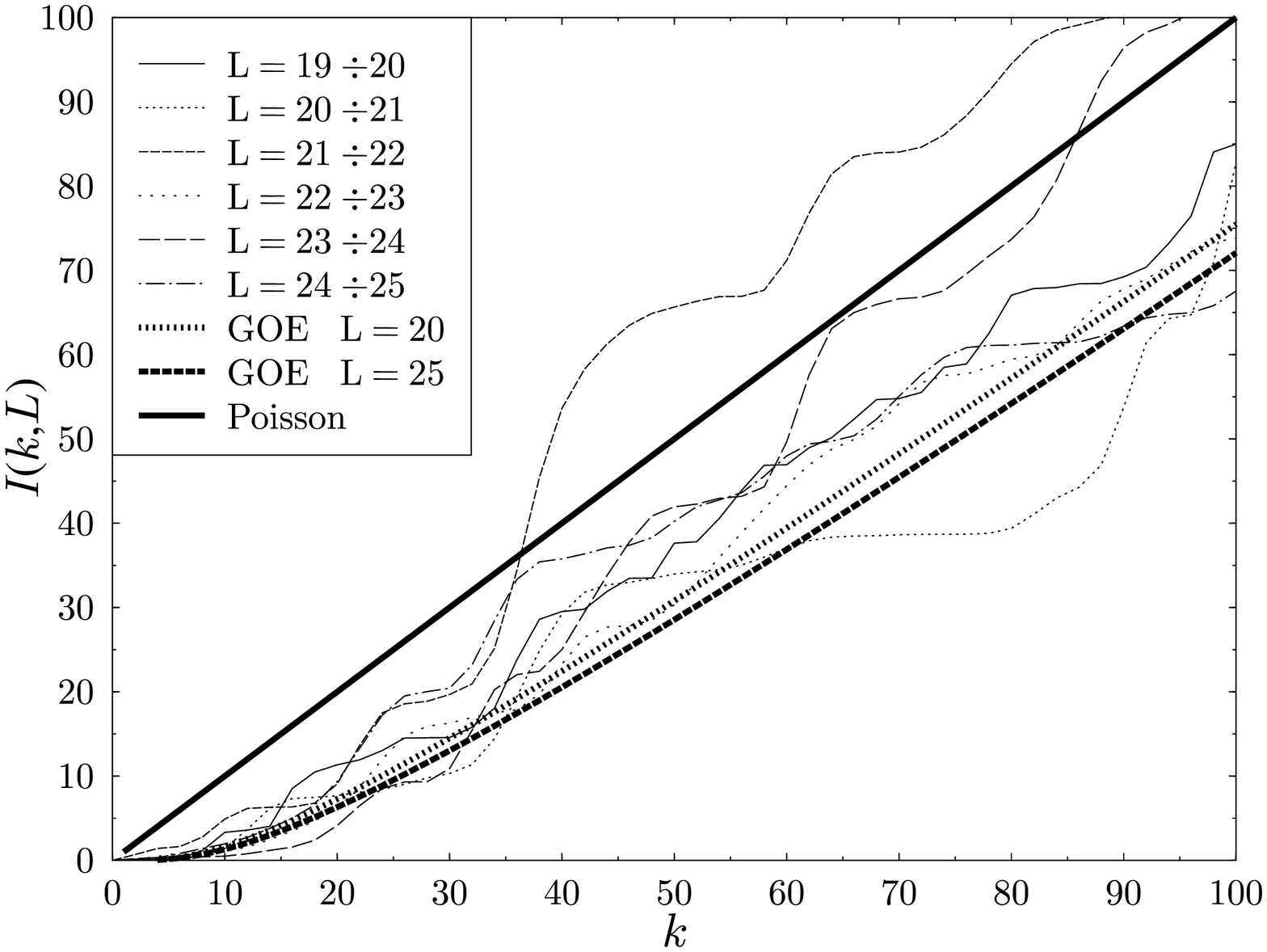,width=9cm} &
      \psfig{figure=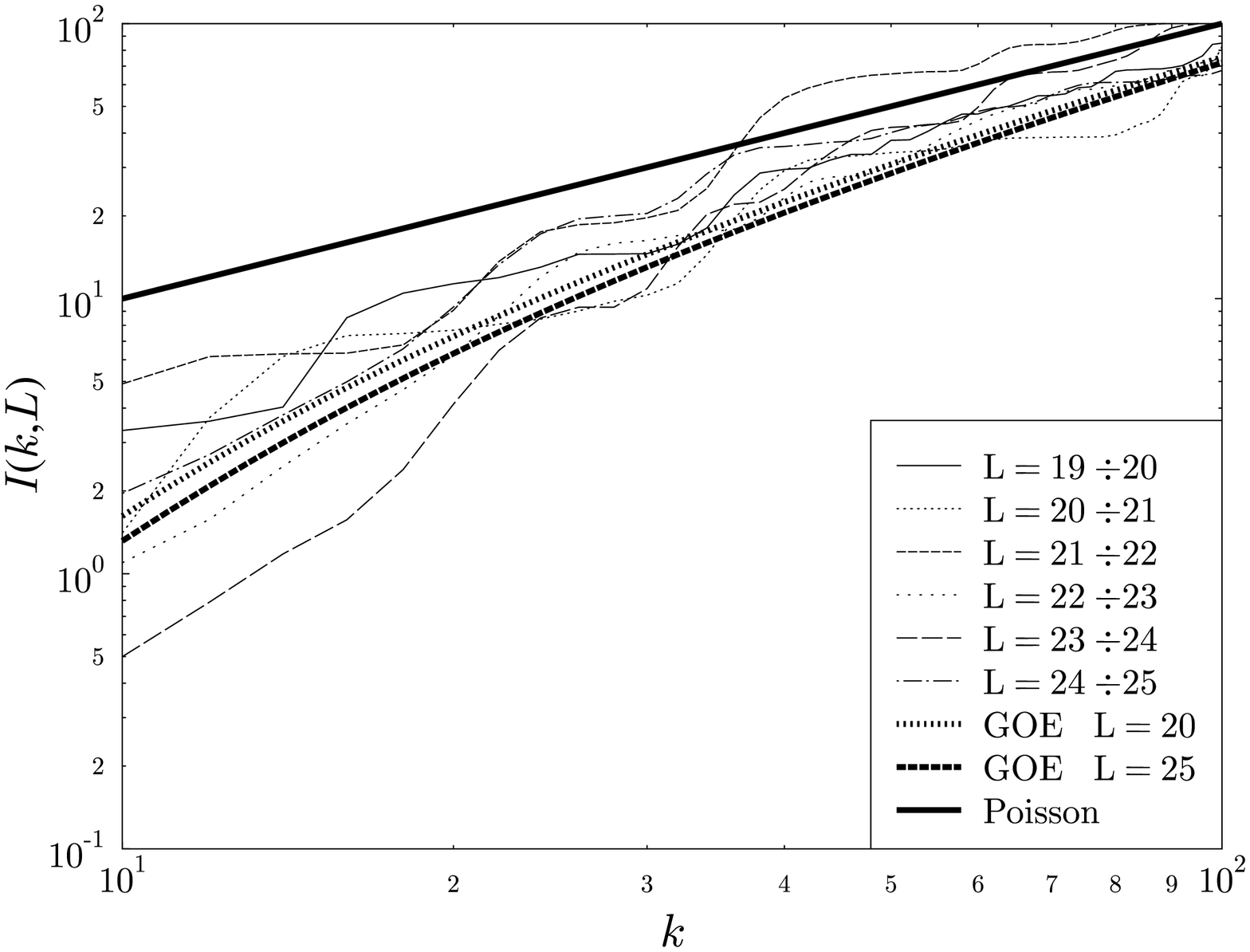,width=9cm}
    \end{tabular}
    \caption{\protect \parbox[t]{10cm}{The function $I(k, L)$ for the 
        hyperbola billiard. For convenience we present the results on
        linear-linear and log-log scales.}}
    \label{fig:ickl-hyperbola}
  \end{center}
\end{figure}

The desymmetrized hyperbola billiard is a planar billiard that is
defined by the curves $x=0, \; y=0, \; y=x, \; y=1/x$. The billiard
was throughly studied by Sieber \cite{MartinThesis}. The compliance of
$K_{cl}$ with GOE for the desymmetrized hyperbola billiard was already
demonstrated numerically in \cite{argaman}. There a database of 
about 100000 POs with length $L < 25$
(not including time reversal) has been used. 
Here we wish merely to demonstrate our
numerical procedure using the same database \cite{ThnaksMartim} and to
illustrate the utility of $I(k, L)$. The results are shown in
figure~\ref{fig:ickl-hyperbola}, where we choose a few $L$ intervals
with $\delta L = 1$. The overall $k$-dependence of the numerical data agrees
quite well with the theoretical curves of GOE and is markedly
different from the Poissonian (no correlations) case.  This is a clear
indication of the existence of classical PO-correlations that
indeed conspire to give the correct GOE form factor.  The agreement of
numerical and theoretical data is also an indication to the correct
$L$ dependence, at least on average. The changes in $L$ of the
theoretical curves (GOE) between the lower and upper values of the
$L$'s considered are minute in comparison to the numerical dispersion,
and therefore it is impossible to reach any definite conclusion concerning 
the detailed $L$ dependence.

\subsection{The 2D Sinai billiard}
\label{subsec:sinai2d}

\begin{figure}[htb]
  \begin{center}
    \leavevmode
    \begin{tabular}{cc}
      \psfig{figure=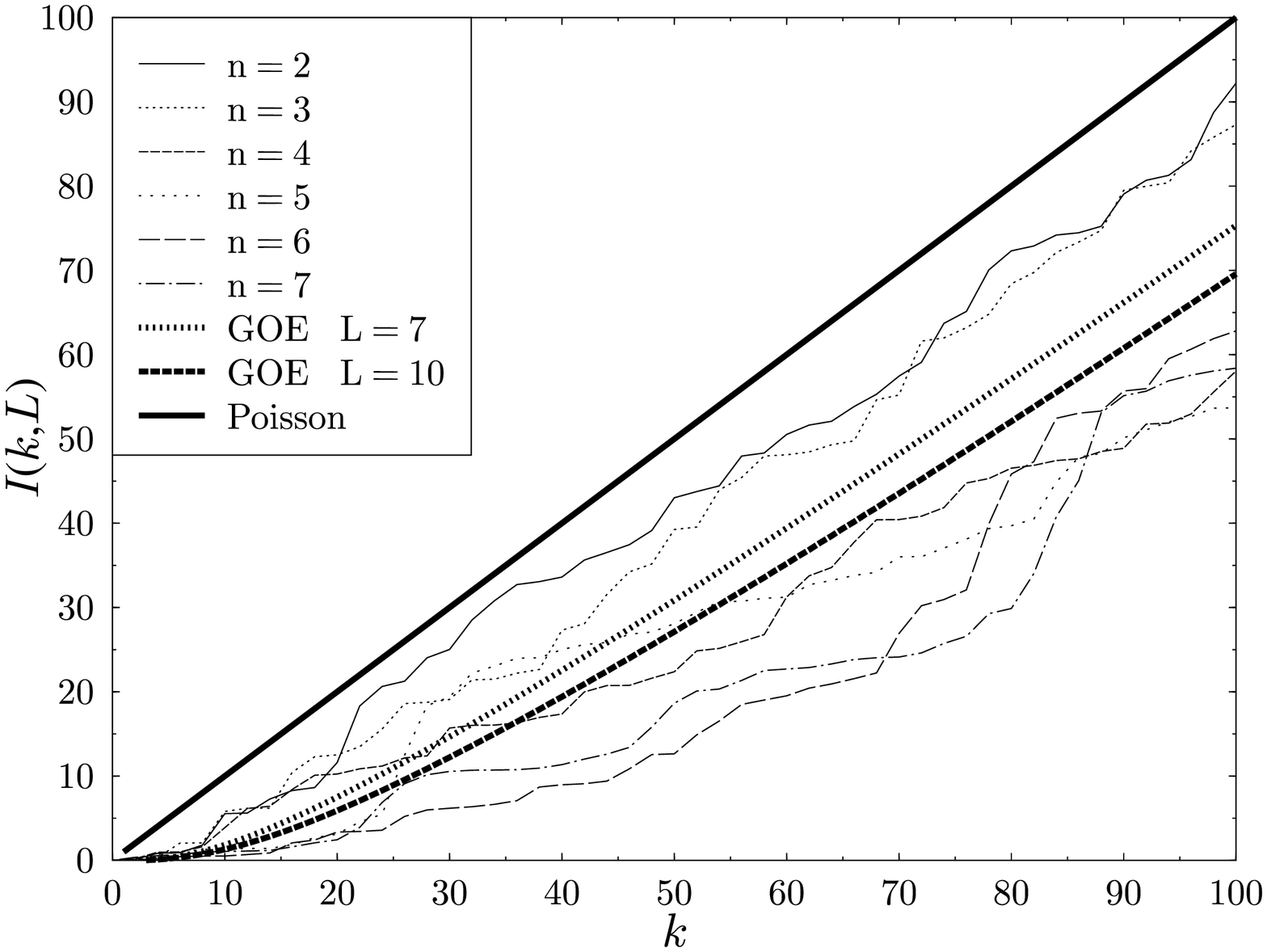,width=9cm} &
      \psfig{figure=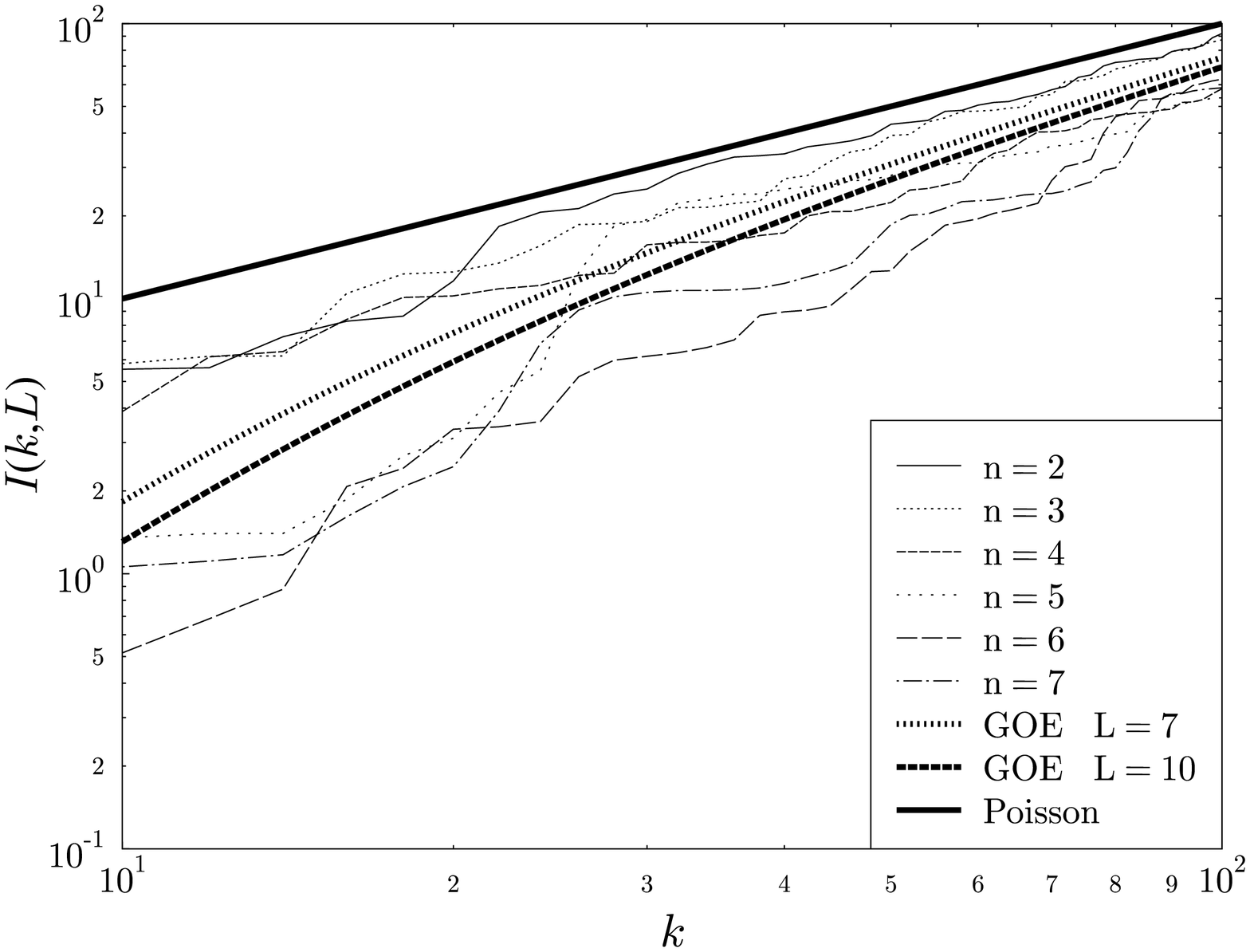,width=9cm}
    \end{tabular}
    \caption{\protect\parbox[t]{10cm}{The function $I(k, L)$ for the 2D 
        quarter Sinai billiard. POs were considered in the
        length interval $7 \leq L \leq 10$.}}
    \label{fig:ickl-sinai2d}
  \end{center}
\end{figure}

Here we considered the 2D Sinai billiard with edge of length $2$ and 
an inscribed circle of radius $R=0.5$. The billiard was desymmetrized into 
quarter and POs were computed by Schanz and Smilansky
\cite{SS95+Thnaks}. The database consisted of about 5000 POs 
in the interval $L < 10$ (not including time reversal and $x
\leftrightarrow y$ reflection symmetries). Our numerical results of
$I(k, L)$ were computed in the interval $7 \leq L \leq 10$. We found
it very useful to separate the POs into classes according
to the number $n$ of their bounces with the circle. The actual
justification for this separation into classes is a-posteriori, but
following is also an a-priori motivation. Each bounce with the
inscribed disc significantly increases (on average) the instability of
the orbit. Hence, POs with different $n$'s have in general
very different amplitudes. Therefore, if we calculate $C(k, L)$ with
the amplitudes $A_j$ taken into account, then the relatively few
POs with 1--2 bounces dominate and overwhelm the much
smaller contributions of the other POs (which are much
more numerous). On the other hand, if we ignore $A_j$'s (expect sign)
and replace them by some average value, then orbits with small number
of bounces will be effectively ignored. (This state of affairs is
valid when $L$ is relatively small and the $n=1$ class dominates.  In
general, the important factor is the multiplication of the density
$d_{n}(L)$ of orbits with $n$ bounces near length $L$ with the typical
amplitude $A_{n}(L)$.) In order to extract as much information as
possible, it is imperative to separate the POs into
classes according to $n$, such that all of the $A_j$ in the same class 
are of similar magnitude. The numerical results are shown in figure
\ref{fig:ickl-sinai2d}. Similarly to the case of the hyperbola
billiard PO-correlations are clearly observed, and the general
appearance is similar to the predictions of GOE that correspond to the
length interval investigated. The numerical dispersion is, however,
too large to allow a detailed test of the $L$-dependence. 

The numerical results indeed indicate that the division into classes
according to $n$ is very useful as suggested above. It confirms the 
idea that there are non-trivial correlations between POs within the 
same class. This information is very valuable in the
quest for the classical origin of PO-correlations. We shall
exploit it in the case of the 3D Sinai billiard (see below). The
existence of inter-class correlations as was shown above does not
exclude the possibility of having intra-class correlations, but we did
not investigate this issue.

\subsection{The 3D Sinai billiard}
\label{subsec:sinai3d}

\begin{figure}[htb]
  \begin{center}
    \leavevmode
    \begin{tabular}{ccc}
      \psfig{figure=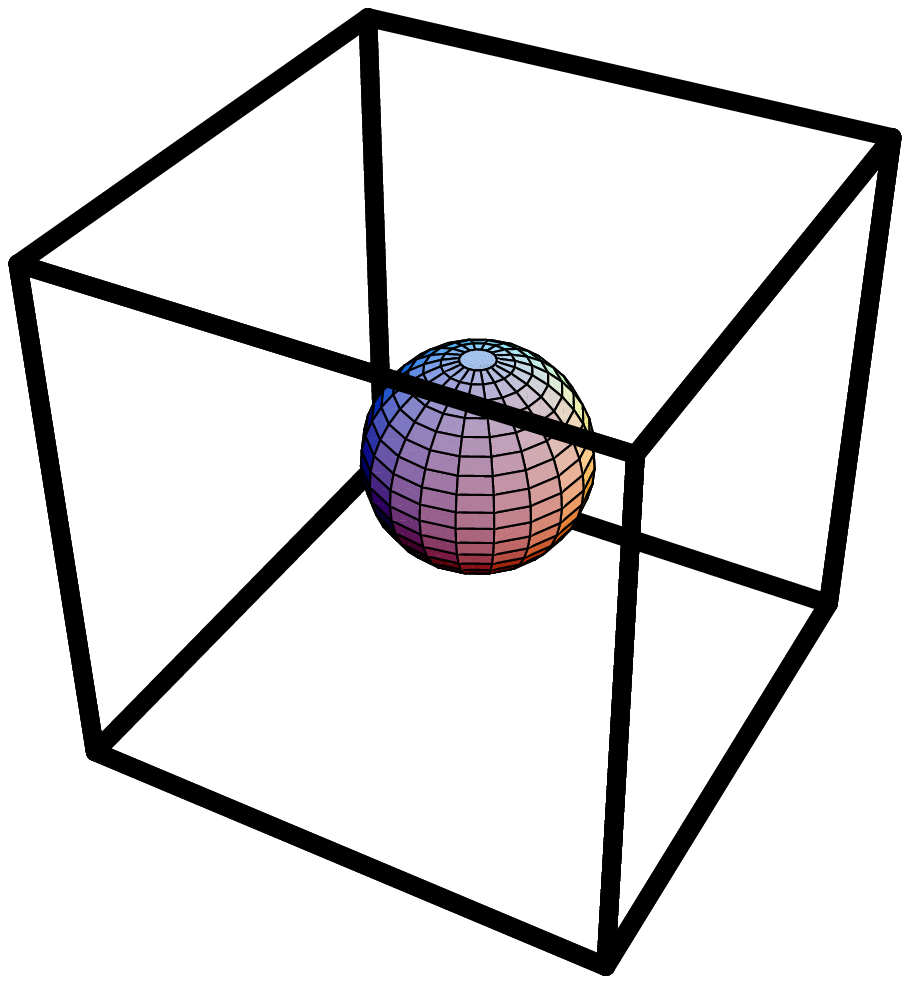,width=5cm} &
      \psfig{figure=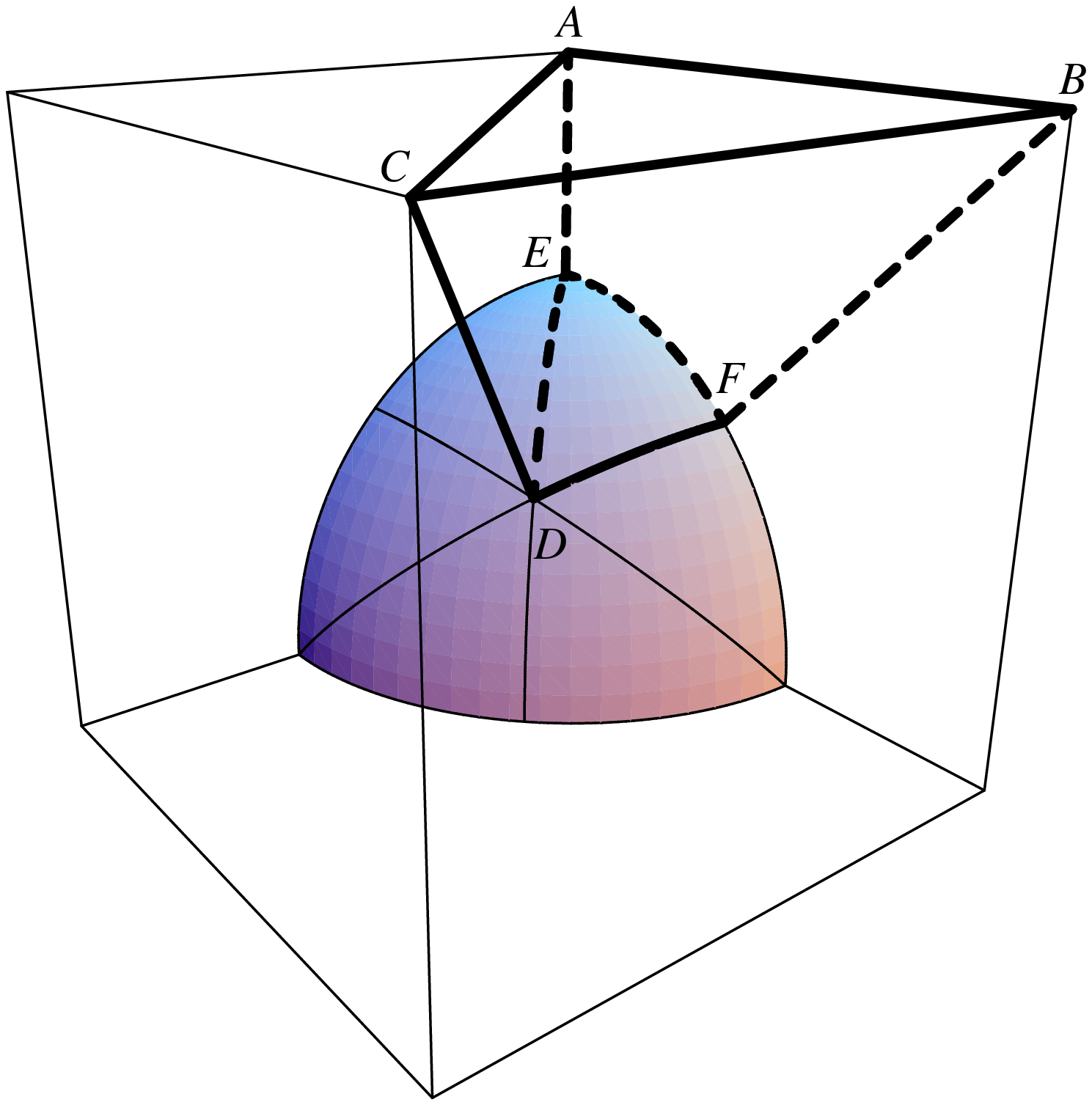,width=5cm} &
      \psfig{figure=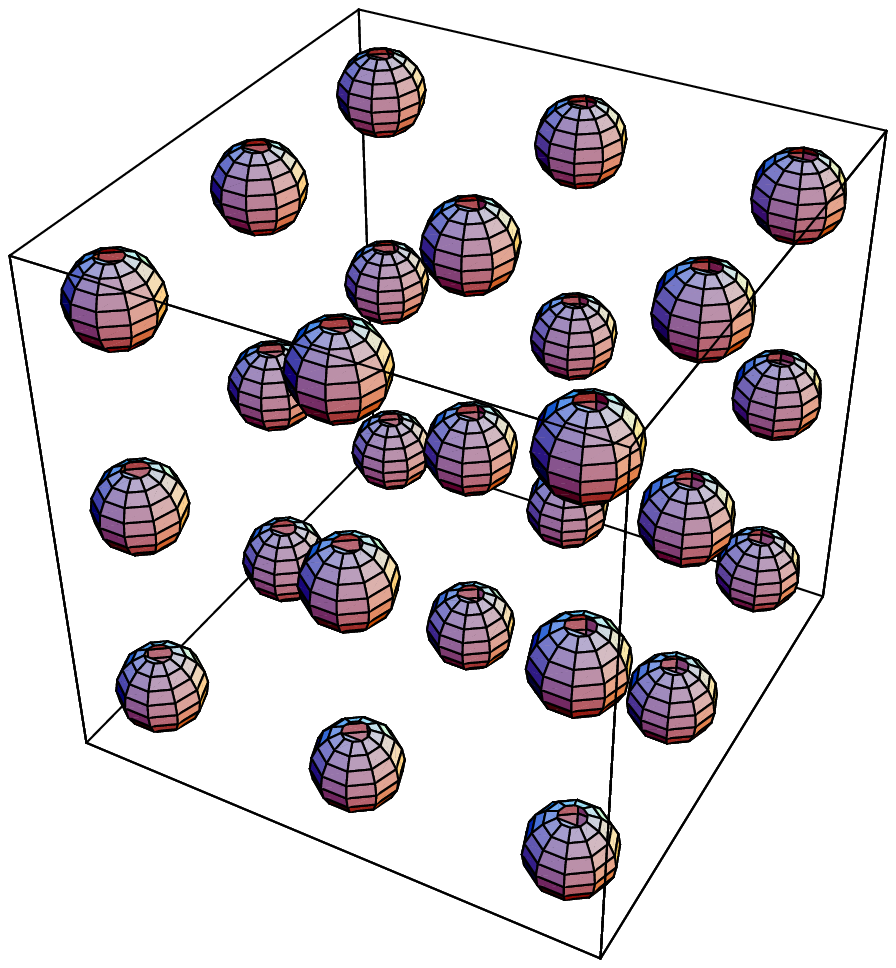,width=5cm} 
    \end{tabular}
    \caption{The 3D Sinai billiard. Left: original billiard, 
      middle: desymmetrized billiard, right: unfolded billiard.}
    \label{fig:sb3d}
  \end{center}
\end{figure}

In this subsection we investigate the classical PO-correlations of
the 3D Sinai billiard. The 3D Sinai billiard is the natural extension
of the familiar 2D Sinai billiard, and is defined as the free space
between a cubic 3-torus and an inscribed sphere. It is presented in
figure~\ref{fig:sb3d}. We consider a torus of edge $1$ and sphere
of radius $R=0.2$. The billiard  is desymmetrized into its fundamental
domain (1/48 of the original billiard).
The main reason for investigating this system is that we had an
extensive database available that consisted of $7 \cdot 10^6$ POs 
(not including time-reversal conjugates). This allows us to
perform meaningful statistical checks. The 3D Sinai billiard is
plagued with non-generic neutral families of POs
(``bouncing-balls'') which were already mentioned in connection with
the mixed boundary conditions systems (subsection~\ref{subsec:mbc}).
As was discussed there, we should resort to the trace formula
(\ref{eq:mbcft}) that avoids the bouncing-balls, and takes into
account only the generic, isolated and unstable POs.
Consequently, the classical spectral density is defined with the
coefficients $\tilde{A}_j$ rather than with $A_j$.

For a generic billiard system with time reversal symmetry one expects
that the quantal form factor will satisfy $K_{qm} \approx K_{GOE}
\approx g_{\mbox{eff}}K_D(t)$ where $g_{\mbox{eff}} \approx 2$. The
similarity $K_{qm} \approx K_{GOE}$ should hold globally while the
similarity $K_{qm}\approx g_{\mbox{eff}} K_D$ holds for relatively
short times. Also for the 3D Sinai billiard with mixed boundary
conditions it has been observed \cite{HarelPhD} that globally $K_{qm}
\approx K_{GOE}$, provided that $K_{qm}$ is normalized such that it
approaches $1$ for $L \rightarrow \infty$. However, for short times
$K_{GOE} < K_{qm} < g_{\mbox{eff}}K_D$. The deviation from GOE for
short times is a non-generic feature of the system and will not
concern us here. The deviation from the diagonal approximation will
be discussed in the sequel in terms of classical correlations.

\begin{figure}
  \begin{center}
    \leavevmode
    \begin{tabular}{cc}
      \psfig{figure=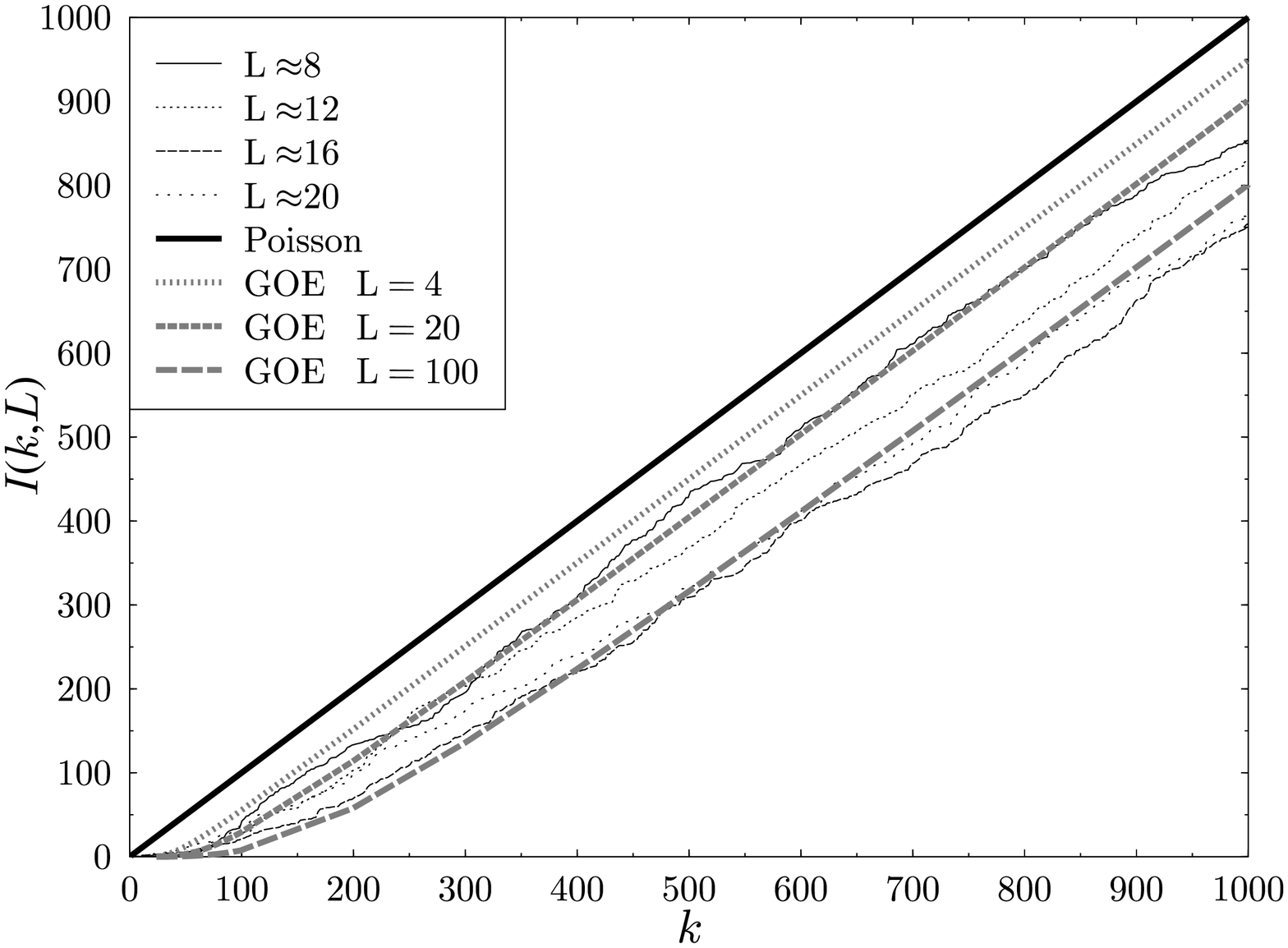,width=9cm} &
      \psfig{figure=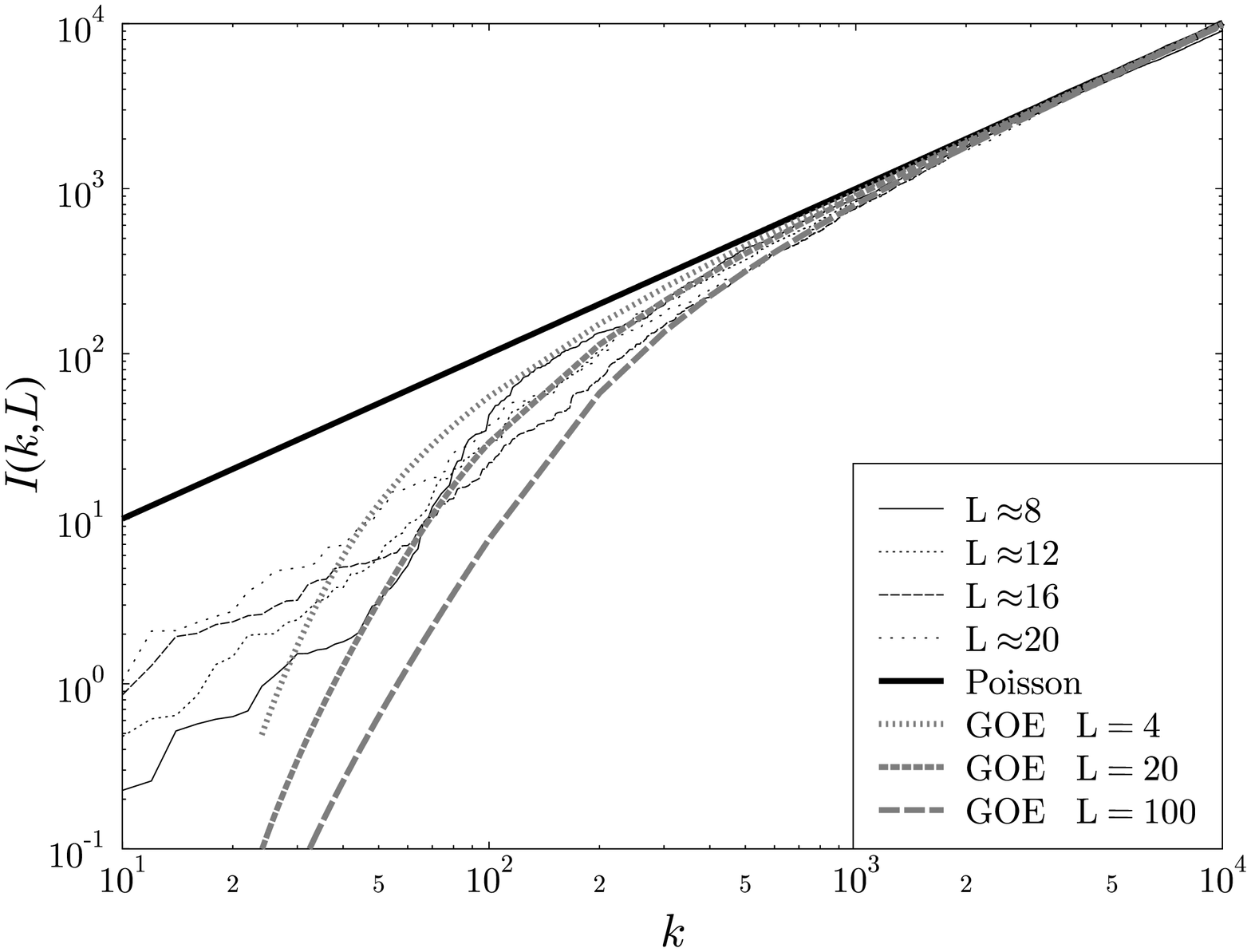,width=9cm} \\
      \psfig{figure=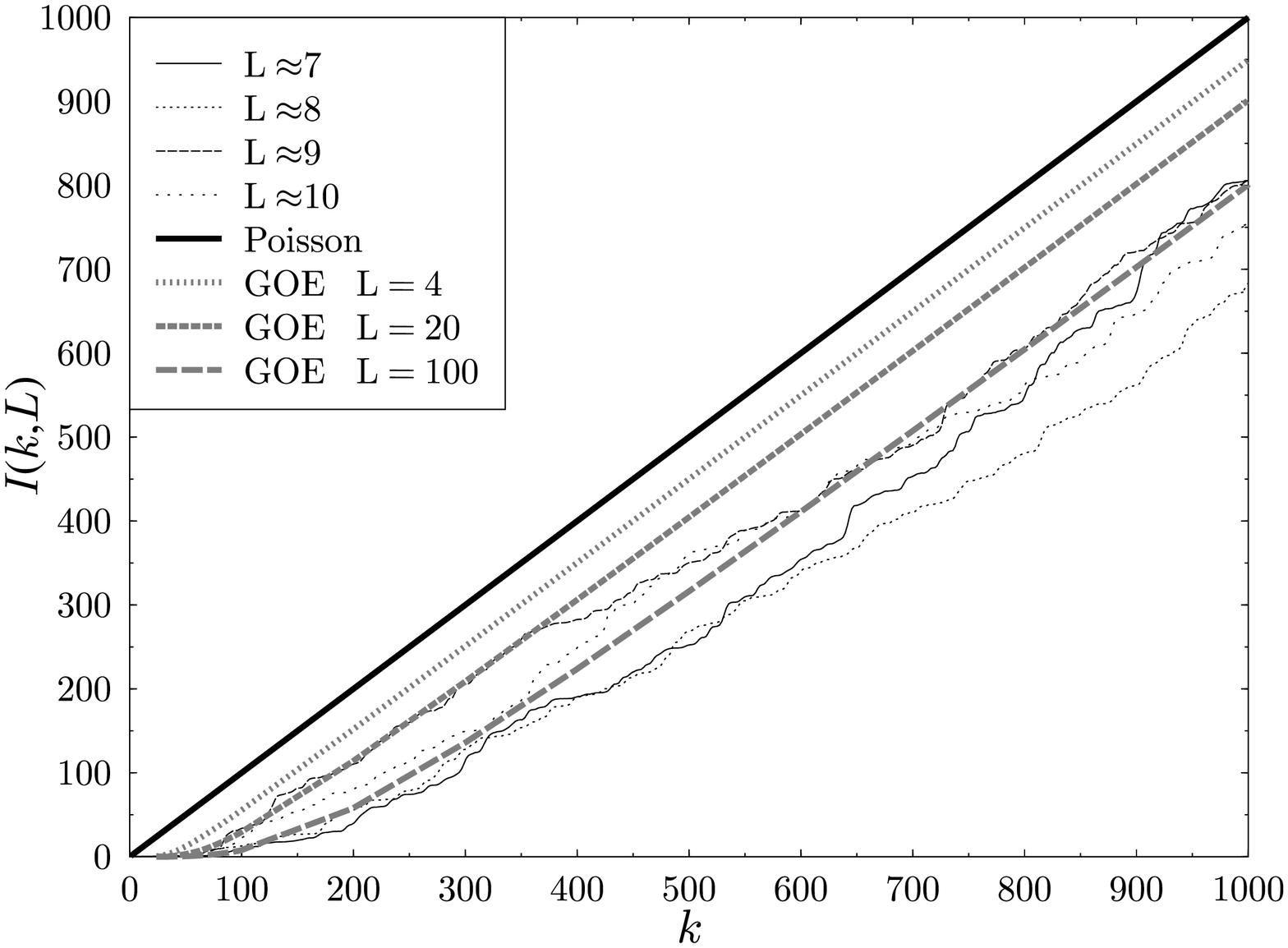,width=9cm} &
      \psfig{figure=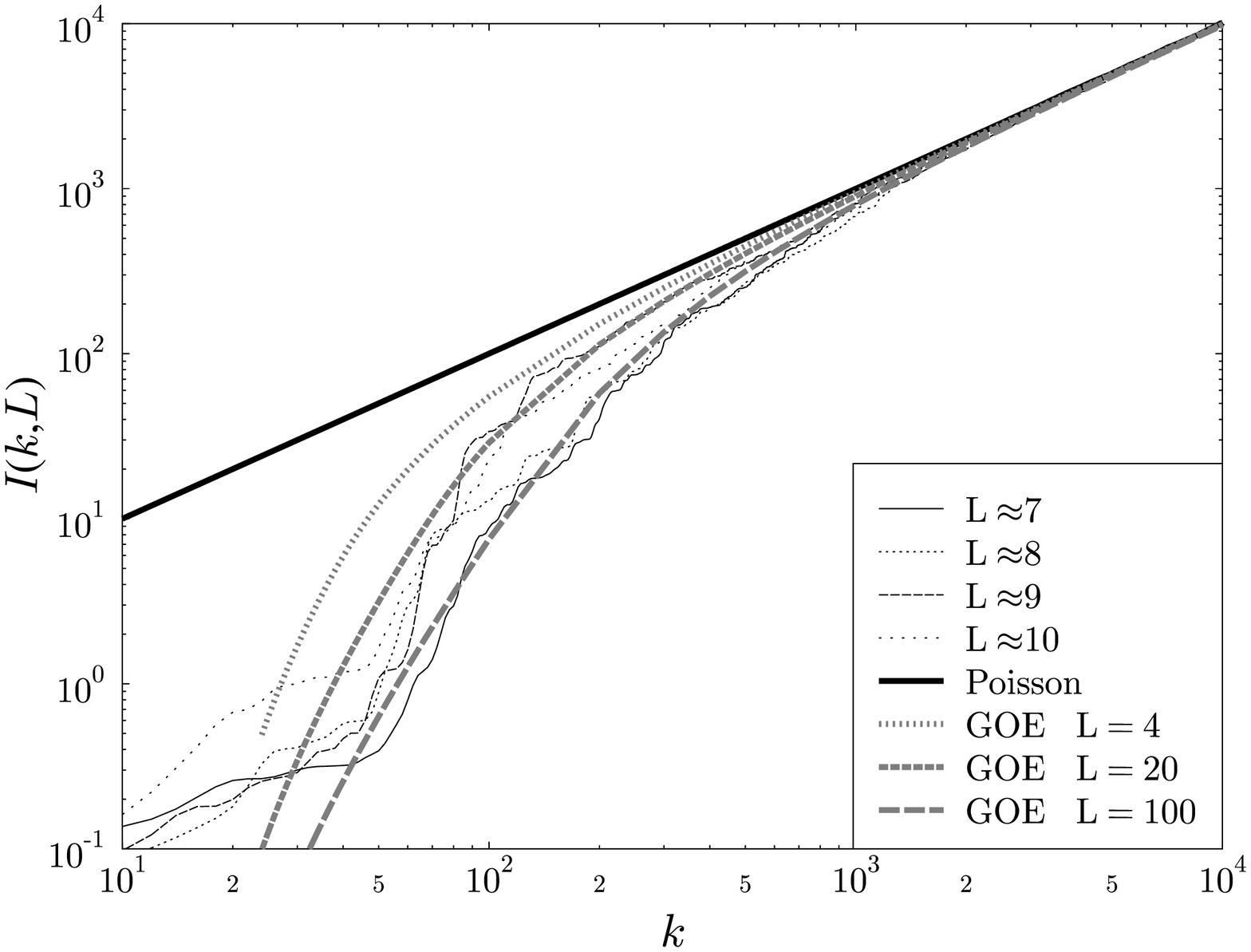,width=9cm} \\
      \psfig{figure=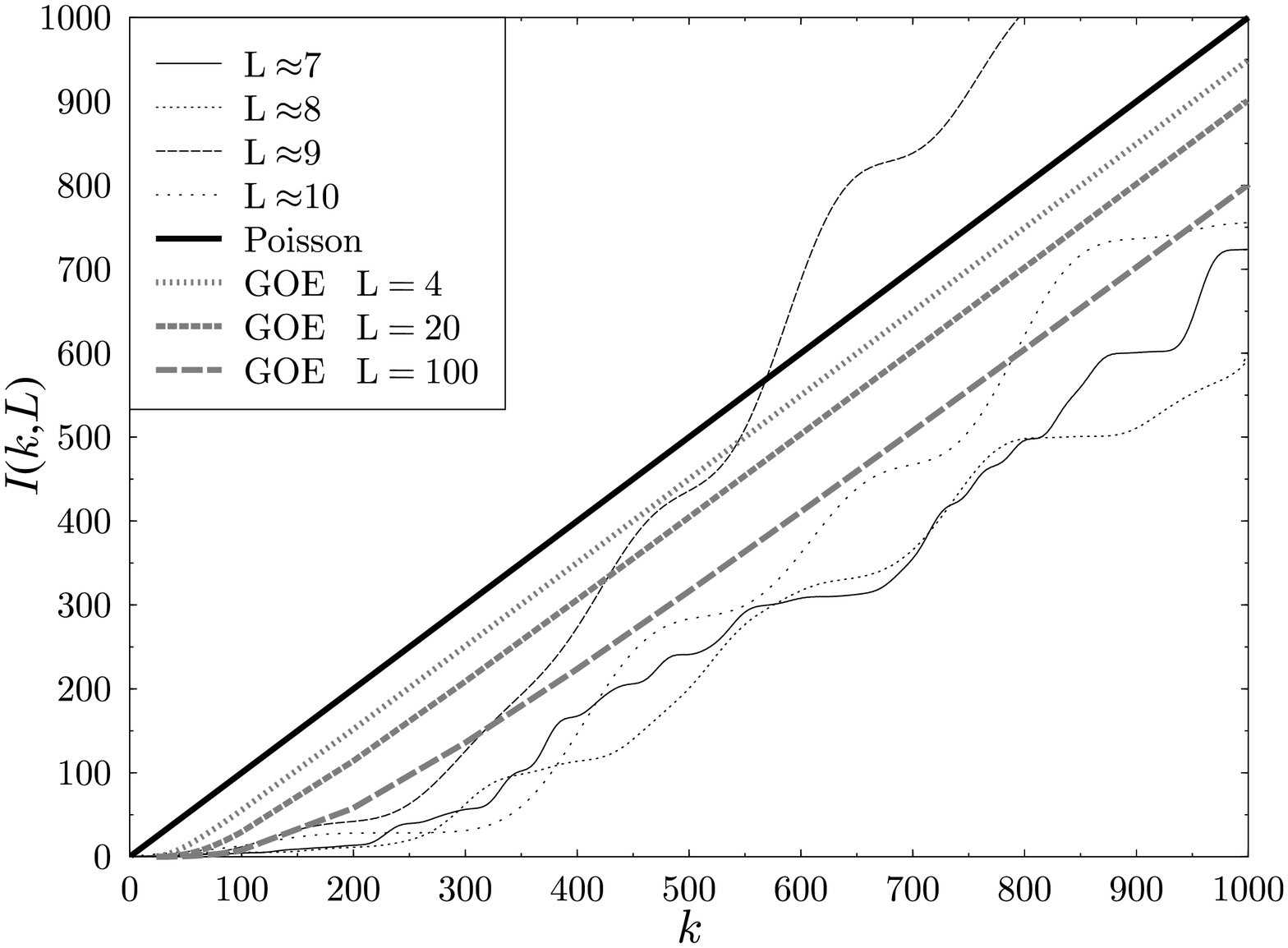,width=9cm} &
      \psfig{figure=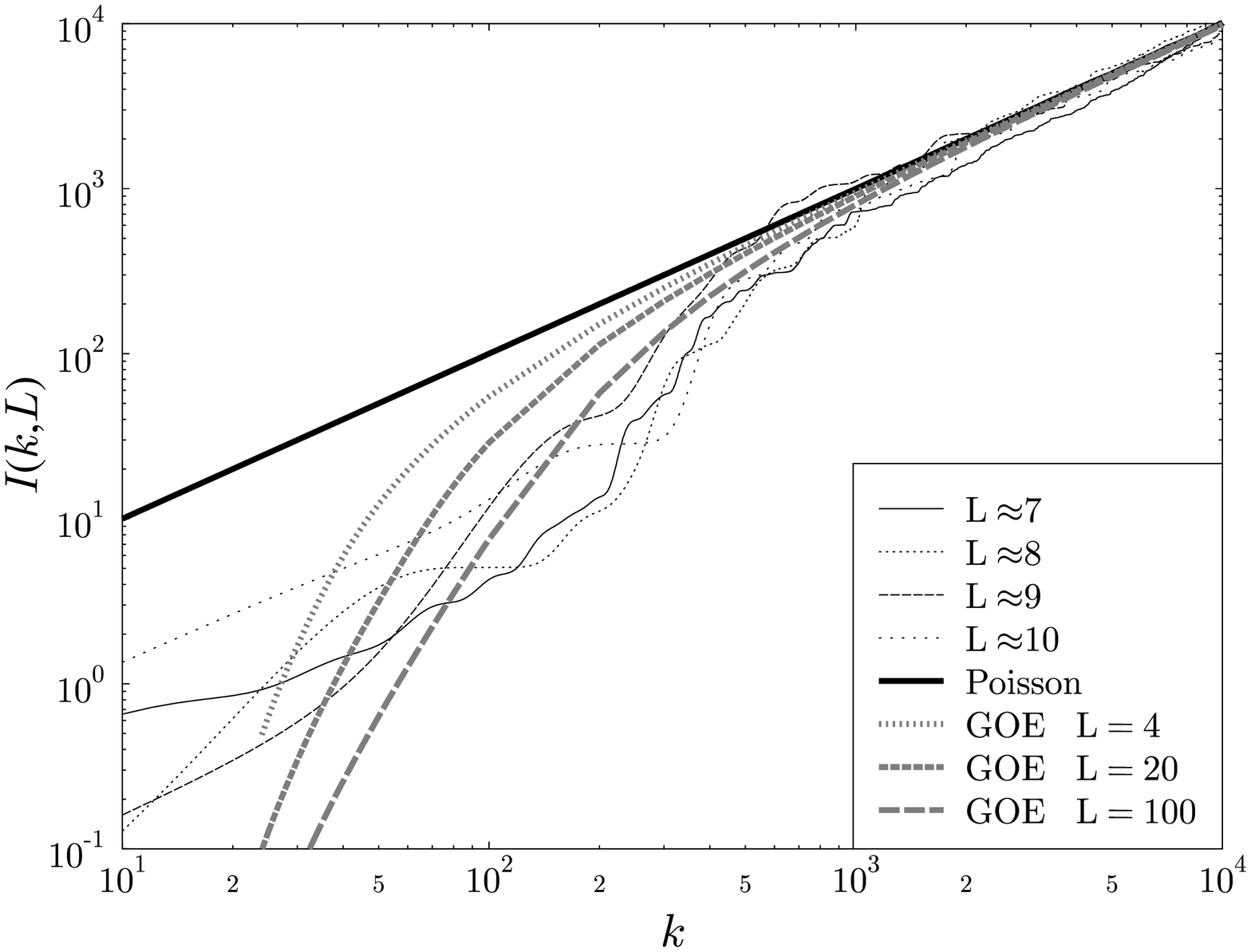,width=9cm}
    \end{tabular}
    \caption{The integrated correlation factor $I(k, L)$ for the 
      3D Sinai billiard with $R=0.2$ (desymmetrized).
      POs are classified according to their number of
      bounces from the sphere. Top: $n=1$, middle: $n=2$, bottom:
      $n=3$. The notation ``$L \approx 8$'' means a window in $L$ near
      $L=8$. On the left we present the data in linear scale, and on
      the right on log-log scale. The GOE predictions are truncated
      for $k$ smaller than the smallest eigenvalue $k_1 \approx
      23.5$.}
    \label{fig:ickl-sb3d-n123}
  \end{center}
\end{figure}

\begin{figure}
  \begin{center}
    \leavevmode
    \begin{tabular}{cc}
      \psfig{figure=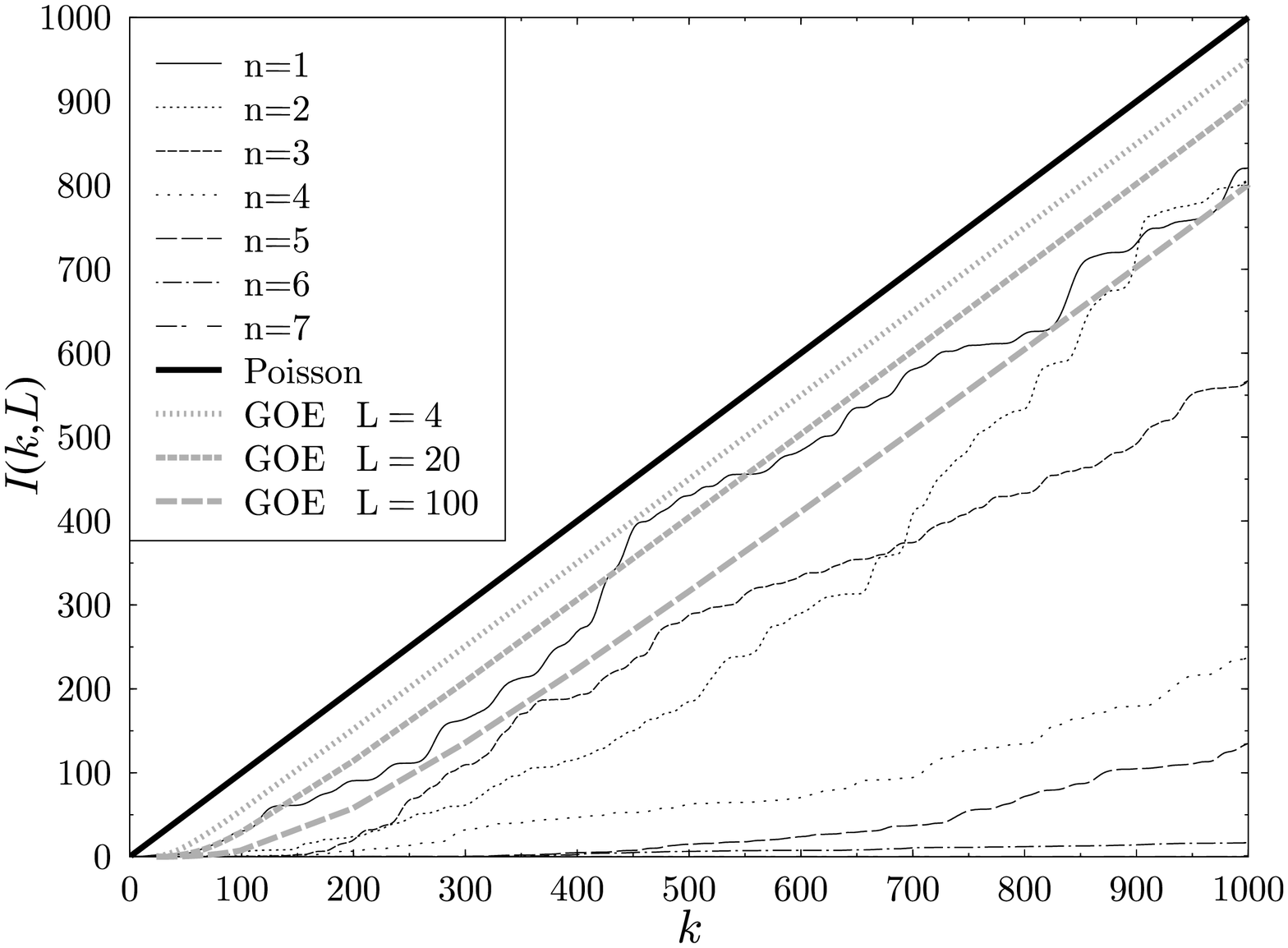,width=9cm} &
      \psfig{figure=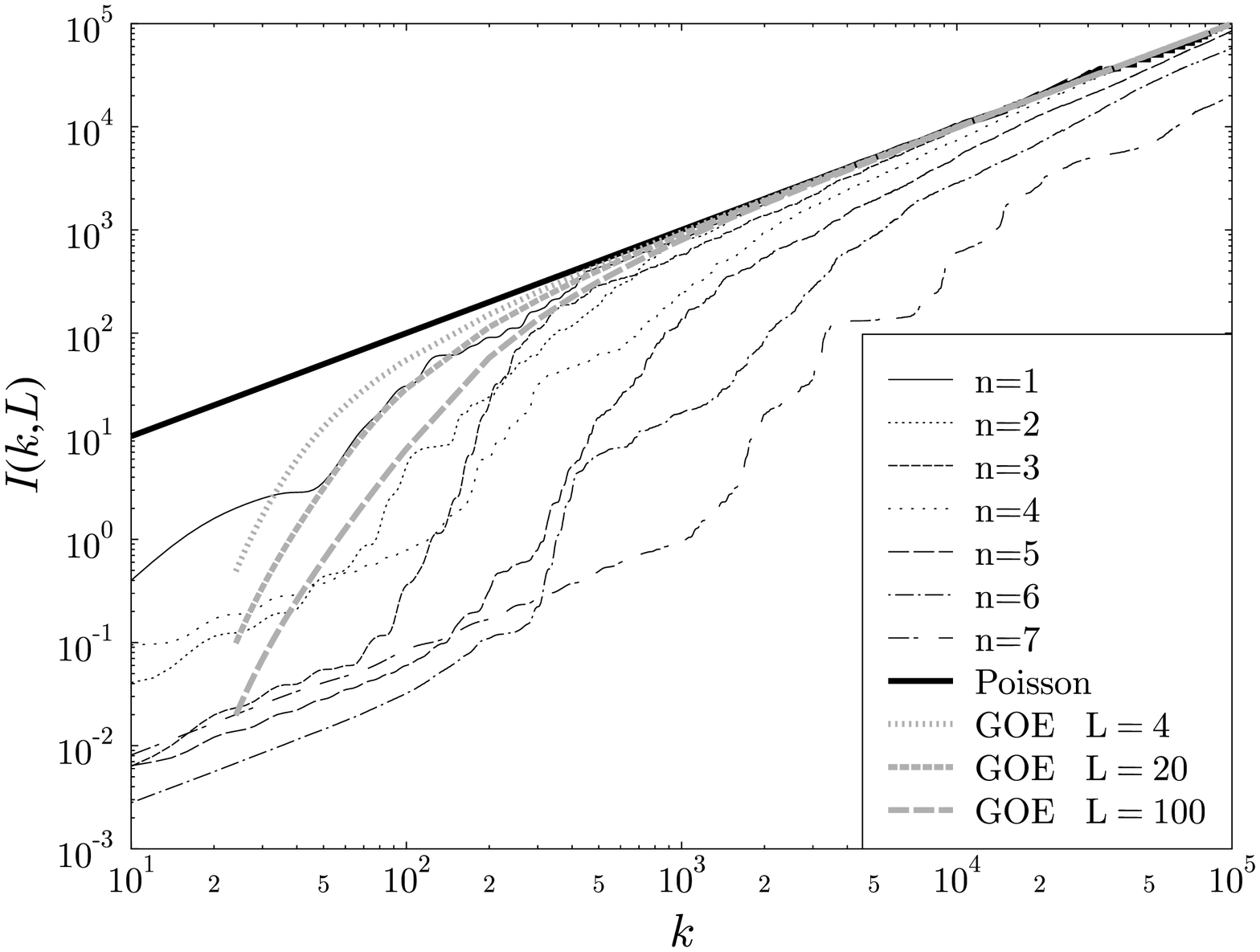,width=9cm}
    \end{tabular}
    \caption{The integrated correlation factor $I(k, L)$ for 3D Sinai 
      billiard with $L=5$ fixed and $n$ varied.}
    \label{fig:ickl-sb3d-l5}
  \end{center}
\end{figure}

As for the 2D case, we classify the POs according to the
number of their bounces $n$ from the inscribed sphere. The a-priori
reason is that the amplitudes are very strongly correlated with $n$
(much more than in the 2D case) and the division into the $n$-classes
due to this hierarchy is plausible. In figure~\ref{fig:ickl-sb3d-n123}
we present the function $I(k, L)$ for $n = 1, 2, 3$ and various
lengths.
The numerical results leave no doubt about the existence of classical 
PO-correlations. All of the data curves are strikingly different from the
Poisson (lack of correlations) case. The division into classes turns
out to be very useful in revealing very prominent correlations.
Closer look at the data shows that the numerical correlations are
generally weaker than GOE for low $k$'s ($k \lesssim 30$) and are
stronger otherwise. One might interpret this discrepancy as due to a
non-generic effect (that prevails at least for low $L$'s), but
nevertheless we bear in mind that in the $L$-regime in which we worked
the $n=1$ term was dominant, and its correlations are not so different
from the universal ones. Also, there is no clear and systematic
$L$-dependence. The numerical data exhibit large dispersion that
prevents us from concluding anything definite concerning the
$L$-dependence.
It is therefore useful to examine the correlations for a fixed $L$ and
various $n$'s. This is done in figure~\ref{fig:ickl-sb3d-l5}, where we
fixed $L=5$ and examined the correlations for $n = 1, \ldots, 7$.
This figure clearly indicates the systematic increase of correlations
with increasing $n$.

\begin{figure}
  \begin{center}
    \leavevmode
    \begin{tabular}{cc}
      \psfig{figure=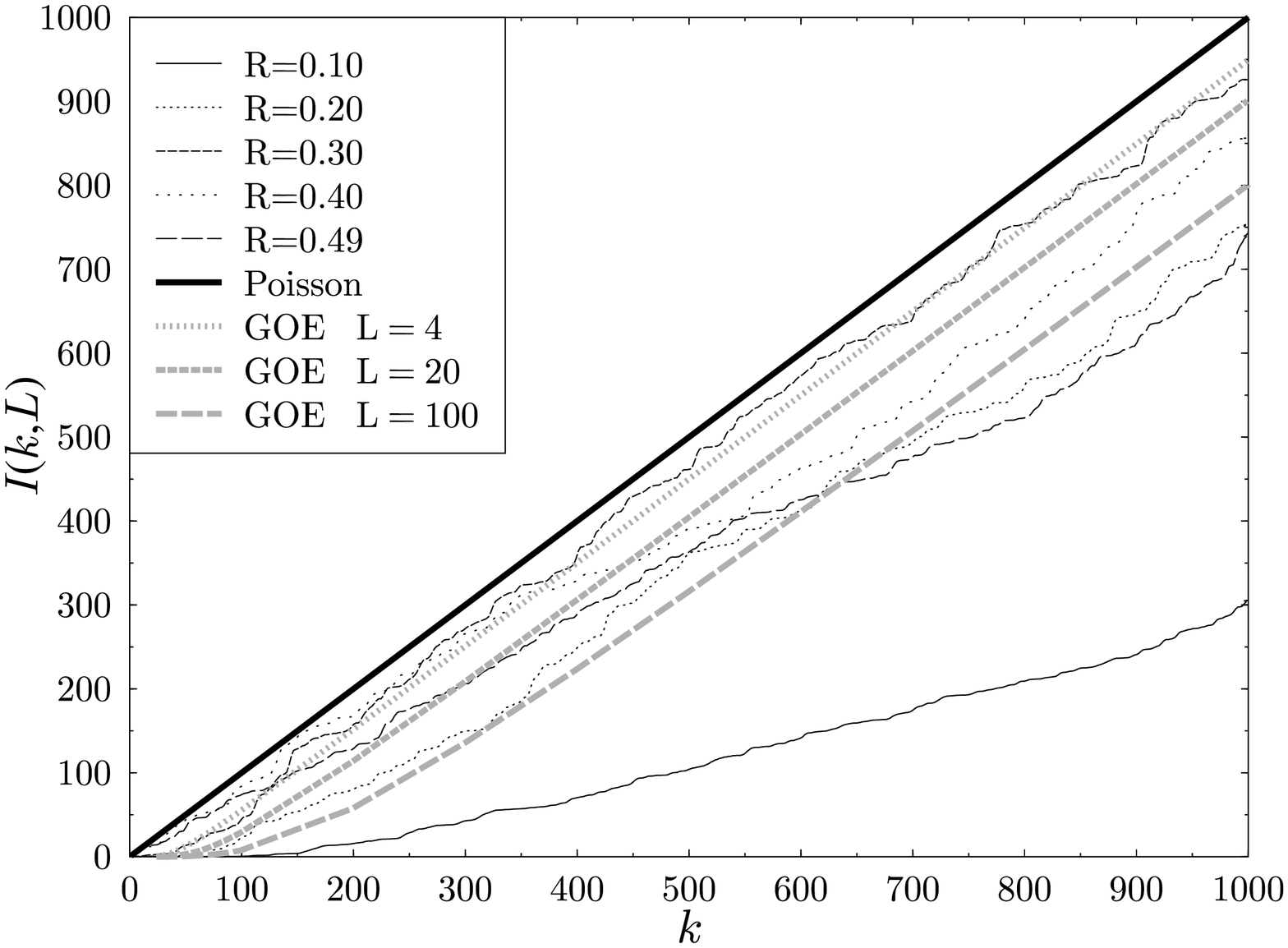,width=9cm} &
      \psfig{figure=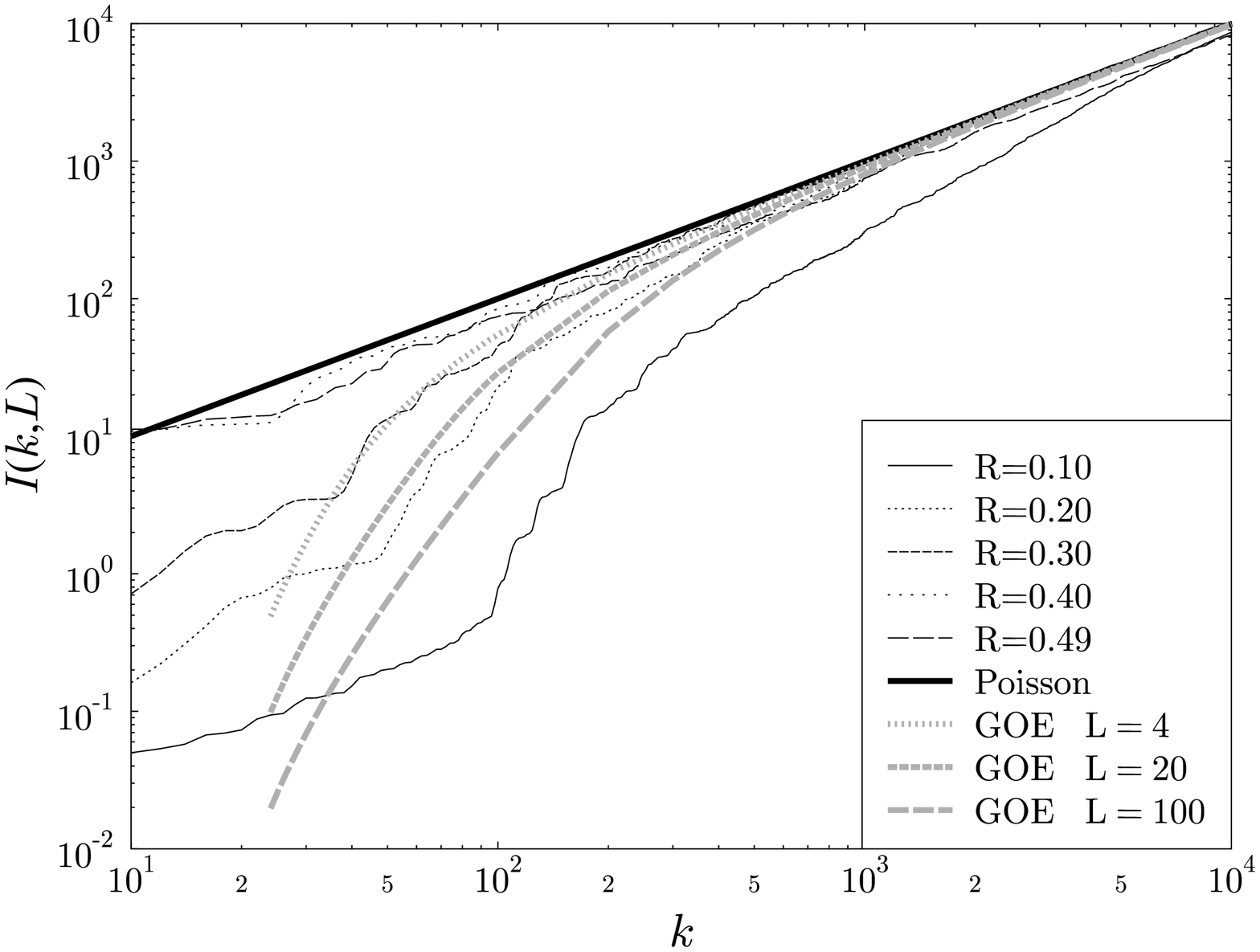,width=9cm}
    \end{tabular}
    \caption{The integrated correlation factor $I(k, L)$ for 3D Sinai 
      billiard with $L=10$ and $n=2$ fixed and $R$ varied.}
    \label{fig:ickl-sb3d-r}
  \end{center}
\end{figure}

\begin{figure}
  \begin{center}
    \leavevmode
    \psfig{figure=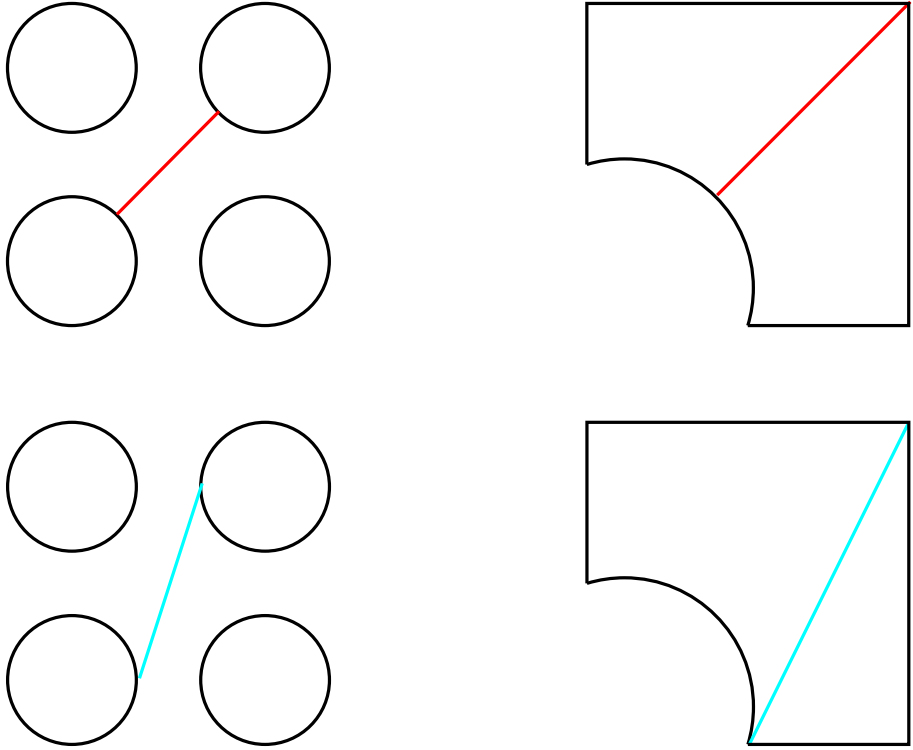,height=10cm} 
    \vspace*{1cm}
    \caption{Two different POs with the same coding. 
      The POs visit the same discs, but correspond to two
      different symmetry elements. The illustration is for the 2D
      Sinai billiard, but the same principle applies also to the 3D
      Sinai billiard.}
    \label{fig:sb2d-desym}
  \end{center}
\end{figure}

To get a hint about the source of the PO-correlations, we have
fixed $L=10$ and $n=2$ and varied the radius of the inscribed sphere.
The results are shown in figure~\ref{fig:ickl-sb3d-r}.
We clearly observe the growth of the correlations as $R$ decreases.
This effect is qualitatively explained by the following argument. 
Given a PO of the full 3D Sinai billiard (on the torus),  
it can be described by listing the locations of the centers 
of the spheres from which it bounces.   
One should use for this purpose the unfolded representation 
of the billiard (see figure~\ref{fig:sb3d}).
This list of ``addresses'' defines the code word $W$ which is known to 
be unique \cite{Bun95,Sch96}. 
This uniqueness property does not hold for the desymmetrized Sinai 
billiard. Here, the code word $W$ does not determine the PO uniquely.
Periodicity is required in the desymmetrized domain, hence one has to
determine {\em additionally} the symmetry element $\hat{g}$ according 
to which the PO closes on itself in the unfolded representation. 
This is illustrated in figure \ref{fig:sb2d-desym}.
For the 3D Sinai billiard, the symmetry group is $O_h$
\cite{TinkhamBook} which contains 48 elements. Therefore, there are up
to 48 POs that visit the same spheres in the unfolded
domain. We denote this set as $\gamma(W)$. If the radius $R$ is small
enough, the lengths of all of the orbits in $\gamma(W)$ will be
similar, and the orders of magnitude of their instability amplitudes
are expected to be the same. The signs of the amplitudes within
$\gamma(W)$ will differ, however, according to the parity of the
symmetry element. Hence we expect a better and better cancelations
within each $\gamma(W)$ as $R \rightarrow 0$. This qualitatively
explains the effect seen in figure~\ref{fig:ickl-sb3d-r}.

\begin{figure}
  \begin{center}
    \leavevmode
    \begin{tabular}{cc}
      \psfig{figure=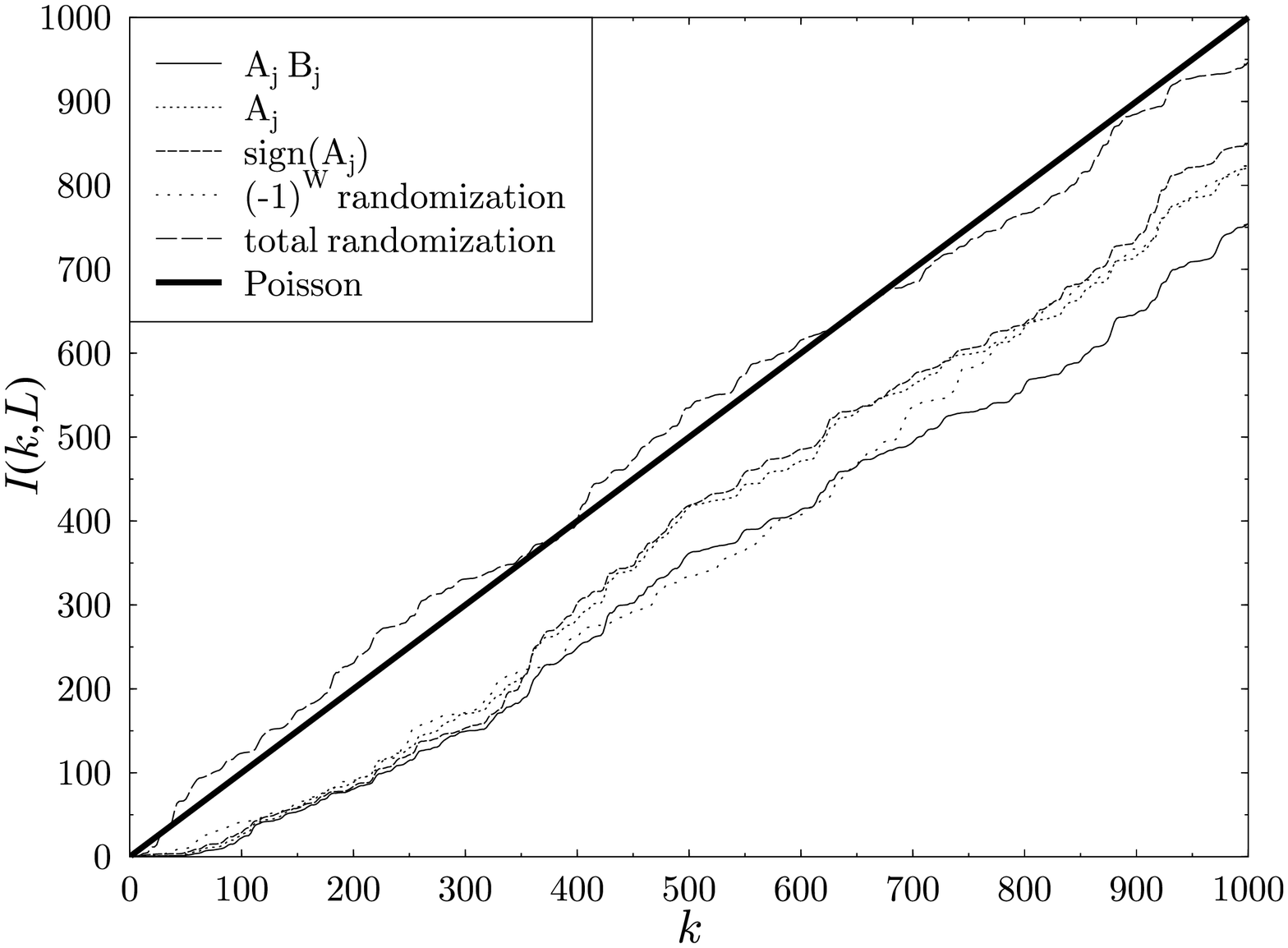,width=9cm} &
      \psfig{figure=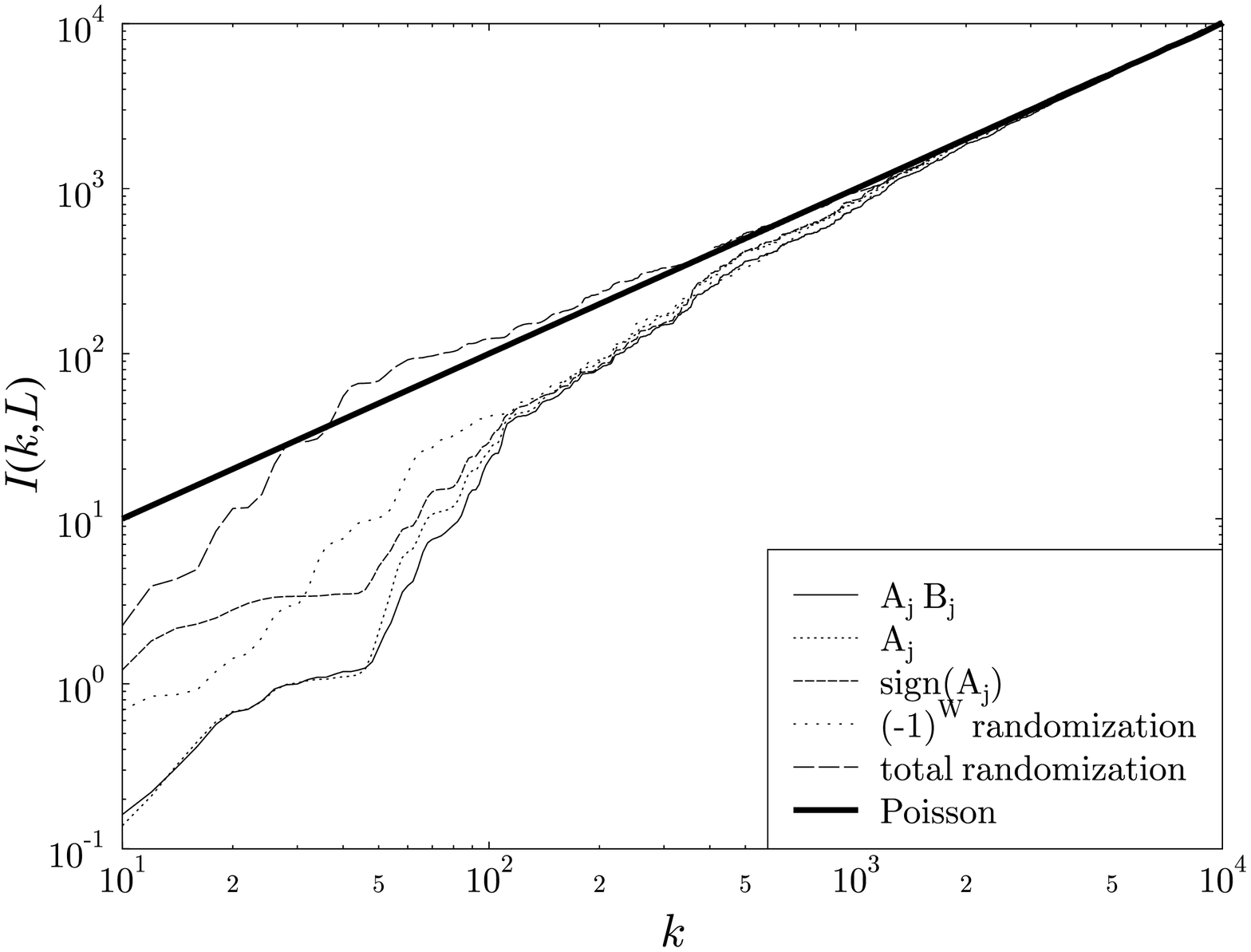,width=9cm} \\
      \psfig{figure=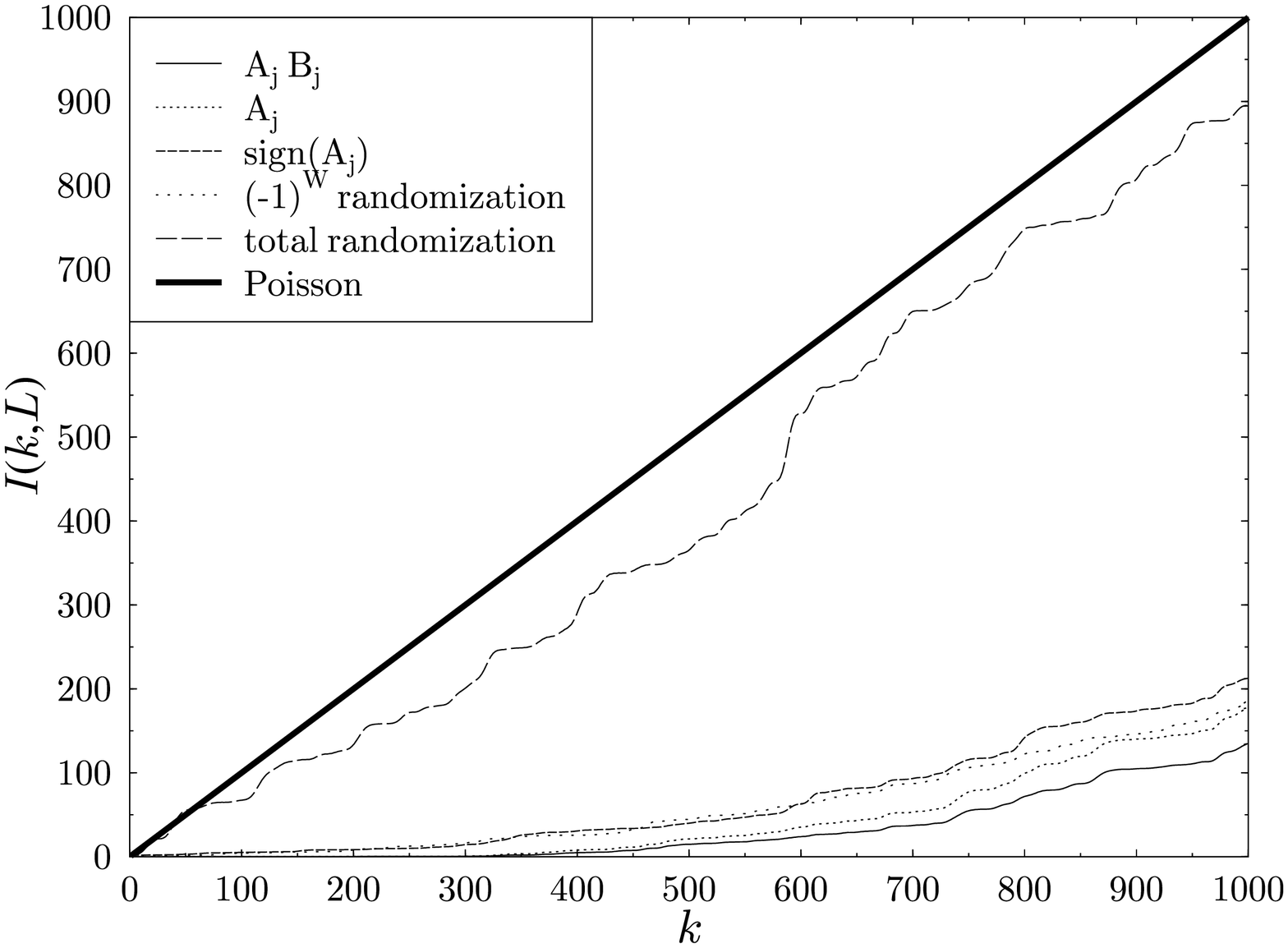,width=9cm} &
      \psfig{figure=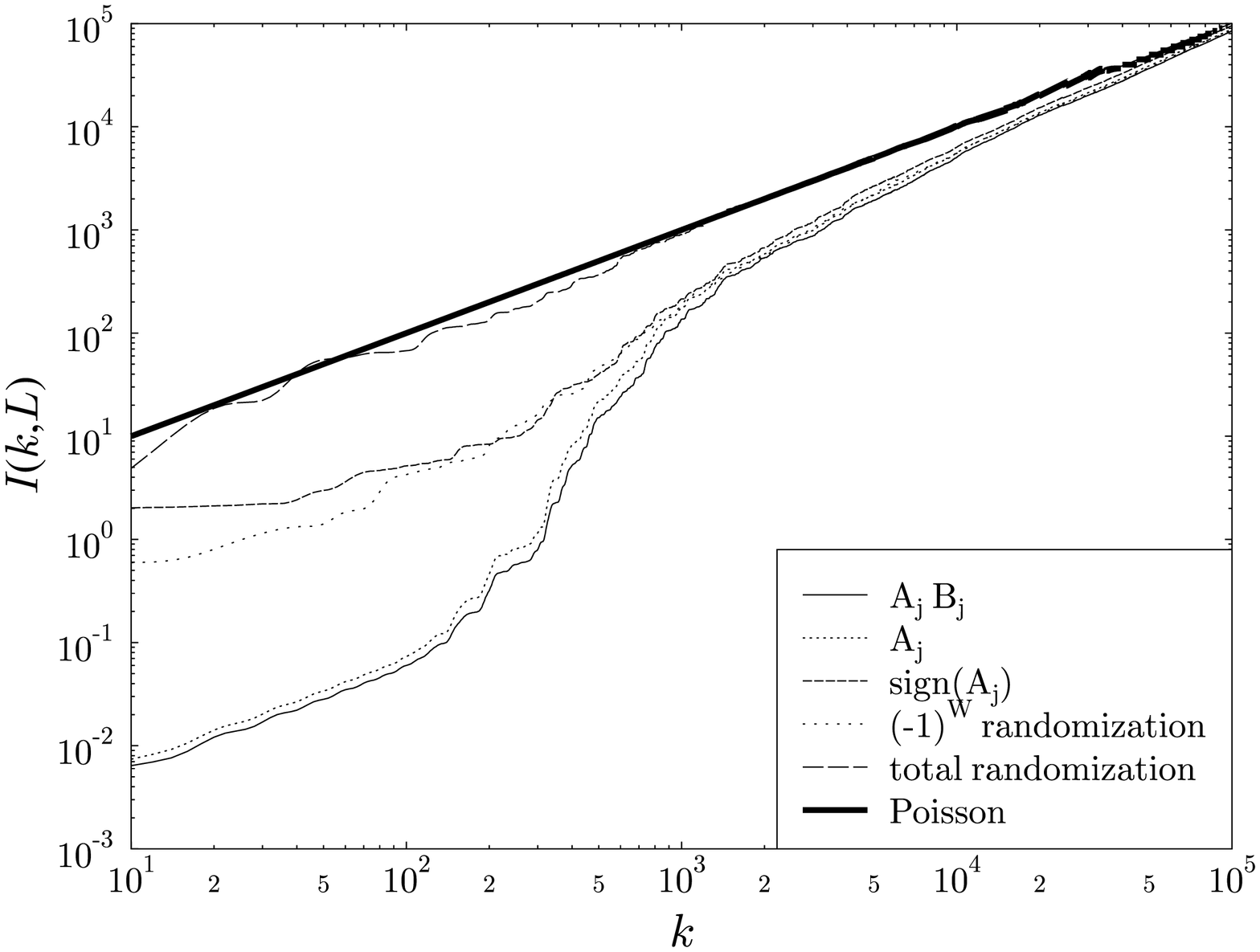,width=9cm}
    \end{tabular}
    \caption{Controlled breaking of correlations. See text for details. 
      In both cases $R=0.2$. Upper plots: $L \approx 10, n = 2$. 
      Lower plots: $L\approx 5, n = 5$.}
    \label{fig:break-corr}
  \end{center}
\end{figure}

To test this argument we performed a numerical experiment, in which we
gradually destroyed the possible correlations in a controlled way. The
results are displayed in figure~\ref{fig:break-corr}.
Two cases were considered: $R=0.2, L \approx 10, n = 2$ and $R = 0.2,
L \approx 5, n = 5$. We first plotted the full correlations in which
the coefficients $A_j B_j$ are taken into account 
(see (\ref{eq:mbcft})). 
As in previous cases, strong classical PO-correlations are
observed. We next disregarded the $B_j$ factors, and re-calculated
$I(k, L)$ using only $A_j$'s. As clearly seen from the figure, the
changes were minor. Next, we replaced the $A_j$'s with their signs
$\pm 1$, disregarding their magnitude. Correlations were somewhat
diminished, but one can say with full confidence that strong and
significant correlations still prevail. This proves, that the
coefficients are responsible to part of the correlations, but more
importantly, it proves that there are correlations between the
actions, regardless of the coefficients. This gives a justification to
our point of view taken in the theoretical part, in which we regarded
the $A_j$'s as ``slowly varying''. The following step was to perform a
controlled randomization of the $\pm 1$ coefficients between
$\gamma(W)$ sets. We picked each set of orbits $\gamma(W)$ and either
kept the original signs or reversed all of them. This means, that we
kept the relative sign within each $\gamma(W)$. The definite result
was that PO-correlations persisted! This proves, that the
correlations between POs are largely due to correlations within
the sets $\gamma(W)$. To complete the argument, we randomized 
also the signs within each $\gamma(W)$, and indeed correlations were
completely destroyed. The overall conclusion is that we were able to
identify the grouping of orbits into sets with the same code word $W$
but with different symmetry $\hat{g}$ as a prominent source of the 
classical PO-correlations in the 3D Sinai billiard.

\begin{figure}
  \begin{center}
    \leavevmode 
    \psfig{figure=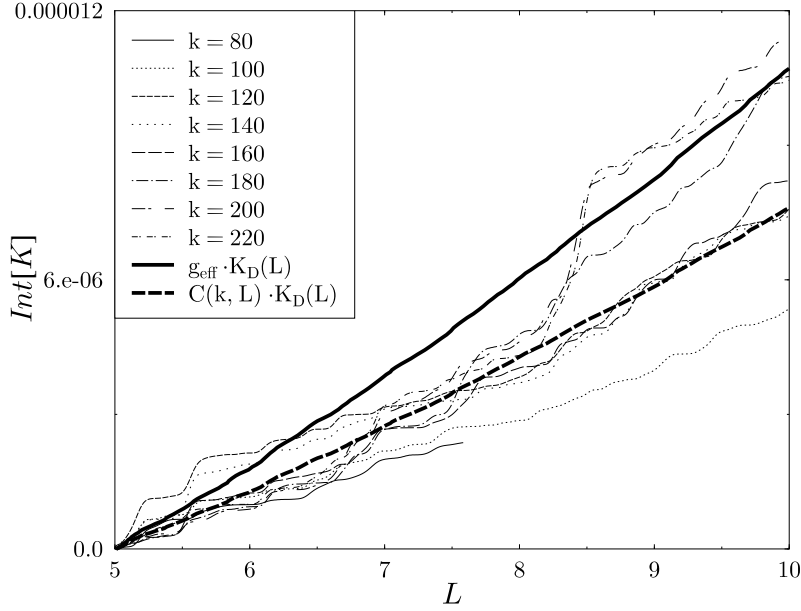,width=12cm}
    \caption{Comparison between the quantal and the 
             classical form factors for
             3D Sinai billiard with $R=0.2$.}
    \label{fig:miktqm} 
  \end{center}
\end{figure}

The above investigation of PO-correlations was purely classical.
It is therefore very interesting to enquire whether the quantal levels
``feel'' the observed classical PO-correlations. To do this, we
compared $K_{qm}(k, L)$ (based on the left hand side of
(\ref{eq:mbcft})) to $g_{\mbox{eff}}K_{D}(L)$ (based on the right hand
side of the same equation). The quantal from factor $K_{qm}$ was
computed from the exact quantal levels and velocities, and
$g_{\mbox{eff}}K_{D}(L)$ was computed from the classical PO data 
(we did not assume any sum rule). The results are shown in
figure~\ref{fig:miktqm}. Note, that we integrated the resulting form
factors in $L$ to diminish fluctuations, and started the integration
at $L=5$ where the density of POs is already very high.
Evidently, the classical diagonal approximation $g_{\mbox{eff}}K_{D}$
is consistently and significantly larger than $K_{qm}$ in the $(k, L)$
regime that we checked. Strong classical non-trivial correlations are
therefore needed in order to ``bend'' $g_{\mbox{eff}}K_D$ down to the
correct $K_{qm}$. Indeed, in the relevant $(k, L)$ domain, the
classical correlation factor $C(k, L)$ for the dominant $n=1$
contribution is $\approx 0.7 g_{\mbox{eff}}$ (see figure
\ref{fig:ickl-sb3d-n123}). Quantitatively, if we multiply
$g_{\mbox{eff}}K_D$ by the numerical factor $0.7$, then we get on 
the average compliance with
the quantum results, as seen in figure~\ref{fig:miktqm}. Therefore, we
have demonstrated a case in which classical correlations are clearly
manifest in the quantum mechanical spectral statistics. 

%
%

\rule{18cm}{.01in} \\

{\bf\Huge  Appendices}


\def\thesection{\Alph{section}}
\def\theequation{\Alph{section}.\arabic{equation}}
\setcounter{section}{0}

\section{Effect of Deformation on PO-s Statistics}  
\setcounter{equation}{0}

Consider a billiard and a particular PO. Assume
that the boundary of the billiard in the vicinity 
of {\em one} hitting point is deformed, so that the 
normal shift is $b$. Assuming no bifurcation, the
length of the PO will become 
$L_j \rightarrow L_j+\delta_j$, where 
$\delta_j=2b\cos(\theta_j)$. The incidence angle $\theta_j$ 
is measured with respect to the normal direction 
at the point where the PO hits the deformed surface. 
In order to prove this result let us define 
a coordinate $x$ along the boundary (without loss
of generality we consider a planar 2D system).    
The hitting points are $(x_1,x_2,...,x_n)$. The PO 
constitutes an extremum of the length functional 
${\cal L}(x_1,x_2,...,x_n)$, thus 
$\delta {\cal L}/\delta x_s = 0$ for the PO.          
For a deformation 
$\delta{\cal L}= (\delta {\cal L}/\delta b){\cdot}b
+\sum_s(\delta {\cal L}/\delta x_s){\cdot}
(\delta x_s/\delta b){\cdot}b = 
(\delta {\cal L}/\delta b){\cdot}b$. 
Hence, in order to derive the required result,  
it is legitimate to {\em fix} all the hitting points 
and to consider merely the length change of the chords 
due to the deformation. This final step is trivial.  

We now wish to consider the effect of general 
deformation on the two point statistics. Assuming 
that there are no bifurcations, the new 
lengths are $L_j \rightarrow L_j+\delta_j$. 
Treating the differences $y=\delta_j{-}\delta_i$ 
as {\em realizations of independent random variables} 
with probability distribution $G(y)$, one obtains
$\langle\sum' A_i^*A_j
\delta(x-(L_j{-}L_i)-(\delta_j{-}\delta_i))   
\rangle = 
\int dy G(y) \langle \sum' A_i^*A_j 
\delta(x-y-(L_j{-}L_i))\rangle$.
The prime implies omission of the diagonal terms $i=j$. 
It is implicit that the sum is restricted to some 
$L$-window. From here one easily derive 
the result $C(k)\rightarrow [1-\tilde{G}(k){\cdot}(1{-}C(k))]$, 
where $\tilde{G}(k)$ is the FT of $G(y)$.
     
The case of symmetry breaking due to deformation
that preserves family-structure should be treated
somewhat differently. Here it is convenient to 
write the change of length as 
$L_j \rightarrow L_j+\delta_{\alpha r}$, where 
$\alpha$ identifies families of the {\em folded}
system and $r$ distinguish between families 
of the {\em unfolded} system. {\em Disregarding 
cross-correlations between families} the double 
sum can be written as 
$\langle\sum_{\alpha}\sum_{ij}\sum_{rs} A_i^*A_j \delta
(x-(L_j{-}L_i)-(\delta_{\alpha r}{-}\delta_{\alpha s}))   
\rangle$.   
The $ij$ summation is over POs of the {\em folded} 
system that belong to the $\alpha$ family. 
This sum can be manipulated as before, note 
that it is essential to separate the $r\ne s$ from
the $r=s$ terms, yielding  
$ \int dy \langle \sum_{ij}\sum_{rs} A_i^*A_j 
\delta(x-y-(L_j{-}L_i))\cdot
[\delta_{rs}\delta(y)+(1{-}\delta_{rs})G(y)] \rangle
=  \int dy \langle \sum_{ij} A_i^*A_j 
\delta(x-y-(L_j{-}L_i))\cdot g
[\delta(y)+(g{-}1)G(y)] \rangle $.
From here one easily derive the result 
$C(k)\rightarrow [1+(g{-}1)\tilde{G}(k)]{\cdot}C_0(k)$.

\section{Diagonal Sums for the Chain Problem}  
\setcounter{equation}{0}

We consider a chain that consists of $N$ cells. 
The average escape time out of a cell is 
found as in section V-C, yielding 
$1/t_0=2 \times a_0^{d_0{-}1}/\Omega_0$. Here 
$\Omega_0$ is the volume of a single cell, 
and $a_0^{d_0{-}1}$ is the cross section of 
each of the two `holes' that connect the cell
to the other cells. Let $x$ be an integer 
coordinate that identifies the cells, and 
let $p(x)$ be the probability for finding 
the particle in the $x$-cell. This probability 
satisfies the equation
\begin{eqnarray}
\frac{d p(x)}{dt} \ = \ -\frac{1}{2t_0}
[2p(x)-p(x{-}1)-p(x{+}1)] \ \approx \
-\frac{1}{2t_0}\frac{\partial^2}{\partial x^2}p(x)
\ \ \ \ .
\end{eqnarray}
Above the difference equation is approximated by 
a diffusion equation. This approximation is valid 
on time scales such that $p(x)$ is smeared over 
many cells.

For an open chain of $n$ cells ($n<N$) we solve this 
equation with the boundary conditions $p(n/2)=p(-n/2)=0$, 
meaning that there is a zero probability to return once 
the particle escapes from the endpoints (`absorption').
We denote by $p(x|x')$ the solution, where $x'$ is 
the cell where the particle is initially prepared.  
In order to solve the diffusion equation one should 
define an appropriate set of orthonormal functions, 
which is in the present case:
\begin{eqnarray}
\varphi_s(x) \ = \ \left\{ \matrix{
\sqrt{2}\cos(\pi s x/n) & \ \ \ \mbox{for} \ \ s=1,3,5,... \cr
\sqrt{2}\sin(\pi s x/n) & \ \ \ \mbox{for} \ \ s=2,4,6,... }
\right.
\end{eqnarray}
Then we obtain the solution
\begin{eqnarray}
p(x|x') \ = \ \sum_{s=1}^{\infty}
\mbox{e}^{-(\pi^2 s^2 t)/(2n^2 t_0)}
\ \varphi_s(x')\varphi_s(x)
\end{eqnarray}
The classical probability to return is simply
\begin{eqnarray}
p_n(t) \ = \ \int_{-n/2}^{n/2}p(x|x)dx \ = \ 
\sum_{s=1}^{\infty}\exp\left(-\frac{\pi^2 s^2 t}{2 n^2 t_0} \right) 
\ \  \ .
\end{eqnarray}
If the chain were {\em closed} then 
$p(x|x')\rightarrow 1/n$ for $t\rightarrow \infty$, 
and consequently one would obtain the 
asymptotic ergodic result $p_n(t)\rightarrow 1$.
However, for an {\em open} chain the particle will escape 
eventually via the endpoints-holes, leading to 
the asymptotic result $p_n(t)\rightarrow 0$. 
Specifically, for very long times $n^2t_0\ll t$ 
one obtains $p_n(t)\approx\exp(-\pi^2t/(2n^2t_0))$. 
For intermediate times ($t\ll n^2t_0$ but still $t_0\ll t$), 
one obtains $p_n(t)\approx n/\sqrt{2\pi(t/t_0)}$. 
The latter result is not sensitive to the 
endpoint conditions and is valid also for a closed chain.
For short times ($t$ of the order $t_0$) the 
above solution is valid qualitatively but not  
quantitatively. In particular, for $n=1$ one should
obtain the one-cell result $p_1(t)=\exp(-t/t_0)$.    

We turn now to the estimate of $K_n(t)$ which is 
the diagonal sum that corresponds to a class of POs 
that explore $n$ specified cells. 
For a {\em closed chain} there are $N$ classes that 
correspond to each $n<N$, and one class that correspond 
to $n=N$. The sum $K_N(t)+\sum_{n<N}NK_n(t)$  
should be equal to $tP_{cl}(t)$.  
For $t\ll N^2t_0$ we have the result   
$P_{cl}(t)=p_N(t)=N/\sqrt{2\pi(t/t_0)}$, and  
furthermore the first term $K_N(t)$ is negligible. 
For any finite $t$ we may take the limit 
$N\rightarrow\infty$ thus obtaining the sum-rule
\begin{eqnarray} \label{e_rule1}  
\sum_{n=1}^{\infty}K_n(t) \ = \ \frac{t}{\sqrt{2\pi(t/t_0)}}
\ \ \ . 
\end{eqnarray}
For an {\em open chain} that consists of $n$ cells
there are $(n{-}n'{+}1)$ classes of POs that 
explore $n'$ cells. Thus we obtain the sum rule
\begin{eqnarray} \label{e_rule2} 
\sum_{n'=1}^{n} (n{-}n'{+}1) K_{n'}(t) \ = \ t \cdot p_n(t)
\ \ \ . 
\end{eqnarray}
The sum rule  (\ref{e_rule1}) is actually a 
special case of (\ref{e_rule2}), it is obtained 
by considering finite $t$, taking the limit 
$n\rightarrow\infty$ and neglecting $(-n'+1)$. 
Using the above sum rules we may obtain an 
explicit expressions for the functions $K_n(t)$.

At this stage it is useful to note the scaling 
properties of $K_n(t)$. 
Looking on POs of length $t$, the probability 
for finding a PO that explores $n$ cells
should be proportional to $K_n(t)$.
It is obvious that this 
probability should depend on the scaled 
variable $n/\sqrt{t/t_0}$. Taking into account 
the sum-rule (\ref{e_rule1}), 
one obtains the scaling relation 
\begin{eqnarray} \label{e_fx}
K_n(t) \ = \ \frac{t_0}{\sqrt{2\pi}} 
f\left(\frac{n}{\sqrt{t/t_0}}\right)
\ \ \ \ \ .
\end{eqnarray}   
Above, $f(\tau)$ is the scaled probability function. 
It should have the normalization $1$. 
Substitution of (\ref{e_fx}) into (\ref{e_rule2}), 
and changing from summation over $n$ to integration 
over $x=n/\sqrt{t/t_0}$ one obtains 
$\int_0^x(x{-}x')f(x')dx'=\sqrt{2\pi}
\sum_{s=1}^{\infty}\exp(-\pi^2 s^2/(2x^2))$. 
Hence
\begin{eqnarray} 
f(x) \ = \ \frac{d^2}{dx^2}\left[\sqrt{2\pi} \sum_{s=1}^{\infty}
\exp\left( -\frac{1}{2x^2}(\pi s)^2 \right) \right]
\ = \ \frac{d^2}{dx^2} F(x)
\ \ \ \ \ ,
\end{eqnarray} 
where, by Poisson summation
\begin{eqnarray} 
F(x) \ = \ -x + \sqrt{\frac{\pi}{2}} 
\sum_{s=-\infty}^{\infty}
\exp\left( -\frac{1}{2x^2}(\pi s)^2 \right) 
\ = \ 2x \sum_{r=1}^{\infty}
\exp\left( -2 x^2 r^2 \right) 
\ \ \ \ \ .
\end{eqnarray} 
The Function $F(x)$ satisfies $F(0)=\sqrt{\pi/2}$ 
and $F(\infty)=0$ and the corresponding derivatives 
are  $F'(0)=-1$ and  $F'(\infty)=0$. It follows 
that $\int_0^{\infty}f(x)dx = 1$ while 
$\int_0^{\infty}xf(x)dx = \sqrt{\pi/2}$.
The function $f(x)$ is plotted in Fig.\ref{f_explore}. It is
peaked around $x \sim 1.2$ and it vanishes for either 
$x\rightarrow 0$ or $x\rightarrow\infty$.  


\section*{Acknowledgments}

We thank Michael Berry, Thomas Dittrich, Bruno Eckhardt, 
Shmuel Fishman, Martin Gutzwiller, Jon Keating, Holger Schanz
and Martin Sieber for helpful discussions.
The research reported here was supported in part by  
the Minerva Center for Nonlinear Physics of Complex systems, 
and by the Israeli Science Foundation.


\begin{thebibliography}{99}
\bibitem{r_SCTF} 
To obtain (\ref{e2_6}) one should 
use the SCTF with $k$ replaced 
by $k'$, and multiply both sides 
by $w(k'{-}k)$ where $k$ is the
center of the window. Then one 
should perform a Fourier transform
in the variable $k'$ to the variable $L$.
\end{thebibliography}
\end{document}